\documentclass[11pt]{article}

\usepackage[usenames,dvipsnames,table]{xcolor}
\usepackage[utf8]{inputenc}

\pdfoutput=1 
\usepackage{jheppub} 

\usepackage{graphicx}
\usepackage{amsmath, amssymb}

\newcommand{\be}{\begin{eqnarray}}
\newcommand{\ee}{\end{eqnarray}}
\newcommand{\bea}{\begin{eqnarray}}
\newcommand{\eea}{\end{eqnarray}}

\newcommand{\nn}{\nonumber}
\newcommand{\bn}{\begin{enumerate}}
\newcommand{\en}{\end{enumerate}}

\def\Tr{\mathop{\mathrm{Tr}}\nolimits}

\def\ga{\alpha}

\def\Gd{\Delta}

\def\gs{\sigma}

\title{Flips, dualities and symmetry enhancements}

\author[a,b]{Chiung Hwang}
\author[a]{Sara Pasquetti}
\author[a]{Matteo Sacchi}

\affiliation[a]{Dipartimento di Fisica, Universit\`a di Milano-Bicocca \& INFN, Sezione di Milano-Bicocca,
I-20126 Milano, Italy}
\affiliation[b]{Department of Applied Mathematics and Theoretical Physics, University of Cambridge, Cambridge CB3 0WA, United Kingdom}

\emailAdd{chiung.hwang@unimib.it}
\emailAdd{sara.pasquetti@gmail.com} 
\emailAdd{m.sacchi13@campus.unimib.it}

\abstract{We present various 4d $\mathcal{N}=1$ theories enjoying IR global symmetry enhancement. The models we consider have the $USp(2 N)$ gauge group, 8 fundamental, one antisymmetric chirals and various numbers of gauge singlets. By suitably turning on superpotential deformations involving the singlets which break part of the UV symmetry we flow to SCFTs with $E_6$, $SO(10)$, $SO(9)$, $SO(8)$ and $F_4$ IR global symmetry. We explain these patterns of symmetry enhancement following two arguments due to Razamat, Sela and Zafrir. The first one involves the study of the relations satisfied by marginal operators, while the second one relies on the existence of self-duality frames.
}

\begin{document} 

\maketitle
\flushbottom

\section{Introduction}

The 4d $\mathcal{N}=1$ $SU(2)$ gauge theory with 8 fundamental chiral fields
admits 72 dual frames which are rotated into each other by the  action of the coset group $W(E_7)/S_8$ \cite{Spiridonov:2008zr}.
In addition to the original description, there are 35 Seiberg dual frames \cite{Seiberg:1994pq}, 35 Csaki-Schmaltz-Skiba-Terning (CSST) dual frames \cite{Csaki:1997cu} and 1 Intriligator-Pouliot (IP) dual frame \cite{Intriligator:1995ne}. Since the theory has 8 fundamental chirals without any superpotential, it preserves the $SU(8)$ global symmetry. In the Seiberg and CSST dual frames, however, this $SU(8)$ symmetry is broken to $SU(4) \times SU(4) \times U(1)$ in the UV whereas it is restored at the IR fixed point.

In \cite{Spiridonov:2008zr} it was shown that those 72 dualities form an orbit of $W(E_7)/S_8$ and also found that this structure persists for higher rank $USp(2 N)$ theories provided an extra matter in the traceless antisymmetric representation of the gauge group $USp(2 N)$  is added. In the higher rank case 35 frames correspond to the duality
 discussed in \cite{Csaki:1996eu} while the other frames correspond to generalizations of Seiberg and IP dualities.

It is natural to wonder whether it is possible to construct theories which actually display $E_7$, or other enhanced symmetries, rather than being rotated to a dual frame by the $E_7$ Weyl action. The first theories with $E_7$ were constructed in \cite{Dimofte:2012pd}.
This $E_7$ model as well as many other models with enhanced global symmetries can be realised geometrically
by compacftifying the 6d $\mathcal{N}=1$ SCFTs on a Riemann surfaces with fluxes for the global symmetry of the six-dimensional theory \cite{Gaiotto:2015usa,Razamat:2016dpl,Bah:2017gph,Kim:2017toz,Kim:2018bpg,Razamat:2018gro,
Chen:2019njf,Pasquetti:2019hxf,Razamat:2018gbu,Sela:2019nqa}. It is indeed expected that the subgroup of the 6d symmetry preserved by the flux will also be the global symmetry of the resulting 4d $\mathcal{N}=1$ SCFT. It also often turns out that the expected global symmetry is not visible from the UV quiver description  of the SCFT but it emerges in the  IR.

The 6d perspective allows us to make interesting predictions for models with symmetry enhancements in 4d. These can then be tested with a more direct 4d analysis, like the computation of the superconformal index \cite{Romelsberger:2005eg,Kinney:2005ej,Dolan:2008qi}.  On the other hand it is also possible to develop purely 4d QFT strategies to understand and predict symmetry enhancements.

In this note we follow two main strategies to discuss several models with $SU(2)$ gauge group and 8 chirals and  various amounts of singlets displaying $E_6$, $SO(10)$, $SO(9)$, $SO(8)$ and $F_4$ symmetry in the IR\footnote{Turning on suitable complex masses, these models flow to the WZ models with same global symmetry discussed in \cite{Razamat:2016gzx}.}.
We also construct higher rank versions of these models, showing that $\mathcal{N} = 1$ $USp(2N)$ theories  with 8 fundamental and one antisymmetric chirals, with various selections of singlets and superpotentials, display  $E_6$, $SO(10)$, $SO(9)$, $SO(8)$ and $F_4$ symmetries  in the IR.

The first strategy relies on the  relation between the symmetry enhancement and the chiral ring relations of marginal operators, which was discussed in \cite{Razamat:2018gbu} and can be summarised as follows.
As observed in \cite{Beem:2012yn}  conserved currents  and marginal operators contribute to the order $pq$ of the superconformal index:
\begin{align}
I = \dots+(\chi_\text{ind-mar}-\chi_\text{cur}) pq+\dots
\end{align}
where $\chi_\text{cur}$ and $\chi_\text{ind-mar}$ denote the characters of the representations of the conserved current and of the independent marginal operators with 
\begin{align}
\chi_\text{ind-mar}= \chi_\text{mar}- \chi_\text{rel}
\end{align}
 where $\chi_\text{rel}$ is  the character of the representation of the relations satisfied by the marginal operators. Let us assume there are marginal operators of the form $\mathcal O^{(1)} \mathcal O^{(2)}$, with $\mathcal O^{(1)} \neq \mathcal O^{(2)}$, in the representation $R^{(1)} \times R^{(2)}$ satisfying a relation
\begin{align}
\left.\mathcal O^{(1)} \mathcal O^{(2)}\right|_\tau = 0
\end{align}
where $\tau$ is the representation of the relation. For simplicity, it is assumed that $\mathcal O^{(1)} \mathcal O^{(2)}$ is neutral under any $U(1)$ global symmetry. Thus, the index contribution of $\mathcal O^{(1)} \mathcal O^{(2)}$ is given by
\begin{align}
(\chi_{R^{(1)}} \chi_{R^{(2)}}-\chi_\tau) pq \,.
\end{align}
One can remove those marginal operators by flipping either $\mathcal O^{(1)}$ or $\mathcal O^{(2)}$. Say we flip $\mathcal O^{(2)}$ by an extra flipping field $F$ in the representation $\overline R^{(2)}$. Then, we have additional contributions originating from $F$ and its supersymmetric partner $\overline \psi_F$ (where $\overline \psi_F$ denotes the fermionic  partner of the scalar $\overline F$):

\begin{align}
\nn F \mathcal O^{(2)} \quad &\sim \quad \chi_{\overline R^{(2)}} \chi_{R^{(2)}} pq \,, \\
\nn \mathcal O^{(1)} \overline \psi_F \quad &\sim \quad -\chi_{R^{(1)}} \chi_{R^{(2)}} pq \,, \\
F \overline \psi_F \quad &\sim \quad -\chi_{\overline R^{(2)}} \chi_{R^{(2)}} pq \,,
\end{align}
which cancel the contribution of $\mathcal O^{(1)} \mathcal O^{(2)}$ and leave $-\chi_\tau pq$. The remaining contribution $-\chi_\tau pq$, which used to be  the chiral ring relation of marginal operators,  now  joins the current.
 Thus, the total current multiplet contribution is enlarged to $-(\chi_\text{current}+\chi_\tau) pq$ and may form the adjoint representation of an enhanced global symmetry.

In section \ref{sec2}
we will extend this method and consider rank one models where only part of the marginal operators are removed. This has the effect of breaking the UV manifest symmetry to a subgroup, but we will gain a variety of interesting IR enhancements.\\

The second strategy was proposed in  \cite{Razamat:2017wsk} and
relies on the interplay between self-dualities and enhanced symmetries.
With ``self-duality" we mean that the dual theory has exactly the same gauge group, the same matter content (including gauge singlets) and the same superpotential of the original theory, but the two theories are related by a non-trivial map on the operator spectrum and on the global symmetries. The existence of self-dual frames implies that the theory is invariant in the IR under a larger set of transformations than the manifest UV symmetry group. In favorable situations, these additional transformations can lead to a symmetry enhancement in the IR. More precisely, if $G_{IR}$ is the IR  symmetry group, then we should be able to identify as many equivalent frames of the theory as the dimension of the Weyl group of $G_{IR}$. In the case in which part of this symmetry is enhanced in the IR, that is $G_{UV}\subset G_{IR}$, then we expect the transformations $W(G_{IR})/W(G_{UV})$ to come from non-trivial self-dualities of the theory while  the transformations of $W(G_{UV})$ are trivial invariances of the UV Lagrangian.

Reversing the argument, whenever we have self-duality frames we might expect an enhanced symmetry.
For example for the $USp(2 N)$ theories with 8 fundamental and one antisymmetric chiral fields mentioned above, by adding extra singlets, we might find a subset of the 72 dual frames which are actually self-dualities in the sense specified above and provide the missing frames to account for an enhanced IR symmetry.
Following this strategy   \cite{Razamat:2017wsk} constructed an $SU(2)$ theory with $E_6$ global symmetry
and \cite{Razamat:2017hda} a $USp(4 N)$ model with $E_7\times U(1)$ global symmetry, which for $n=1$ is related to the model of \cite{Dimofte:2012pd}.

In section \ref{sec3} we will apply this line of reasoning to the models of section \ref{sec2}, listing  the extra self-duality frames  accounting  for various types of enhancement and checking the superconformal index. 
We will also discuss various deformations which  
break some of the manifest UV symmetries leading to further interesting IR enhancement.
In \ref{hrt} we will present the higher rank version of these models involving $USp(2N)$ gauge groups.
In the appendix \ref{app1} we quickly revise the action of the  Seiberg,  CSST and IP dualities 
on the global symmetries. In appendix \ref{feusp} we will discuss the IR behavior of the $FE[USp(4)]$ theory, which is part of the family of $FE[USp(2N)]$ theories discussed in  \cite{Pasquetti:2019hxf}, arguing that it enjoys $SO(10)\times U(1)^2$ global symmetry.
Finally in appendix \ref{appPL} we study the  plethystic logarithm of the index to extract the relation satisfied by marginal operators. \\

\section{Flips, relations and enhanced global symmetries}
\label{sec2}

In this section we apply the first strategy to look for models with interesting symmetry enhancement patterns.
We begin by revisiting the discussion of the $E_7$ model of \cite{Razamat:2017hda} from this perspective.

Let's consider the  $USp(4)$ gauge theory with 8 fundamental chirals $Q_i$, $i=1,\ldots, 8$ one traceless antisymmetric chiral $X$ and a singlet $x_2$ with $\mathcal{W}=x_2 X^2$.
Its superconformal index is given by\footnote{We unrefine the fugacity for the $U(1)$ symmetry associated to the antisymmetric chiral.}
\begin{align}
I &= 1+{\bf 28} \, t^{-\frac12} (pq)^{\Delta_Q}+{\bf 28} \, t^\frac12 (pq)^{1-\Delta_Q}+t^{-2} (pq)^{1-\Delta_A} \nonumber \\
&\quad +({\bf 336}+{\bf 70}) \, t^{-1} (pq)^{2 \Delta_Q}+{\bf 28} \, t^{-\frac12} (p+q) (pq)^{\Delta_Q} \nonumber \\
&\quad +({\bf 378}+{\bf 336}-{\bf 63}-1) \, pq+\dots
\end{align}
where $\Delta_Q$ is the $R$-charge of the 8 fundamental chirals and $\Delta_A$ is that of the antisymmetric chiral. $\Delta_Q$ and $\Delta_A$ satisfy a relation $4 \Delta_Q+\Delta_A = 2$ which comes from requiring the existence of a non-anomalous R-symmetry. The first three terms are the contributions of the chiral ring generators\footnote{Throughout the paper $\Tr_g$ will denote the trace over the gauge indices. Since all the gauge groups that we will consider are of $USp(2N)$ type, these traces have to be intended with the insertion of the two-index totally antisymmetric tensor $J=\mathbb{I}_N\otimes i\,\gs_2$.}
\begin{align}
m_{0,ij} \equiv \Tr_{g} \left(Q_{i} Q_{j}\right) , \qquad m_{1,ij} \equiv \Tr_{g} \left(Q_{i} A \, Q_{j}\right) , \qquad x_2
\end{align}
respectively, where the first two are in the antisymmetric representation of the $SU(8)$ flavor symmetry while the last one is a singlet. In addition, one can see the current multiplet contribution $-({\bf 63}+1) \, pq$, which reflects the $SU(8) \times U(1)_t$ global symmetry. In this theory, the marginal operators are given by
\begin{align}
m_{0,ij} m_{1,kl}
\label{marginalusp4}
\end{align}
satisfying the relation
\begin{align}
m_{0,[ij} m_{1,kl]} = 0 \,,
\label{relationsusp4}
\end{align}
which can be explained as  follows \cite{Razamat:2017hda}. Consider the object $Q_i^aQ_j^bQ_k^cQ_l^dX_{ef}$, where both the $USp(4)$ gauge indices $a$, $b$, $c$, $d$, $e$, $f$ and the $SU(8)$ flavor indices $i$, $j$, $k$, $l$ are not contracted. We want to show that if we antisymmetrize all the flavor indices, there is no way of contracting the gauge indices to make a gauge invariant object. This is because $Q_{[i}^aQ_j^bQ_k^cQ_{l]}^d$, where all the flavor indices are antisymmetrized, transforms in the fourth antisymmetric power of the fundamental representation of $USp(4)$ (since the $Q$s are bosons), which is just a singlet. Hence, there is no way of multiplying this by the antisymmetric $A_{ef}$ and contracting the gauge indices so to make a non-vanishing gauge invariant object.
Thus, those in \eqref{marginalusp4} subject to the relation \eqref{relationsusp4} are $28 \times 28-70 = 378+336$ independent marginal operators.

 One can remove the marginal operators by flipping either $m_0$ or $m_1$, where the two choices are merely related by a duality. Once the marginal operators are removed, the 70 relations among them now join the current. Thus, the total number of conserved currents is 63+1+70 = 133+1, which form the adjoint representation of $E_7 \times U(1)$. The model  where $m_0$ is flipped by $M_0$ is exactly the model with  $E_7 \times U(1)$ global symmetry
 found in \cite{Razamat:2017hda}.\\

One may wonder whether we can analogously construct an $SU(2)$ model with $E_7$ symmetry. 
If we look at the index of the $SU(2)$ theory with 8 chirals and $\mathcal{W}=0$:
\begin{align}
I &= 1+{\bf 28} \, (pq)^\frac12+{\bf 28} \, (p+q) (pq)^\frac12+({\bf 336}-{\bf 63}) \, pq+\dots,
\label{indexpuresu2w8}
\end{align}
we note two things: first, the global symmetry is $SU(8)$ without additional $U(1)$ because there is no antisymmetric matter for the $SU(2)$ theory; second, the only chiral ring generators are the mesons $m_{0,ij} = \Tr_{g} \left(Q_{i} Q_{j}\right)$, which transform in the antisymmetric representation of such $SU(8)$ flavor symmetry. The marginal operators in this theory are given by 
\begin{align}
m_{0,ij} m_{0,kl}
\label{marginalsu2}
\end{align}
subject to the relation
\begin{align}
m_{0,[ij} m_{0,kl]} = 0 \,.
\label{relationssu2}
\end{align}
One way to see where these relations originate is along the lines of the argument we used for the $USp(4)$ gauge theory. If we consider the combination $Q_{[i}^aQ_j^bQ_k^cQ_{l]}^d$ where all the $SU(2)$ gauge indices are not contracted while the $SU(8)$ flavor indices are antisymmetrized, it should transform in the fourth antisymmetric power of the fundamental representation of $SU(2)$, which doesn't exist. Hence, we can't construct a gauge invariant object out of it.  Thus, those in \eqref{marginalsu2} subject to the relation \eqref{relationssu2} are $\frac{28 \cdot 29}{2}-70 = 336$ independent marginal operators. 

One may attempt to remove the marginal operators by flipping $m_0$ such that the 70 relations are translated into 70 conserved currents as above. However, even though we introduce new flipping fields, say $M_{0,ij}$, which flip $m_{0,ij}$ such that the original marginal operators are removed, $M_{0,ij}$ provide new marginal operators $M_{0,ij} M_{0,kl}$ subject to the same number of relations $M_{0,[ij} M_{0,kl]} = 0$. Thus, there is no change in the contributions of the relation and the current multiplet. This is because the assumption $\mathcal O^{(1)} \neq \mathcal O^{(2)}$ in the above argument fails to hold.
Indeed, a similar situation happens for higher odd ranks; there are operators of $R$-charge 1 whose squares give marginal operators which cannot be removed by the flipping of the operators of $R$-charge 1. Therefore, one cannot obtain an $E_7$ model for odd-rank theories, at least in this way.
\\

Since in $SU(2)$  models we cannot achieve the symmetry enhancement by  removing  completely the  marginal operators we can try to introduce
 flips of operators which break the $SU(8)$ global symmetry into subgroups and only partially remove the marginal operators. We will see that while those partial flips reduce the manifest global symmetry in the UV,  they eventually lead to intriguing patterns of symmetry enhancements in the IR.

Since the partial flips break the UV global symmetry, they can be organised along the line of the symmetry breaking pattern of the $SU(8)$ global symmetry. The maximal subgroups of $SU(8)$ are:
\begin{gather}
\begin{gathered}
\label{eq:maximal subgroups}
SU(7) \times U(1) \,, \\
SU(6) \times SU(2) \times U(1) \,, \\
SU(5) \times SU(3) \times U(1) \,, \\
SU(4) \times SU(4) \times U(1) \,.
\end{gathered}
\end{gather}
The $R$-charges of the chiral fields in the original theory preserving $SU(8)$ are determined by the anomaly condition and do not change along the RG-flow because there is no $U(1)$ that can be mixed with $U(1)_R$. On the other hand, if the UV symmetry is broken as in \eqref{eq:maximal subgroups} by introducing flipping fields we do have a $U(1)$ symmetry. If this $U(1)$ is not mixed with $U(1)_R$, the $R$-charges of the operators do not change and we cannot remove the marginal operators for the same reason we explained before for the $E_7$ case. Thus, nothing interesting happens in this case.
On the other hand if the $U(1)$ is mixed with $U(1)_R$, the contributions charged under $U(1)$ at order $pq$ before the flip  won't appear at order $pq$ anymore after the flip. Thus, only the $U(1)$-neutral contributions will remain at order $pq$ and, among these, those with negative sign that used to correspond to relations before the flip are of our interest
since they may now combine to the flavor current.
Thus, when we will decompose operators, relations and currents using the branching rules for the  \eqref{eq:maximal subgroups} cases, we will only look at the $U(1)$-neutral contributions.
As we are going to discuss below only  the symmetry breaking to $SU(6) \times SU(2) \times U(1)$ has neutral sectors suitable for our discussion
and leads to $E_6\times U(1)$ IR global symmetry.

We can then  further break  the  $SU(6) \times SU(2) \times U(1)$ symmetry. We may consider either the  breaking of $SU(6)$ or $SU(2)$. The breaking of the latter into $U(1)$, however, doesn't have $U(1)$-neutral relations, thus, we only need to consider the breaking of $SU(6)$ into: 
\begin{gather}
\nn SU(5) \times U(1) \,, \\
SU(4) \times SU(2) \times U(1) \,, \\
\nn SU(3) \times SU(3) \times U(1)\,.
\end{gather}
In this case  only the  $SU(4) \times SU(2) \times U(1)$ breaking has $U(1)$-neutral sectors and as we will see leads to  $SO(10)\times U(1)^2$ IR global symmetry.

Lastly, we consider the breaking of $SU(4)$, which includes
\begin{gather}
\nn SU(3) \times U(1) \,, \\
SU(2) \times SU(2) \times U(1)
\end{gather}
where only $SU(2) \times SU(2) \times U(1)$ has $U(1)$-neutral relations and as we will see  leads to  $SO(8)\times U(1)^3$ IR global symmetry.\\

\subsection{$SU(8) \rightarrow SU(6) \times SU(2) \times U(1)$ breaking and $E_6 \times U(1)$ enhancement}\label{m6}

We begin by decomposing $SU(8)$ representations in terms of  $SU(6) \times SU(2) \times U(1)_v$ (we use the branching rules of \cite{Yamatsu:2015npn}):
\begin{itemize}
\item Conserved currents:
\begin{align}
{\bf 63} = ({\bf 1},{\bf 1})_0+({\bf 1},{\bf 3})_0+({\bf 6},{\bf 2})_4+({\bf \overline 6},{\bf 2})_{-4}+({\bf 35},{\bf 1})_0 \,.
\end{align}
\item Relations:
\begin{align}
{\bf 70} = ({\bf 15},{\bf 1})_{-4}+({\bf \overline 15},{\bf 1})_4+({\bf 20},{\bf 2})_0 \,.
\end{align}
\item Independent marginal operators:
\begin{align}
{\bf 336} &= ({\bf 1},{\bf 1})_{-12}+({\bf 6},{\bf 2})_{-8}+({\bf 15},{\bf 1})_{-4}+({\bf 21},{\bf 3})_{-4}+({\bf 70},{\bf 2})_0+({\bf \overline{105}'},1)_4 \,.
\end{align}
\end{itemize}
For each contribution above, the $U(1)$-neutral sectors are as follows.
\begin{itemize}
\item $U(1)$-neutral conserved currents:
\begin{align}
\label{eq:U(1)-neutral currents}
39 = ({\bf 1},{\bf 1})_0+({\bf 1},{\bf 3})_0+({\bf 35},{\bf 1})_0 \,.
\end{align}
\item  $U(1)$-neutral relations:
\begin{align}
\label{eq:U(1)-neutral relations}
40 = ({\bf 20},{\bf 2})_0 \,.
\end{align}
\item $U(1)$-neutral  independent marginal operators:
\begin{align}
\label{eq:U(1)-neutral marginal operators}
140 &= ({\bf 70},{\bf 2})_0 \,.
\end{align}
\end{itemize}
As we explained, the $U(1)$-charged sectors won't appear at order $pq$ once the $U(1)$ is mixed with $U(1)_R$ along the RG-flow. The marginal operators originate from the (symmetric) product of 28 $m_{0,ij} = \Tr_{g} \left(Q_{i} Q_{j}\right)$, which is decomposed into
\begin{align}
\label{eq:constituting operators}
{\bf 28} = ({\bf 1},{\bf 1})_{-6}+({\bf 6},{\bf 2})_{-2}+({\bf 15},{\bf 1})_2 \,,
\end{align}
so we see that the $U(1)$-neutral marginal operators are given by $({\bf 6},{\bf 2})_{-2} \times ({\bf 15},{\bf 1})_2 = ({\bf 20},{\bf 2})_0+({\bf 70},{\bf 2})_0$. Thus, we can remove the marginal operators by flipping either $({\bf 6},{\bf 2})_{-2}$ or $({\bf 15},{\bf 1})_2$. This partial flip breaks $SU(8)$ to $SU(6) \times SU(2) \times U(1)_v$.

In the models where $({\bf 6},{\bf 2})_{-2}$ or $({\bf 15},{\bf 1})_2$ is flipped, the $U(1)$-neutral marginal operators disappear and 39 of the $U(1)$-neutral conserved currents are enlarged by the 40 $U(1)$-neutral relations. Therefore, we have $79 =39+40= 78+1$ conserved currents in total, which may constitute the adjoint representation of $E_6 \times U(1)_v$. In section \ref{basicmodles} we will check this enhancement using the superconformal index.

The $({\bf 1},{\bf 1})_{-6}$ singlet can be either flipped or not without affecting the enhancement.
The models with $({\bf 6},{\bf 2})_{-2}$ or  $({\bf 15},{\bf 1})_2$  flipped are related by the Intriligator-Pouliot duality plus the  flip of $({\bf 1},{\bf 1})_{-6}$.

%

Our analysis actually   holds regardless of the gauge rank. The only change is that there is another $U(1)_a$ in the UV global symmetry acting on the  traceless antisymmetric matter. Hence, we expect  higher rank  models exhibiting $E_6 \times U(1)_v\times U(1)_a$ obtained by partial flips of operators constituting the marginal operators. We will discuss this in section \ref{hrt}.
\\

We close this section with a comment on the other cases in \eqref{eq:maximal subgroups}.
For $SU(7) \times U(1)$ and $SU(5) \times SU(3) \times U(1)$, the 70 relations are decomposed into
\begin{align}
\nn {\bf 70} &= {\bf 35}_{-4}+{\bf \overline{35}}_4 \,, \\
{\bf 70} &= ({\bf 5},{\bf 1})_{-12}+({\bf \overline 5},{\bf 1})_{12}+({\bf 10},{\bf \overline 3})_{-4}+({\bf \overline{10}},{\bf 3})_4
\end{align}
respectively. In both cases, there is no $U(1)$-neutral sector. We therefore exclude those cases. Lastly, let us consider $SU(4)^2 \times U(1)$. In this case, there exist $U(1)$-neutral relations and one may attempt to remove marginal operators to turn those relations into conserved currents. However, this requires the flip of the operators in the representation $({\bf 4},{\bf 4})_0$, which is equivalent to the Seiberg duality. Indeed, those operators corresponding to $({\bf 4},{\bf 4})_0$ have $R$-charge 1, which is not affected by the flip because it is not charged under $U(1)$. Thus, it suffers from the same problem as the complete flip expecting $E_7$ and doesn't lead to any enhancement, which is consistent with the fact that the flipped theory is merely the Seiberg-dual theory.
\\

\subsection{$SU(8) \rightarrow SU(4) \times SU(2)^2 \times U(1)^2$ breaking and $SO(10) \times U(1)^2$ enhancement}\label{m10}

Next, we consider the case where we further break  $SU(6)$ to $SU(4) \times SU(2) \times U(1)_u$ so we decompose 
all the contributions according to the branching rules for $SU(6)\times SU(2)\times U(1)_v$ to  $SU(4) \times SU(2)^2 \times U(1)_u\times U(1)_v$. We list only the neutral contributions:
\begin{itemize}
\item Conserved currents:
\begin{align}
\label{eq:U(1)^2-neutral currents}
\nn ({\bf 1},{\bf 1})_0 &= ({\bf 1},{\bf 1},{\bf 1})_{0,0} \,, \\
({\bf 1},{\bf 3})_0 &= ({\bf 1},{\bf 1},{\bf 3})_{0,0} \,, \\
\nn ({\bf 35},{\bf 1})_0  &= ({\bf 1},{\bf 1},{\bf 1})_{0,0}+({\bf 1},{\bf 3},{\bf 1})_{0,0}+({\bf 15},{\bf 1},{\bf 1})_{0,0}+\dots \,.
\end{align}
\item Relations:
\begin{align}
\label{eq:U(1)^2-neutral relations}
({\bf 20},{\bf 2})_0  = ({\bf 6},{\bf 2},{\bf 2})_{0,0}+\dots \,.
\end{align}
\item Independent marginal operators:
\begin{align}
\label{eq:U(1)^2-neutral marginal operators}
({\bf 70},{\bf 2})_0 = ({\bf 6},{\bf 2},{\bf 2})_{0,0}+({\bf 10},{\bf 2},{\bf 2})_{0,0}+\dots \,.
\end{align}
\end{itemize}
Several $\dots$ indicate the contributions charged under $U(1)_u\times U(1)_v$. Recall that the operators constituting the $U(1)$-neutral marginal operators come from the product of $({\bf 6},{\bf 2})_{-2}$ and $({\bf 15},{\bf 1})_2$
 in \eqref{eq:constituting operators}, which are decomposed into
\begin{align}
({\bf 6},{\bf 2})_{-2} &= ({\bf 1},{\bf 2},{\bf 2})_{-2,-2}+({\bf 4},{\bf 1},{\bf 2})_{1,-2} \,, \nn\\
({\bf 15},{\bf 1})_2 & = ({\bf 1},{\bf 1},{\bf 1})_{-4,2}+({\bf 4},{\bf 2},{\bf 1})_{-1,2}+({\bf 6},{\bf 1},{\bf 1})_{2,2} \label{eq:(15,1)}
\end{align}
under $SU(4) \times SU(2)^2 \times U(1)_u\times U(1)_v$. One can see that the $U(1)^2$-neutral marginal operators are either from $({\bf 1},{\bf 2},{\bf 2})_{-2,-2} \times ({\bf 6},{\bf 1},{\bf 1})_{2,2}= ({\bf 6},{\bf 2},{\bf 2})_{0,0}$ or from $({\bf 4},{\bf 1},{\bf 2})_{1,-2} \times ({\bf 4},{\bf 2},{\bf 1})_{-1,2}=({\bf 10},{\bf 2},{\bf 2})_{0,0}+({\bf 6},{\bf 2},{\bf 2})_{0,0} $.  Therefore we can remove the $U(1)^2$-neutral marginal operators by flipping one of $({\bf 1},{\bf 2},{\bf 2})_{-2,-2}$ and $({\bf 6},{\bf 1},{\bf 1})_{2,2}$ and one of $({\bf 4},{\bf 1},{\bf 2})_{1,-2}$ and $({\bf 4},{\bf 2},{\bf 1})_{-1,2}$.
 For example, we can decide to flip  $({\bf 6},{\bf 1},{\bf 1})_{2,2}$ and $({\bf 4},{\bf 2},{\bf 1})_{-1,2}$  
 so the manifest symmetry is broken to $SU(4) \times SU(2)^2 \times U(1)_u\times U(1)_v$.
Other choices of flipping lead to models  related by dualities.
 
  Once the $U(1)^2$-neutral marginal operators are gone in this way the $24 = ({\bf 6},{\bf 2},{\bf 2})_{0,0}$ $U(1)^2$-neutral relations join the $23 = 2 \times ({\bf 1},{\bf 1},{\bf 1})_{0,0}+({\bf 1},{\bf 1},{\bf 3})_{0,0}+({\bf 1},{\bf 3},{\bf 1})_{0,0}+({\bf 15},{\bf 1},{\bf 1})_{0,0}$ of the $U(1)^2$-neutral conserved currents and we expect  enhanced global symmetry  $SO(10) \times U(1)_u\times U(1)_v$ with $24+23 = 47$ currents.
 In section \ref{basicmodles} we will check this enhancement using the superconformal index. 
 The higher rank version of this model is discussed in section \ref{hrt}.

%

Here the fact that $({\bf 1},{\bf 1},{\bf 1})_{-4,2}$ is not flipped is important to realize $SO(10) \times U(1)_u\times U(1)_v$ because if $({\bf 1},{\bf 1},{\bf 1})_{-4,2}$ is also flipped, the broken UV symmetry is $SU(6) \times SU(2) \times U(1)_v$ rather than $SU(4) \times SU(2)^2 \times U(1)_u\times U(1)_v$ and we end up with the model in the previous subsection, which exhibits $E_6 \times U(1)_v$ rather than $SO(10) \times U(1)_u\times U(1)_v$.
\\

Let's quickly check the other breaking patterns of  $SU(6)$.
For $SU(5) \times U(1)$ and $SU(3) \times SU(3) \times U(1)$, the neutral relations are decomposed into
\begin{align}
\nn ({\bf 20},{\bf 2}) &= ({\bf 10},{\bf 2})_{-3,0}+({\bf \overline{10}},{\bf 2})_{3,0} \,, \\
({\bf 20},{\bf 2}) &= ({\bf 3},{\bf \overline{3}}, {\bf 2})_{-1,0}+({\bf \overline{3}}, {\bf 3},{\bf 2})_{1,0}+({\bf 1},{\bf 1},{\bf 2})_{3,0}+ ({\bf 1},{\bf 1},{\bf 2})_{-3,0}
\end{align}
respectively. In both cases, there is no $U(1)$-neutral sector so we exclude those cases.\\

\subsection{$SU(8) \rightarrow SU(2)^4 \times U(1)^3$ breaking and $SO(8) \times U(1)^3$ enhancement}\label{m8}

Lastly, we consider breaking  $SU(4)$ to $SU(2) \times SU(2) \times U(1)_t$
so we  further decompose the $U(1)^2 $-neutral contributions of $SU(4) \times SU(2)^2 \times U(1)_u\times U(1)_v$ to  $SU(2)^4 \times U(1)_t\times U(1)_u\times U(1)_v$. We list only the $U(1)^3$-neutral  contributions:
\begin{itemize}
\item Conserved currents:
\begin{align}
\nn 2 \times ({\bf 1},{\bf 1},{\bf 1})_{0,0} &= 2 \times ({\bf 1},{\bf 1},{\bf 1},{\bf 1})_{0,0,0} \,, \\
\nn({\bf 1},{\bf 1},{\bf 3})_{0,0} &= ({\bf 1},{\bf 1},{\bf 1},{\bf 3})_{0,0,0} \,, \\
({\bf 1},{\bf 3},{\bf 1})_{0,0} &= ({\bf 1},{\bf 1},{\bf 3},{\bf 1})_{0,0,0} \,, \\
\nn ({\bf 15},{\bf 1},{\bf 1})_{0,0} &= ({\bf 1},{\bf 1},{\bf 1},{\bf 1})_{0,0,0}+({\bf 3},{\bf 1},{\bf 1},{\bf 1})_{0,0,0}+({\bf 1},{\bf 3},{\bf 1},{\bf 1})_{0,0,0}+\dots \,.
\end{align}
\item Relations:
\begin{align}
({\bf 6},{\bf 2},{\bf 2})_{0,0} = ({\bf 2},{\bf 2},{\bf 2},{\bf 2})_{0,0,0}+\dots \,.
\end{align}
\item Independent marginal operators:
\begin{align}
\nn ({\bf 6},{\bf 2},{\bf 2})_{0,0} &= ({\bf 2},{\bf 2},{\bf 2},{\bf 2})_{0,0,0}+\dots \,, \\
({\bf 10},{\bf 2},{\bf 2})_{0,0} &= ({\bf 2},{\bf 2},{\bf 2},{\bf 2})_{0,0,0}+\dots \,.
\end{align}
\end{itemize}
Recall that the operators constituting the $U(1)^2$-neutral marginal operators come from the product of operators  in \eqref{eq:(15,1)}, which are decomposed into
\begin{align}
({\bf 1},{\bf 2},{\bf 2})_{-2,-2} &= ({\bf 1},{\bf 1},{\bf 2},{\bf 2})_{0,-2,-2} \,, \nn\\
({\bf 4},{\bf 1},{\bf 2})_{1,-2} &= ({\bf 2},{\bf 1},{\bf 1},{\bf 2})_{1,1,-2}+({\bf 1},{\bf 2},{\bf 1},{\bf 2})_{-1,1,-2} \,, \nn\\
({\bf 4},{\bf 2},{\bf 1})_{-1,2} &= ({\bf 2},{\bf 1},{\bf 2},{\bf 1})_{1,-1,2}+({\bf 1},{\bf 2},{\bf 2},{\bf 1})_{-1,-1,2} \,, \nn\\
({\bf 6},{\bf 1},{\bf 1})_{2,2} &= ({\bf 1},{\bf 1},{\bf 1},{\bf 1})_{2,2,2}+({\bf 1},{\bf 1},{\bf 1},{\bf 1})_{-2,2,2}+({\bf 2},{\bf 2},{\bf 1},{\bf 1})_{0,2,2}
\label{BRSU4}\end{align}
under $SU(2)^4 \times U(1)_t\times U(1)_u \times U(1)_v$. One can see that the $U(1)^3$-neutral marginal operators come from
\begin{align}
\nn ({\bf 1},{\bf 1},{\bf 2},{\bf 2})_{0,-2,-2} \quad &\times \quad ({\bf 2},{\bf 2},{\bf 1},{\bf 1})_{0,2,2} \,, \\
({\bf 2},{\bf 1},{\bf 1},{\bf 2})_{1,1,-2} \quad &\times \quad ({\bf 1},{\bf 2},{\bf 2},{\bf 1})_{-1,-1,2} \,, \\
\nn ({\bf 1},{\bf 2},{\bf 1},{\bf 2})_{-1,1,-2} \quad &\times \quad ({\bf 2},{\bf 1},{\bf 2},{\bf 1})_{1,-1,2} \,.
\end{align} Therefore, one can again remove the $U(1)^3$-neutral marginal operators by flipping one of the two representations in each line. For example, we can flip: $({\bf 2},{\bf 2},{\bf 1},{\bf 1})_{0,2,2}$, $({\bf 1},{\bf 2},{\bf 2},{\bf 1})_{-1,-1,2}$ and $({\bf 2},{\bf 1},{\bf 2},{\bf 1})_{1,-1,2}$. 
Other choices of flipping lead to models  related by dualities.

Once the $U(1)^3$-neutral marginal operators are gone in this way, $16 = ({\bf 2},{\bf 2},{\bf 2},{\bf 2})_{0,0,0}$  $U(1)^3$-neutral relations join the $15 = 3 \times ({\bf 1},{\bf 1},{\bf 1},{\bf 1})_{0,0,0}+({\bf 1},{\bf 1},{\bf 1},{\bf 3})_{0,0,0}+({\bf 1},{\bf 1},{\bf 3},{\bf 1})_{0,0,0}+({\bf 1},{\bf 3},{\bf 1},{\bf 1})_{0,0,0}+({\bf 3},{\bf 1},{\bf 1},{\bf 1})_{0,0,0}$ $U(1)^3$-neutral conserved currents such that the expected enhanced global symmetry is $SO(8) \times U(1)_t\times U(1)_u \times U(1)_v$, which has dimension $16+15 = 31$.
 In section  \ref{basicmodles} we will  check this enhancement using the superconformal index.  The higher rank version of this model is discussed in section \ref{hrt}.


Here one can also flip either $({\bf 1},{\bf 1},{\bf 1},{\bf 1})_{2,2,2}$ or $({\bf 1},{\bf 1},{\bf 1},{\bf 1})_{-2,2,2}$, but not both, without affecting the enhancement because they cannot make a $U(1)^3$-neutral marginal operator. If both are flipped, however, the broken UV symmetry is $SU(4) \times SU(2)^2 \times  U(1)_u \times U(1)_v$ rather than $SU(2)^4 \times U(1)_t\times U(1)_u \times U(1)_v$ and we end up with the model in the previous subsection, which exhibits $SO(10) \times  U(1)_u \times U(1)_v$ rather than $SO(8) \times U(1)_t\times U(1)_u \times U(1)_v$.
\\

We close by quickly checking the other  breaking patterns of  $SU(4)$.
For $SU(3) \times U(1)$ the neutral relations are decomposed into
\begin{align}
({\bf 6},{\bf 2},{\bf 2})_{0,0} &= ({\bf 3},{\bf 2},{\bf 2})_{-2,0,0}+({\bf \overline{3}},{\bf 2},{\bf 2})_{2,0,0}. 
\end{align}
Since there is no $U(1)$-neutral sector we exclude this case.\\

\section{Flips, self-dualities and symmetry enhancements}
\label{sec3}

In the previous section we observed that by considering  three different partial flips breaking the $SU(8)$ UV global symmetry of the $SU(2)$  theory with 8 fundamental chirals to
\begin{gather}
\nn SU(6) \times SU(2) \times U(1)_v \,, \\
SU(4) \times SU(2)^2 \times U(1)_u\times U(1)_v \,, \\
\nn SU(2)^4\times U(1)_t\times U(1)_u \times U(1)_v \,,
\end{gather}
we expect to  find models exhibiting enhanced IR global symmetries
\begin{gather}
\nn E_6 \times U(1)_v \,, \\
SO(10) \times U(1)_u \times U(1)_v \,, \\
\nn SO(8) \times U(1)_t\times U(1)_u \times U(1)_v
\end{gather}
respectively.  In section \ref{basicmodles} we will check that these models indeed enjoy the expected symmetry looking at the superconformal index expansion and in addition,
following the second strategy discussed in the introduction, we list all the duality frames accounting for the enhanced global symmetry.
In section \ref{deformations} we will consider various deformations, which in particular lead to models with $SO(9)$ and $F_4$ symmetries.\\

We begin by introducing all the fields and their charges. In this section we will work with conventions in which only the $SU(2)^4\times U(1)_t\times U(1)_u\times U(1)_v$ subgroup of the full UV symmetry is explicitly manifest.
This is done by splitting the 8 fundamental chiral fields  into four doublets $Q_1,\ldots, Q_4$, one for each $SU(2)$ flavor symmetry. The singlets we introduced in the previous section to break the $SU(8)$ symmetry will also split accordingly.
All models will have   bifundamental singlets $D_{1}, D_{2}, D_{3}$ coupled as: 
\begin{align}
\mathcal W_0 = \Tr_g \Tr_2 \Tr_3 (D_1 Q_2 Q_3) + \Tr_g \Tr_3 \Tr_1 (D_2 Q_3 Q_1) + \Tr_g \Tr_1 \Tr_2 (D_3 Q_1 Q_2)\,,
\end{align}
where $\Tr_{1,2,3,4}$ denote the traces over the $SU(2)_{1,2,3,4}$ flavor indices.
For some models we will also consider singlets $b_i$ with $i=1,\cdots, 4$ contributing $b_i\Tr_g \Tr_i ( Q_i Q_i)$ to the superpotential. In general we will denote by $\mathcal{T}_i$ with $i=1,2,3$ the theory where  the interactions involving respectively  $b_1$ or $b_1,b_2$ or $b_1,b_2, b_3$ are turned on. We will also denote by $\widehat{\mathcal{T}}_i$ the theory where also the interaction involving $b_4$ is turned on. We use a different notation in this case because this latter interaction is not involved in the symmetry enhancement process but we might need to turn it on to avoid having decoupled fields.

\begin{table}[tbp]
\centering
\begin{tabular}{c|ccccccccc|c}
{}   & $SU(2)_1$ & $SU(2)_2$ & $SU(2)_3$ & $SU(2)_4$ & $U(1)_{x_1}$ & $U(1)_{x_2}$ & $U(1)_{x_3}$  & $U(1)_{R_0}$ \\ \hline
$Q_1$  & $\bf 2$ & $\bullet$ & $\bullet$ & $\bullet$ & $\frac{1}{2}$ & $0$ & $0$ &  $0$ \\
$Q_2$ & $\bullet$ & $\bf 2$ & $\bullet$ & $\bullet$ & $0$ & $\frac{1}{2}$ & $0$ & $0$ \\
$Q_3$ &  $\bullet$ & $\bullet$ & $\bf 2$ & $\bullet$ & $0$ & $0$ & $\frac{1}{2}$ &  $0$ \\
$Q_4$ & $\bullet$ & $\bullet$ & $\bullet$ & $\bf 2$ & $-\frac{1}{2}$ & $-\frac{1}{2}$ & $-\frac{1}{2}$  & $2$ \\
$D_1$  & $\bullet$ & $\bf 2$ & $\bf 2$ & $\bullet$ & $0$ & $-\frac{1}{2}$ & $-\frac{1}{2}$ & $2$  \\
$D_2$ & $\bf 2$ & $\bullet$ & $\bf 2$ & $\bullet$ & $-\frac{1}{2}$ & $0$ & $-\frac{1}{2}$ &  $2$ \\
$D_3$  & $\bf 2$ & $\bf 2$ & $\bullet$ & $\bullet$ & $-\frac{1}{2}$  & $-\frac{1}{2}$ & $0$ & $2$ \\
$b_1$  & $\bullet$ & $\bullet$ & $\bullet$ & $\bullet$ & $-1$ & $0$ & $0$ &  $2$ \\
$b_2$ & $\bullet$ & $\bullet$ & $\bullet$ & $\bullet$ & $0$ & $-1$ & $0$ & $2$ \\
$b_3$  & $\bullet$ & $\bullet$ & $\bullet$ & $\bullet$ & $0$ & $0$ & $-1$ & $2$ \\
$b_4$ & $\bullet$ & $\bullet$ & $\bullet$ & $\bullet$ & $1$ & $1$ & $1$ &  $-2$ \\
\end{tabular}
\caption{The representations of the chiral fields under the symmetry groups.}
\label{tab:charges}
\end{table}

For convenience we will also work in a different basis for the $U(1)^3$ symmetry. Specifically, we will use a parametrization of the abelian symmetries that we will denote by $U(1)_{x_1}\times U(1)_{x_2}\times U(1)_{x_3}$ which is related to the one $U(1)_t\times U(1)_u\times U(1)_v$ of the previous section by the following redefinition of the charges:
\begin{align}
&\mathcal{Q}_t= 2 (\mathcal{Q}_1-\mathcal{Q}_2),\nonumber\\
&\mathcal{Q}_u= 2 (\mathcal{Q}_1+\mathcal{Q}_2-2 \mathcal{Q}_3),\nonumber\\
&\mathcal{Q}_v= 2 (\mathcal{Q}_1+\mathcal{Q}_2+\mathcal{Q}_3)\,,
\label{3u1m}
\end{align}
or equivalently at the level of the fugacities in the index
\begin{align}
&x_1=t^2u^2v^2,\nonumber\\
&x_2=t^{-2}u^2v^2,\nonumber\\
&x_3=u^{-4}v^2\,.
\label{3u1mfug}
\end{align}
The charges of the fields with this new parametrization are as in table \ref{tab:charges}, where we also give a possible choice of UV trial R-symmetry $U(1)_{R_0}$.\\

\subsection{Self-dualities and Enhancements }\label{basicmodles}

\subsubsection{$SO(8) \times U(1)^3$ models}
\label{sec:T0}

\begin{figure}[t]
	\centering
	\makebox[\linewidth][c]{
  	\includegraphics[scale=1.2]{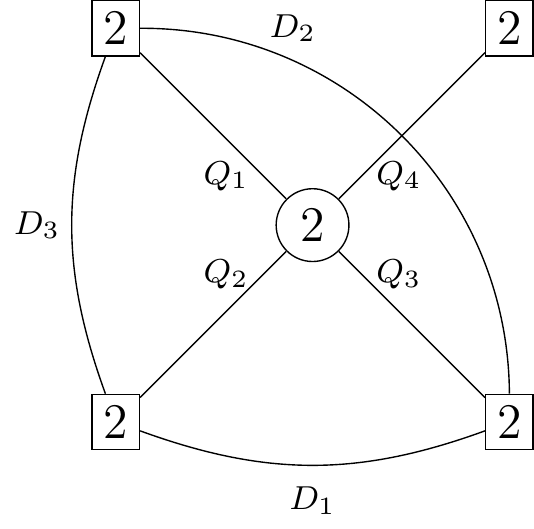} }
  	\caption{Quiver diagram for the $\mathcal{T}_0$ model.}
  	\label{quivernob_fields}
\end{figure}

We start discussing two models enjoying the $SO(8) \times U(1)^3$ enhancement.
In the first model $\mathcal{T}_0$ we introduce only the singlets $D_1, \, D_2, \, D_3$ interacting with $\mathcal{W}_0$. The label for the theory stands for the fact that we don't introduce any of the $b_i$ fields in this case.
The matter content of the theory is summarized in the quiver diagram of figure \ref{quivernob_fields}.

Combining the information on the $U(1)^3$ charges of table \ref{tab:charges} with the redefinition \eqref{3u1m} we can see that the operators flipped by $D_1,D_2,D_3$ in $\mathcal{W}_0$
\begin{align}
\Tr_{g} (Q_1 Q_2) \,, \qquad \Tr_g (Q_2 Q_3) \, , \qquad \Tr_g (Q_1 Q_3)
\end{align}
are precisely the operators $({\bf 2},{\bf 2},{\bf 1},{\bf 1})_{0,2,2}$, $({\bf 1},{\bf 2},{\bf 2},{\bf 1})_{-1,-1,2}$, $({\bf 2},{\bf 1},{\bf 2},{\bf 1})_{1,-1,2}$ of section \ref{m8}. Hence, we expect in this case that 
the manifest $SU(2)^4 \times U(1)^3$ UV global symmetry gets enhanced in the IR to $SO(8) \times U(1)^3$. This can be checked computing the superconformal index of the theory $\mathcal{T}_0$. We first perform $a$-maximization \cite{Intriligator:2003jj} to find the values of the mixing coefficients of the R-symmetry with $U(1)_{x_1}\times U(1)_{x_2}\times U(1)_{x_3}$ corresponding to the superconformal R-symmetry, which we approximate to
\begin{align}
R_1=R_2=R_3\simeq\frac{11}{10}\,.
\end{align}
The superconformal index then reads\footnote{Since none of the $U(1)$ symmetries participates in the enhancement it is equivalent to parametrize them with $U(1)_{x_1,x_2,x_3}$ or $U(1)_{t,u,v}$ when computing the index. We decide to use the latter parametrization. }
\begin{align}
\mathcal{I}_0&=1+ v^{-6}(pq)^{\frac{7}{20}}(1+p+q)+({\bf 8}_vv^{-2}u^{-2}+{\bf 8}_sv^{-2}u\,t^{-1}+{\bf 8}_cv^{-2}u\,t)(pq)^{\frac{9}{20}}(1+p+q)+\nn\\
&+(v^2u^{-4}+v^2u^2t^{-2}+v^2u^2t^2)(pq)^{\frac{11}{20}}+v^{-12}(pq)^{\frac{7}{10}}+({\bf 8}_vv^{-8}u^{-2}+{\bf 8}_sv^{-8}u\,t^{-1}+\nn\\
&+{\bf 8}_cv^{-8}u\,t)(pq)^{\frac{4}{5}}+(({\bf 35}_v+1)v^{-4}u^{-4}+({\bf 35}_s+1)v^{-4}u^2t^{-2}+({\bf 35}_c+1)v^{-4}u^2t^2+\nn\\
&+{\bf 56}_vv^{-4}u^2+{\bf 56}_sv^{-4}u^{-1}t+{\bf 56}_cv^{-4}u^{-1}t^{-1})(pq)^{\frac{9}{10}}+({\bf 8}_vu^{-6}+{\bf 8}_su^{-3}t^3+{\bf 8}_cu^3t^{-3}+\nn\\
&{\color{blue}-({\bf 28}+3)})pq+\cdots\,.
\end{align}
In the expression of the index, each number is the character of an $SO(8)$ representation and, in particular, the term $-({\bf 28}+3)pq$ highlighted in blue reflects the current multiplet, which is in the adjoint representation  of $SO(8) \times U(1)^3$. 
 In this case we also have $72-48=24$ marginal operators. Notice that the fact $R_1 = R_2 = R_3$ implies, among the three $U(1)$ symmetries, only $U(1)_v$ mixes with the $R$-symmetry. Thus, although we have used approximate $R$-charges for the expansion of the index, the terms independent of fugacity $v$ are exact; the terms of order $pq$, corresponding to 24 marginal operators and 31 conserved currents, are such examples.
 
Now according to the argument in \cite{Razamat:2017wsk}, since the size of the Weyl group of $SO(8)$ is given by
\begin{align}
\left|W(SO(8))\right| = 4! \times 2^3
\end{align}
and $\left|W(SU(2)^4)\right| = 2! \, 2! \, 2! \, 2!$ is manifest, we expect $12 = 4! \, 2^3/(2! \, 2! \, 2! \, 2!)$ self-dual frames including the original theory itself. To find the self-dual frames we proceed as follows.
We specialize the $SU(8)$ fugacities vector $\vec{u}$ defined in \eqref{chargesvector} according to the breaking of $SU(8)\to \prod_{i=1}^4 SU(2)_i \times U(1)_t\times U(1)_u \times U(1)_v $:
\begin{align}
u=(v\,u\,t\,y_1,v\,u\,t\,y_1^{-1},v\,u\,t^{-1}y_2,v\,u\,t^{-1}y_2^{-1},v\,u^{-2}y_3,v\,u^{-2}y_3^{-1},pq\,v^{-3}y_4,pq\,v^{-3}y_4^{-1})\,,
\end{align}
where $y_i$ is the $SU(2)_i$ fugactity.
This corresponds to choosing one particular representative in the oribit of the UV $SU(2)^4$ Weyl group.
Now we implement the Seiberg, CSST and IP dualities. As explained in appendix \ref{app1}, these dualities transform the fugacities vector respectively as in eqs.~\eqref{us}, \eqref{ucsst}, \eqref{uip}.
Inspecting the transformed vectors we can identify the self-dual frames.
Those will correspond to frames where we have a collection of  charged chirals with the same R-charge and $U(1)_{t,u,v}$ charges  as in the original frame.
We also checked that the self-dual frames have the same collection of singlets.
In the end we have found that the 12 self-dual frames are realized by
\begin{gather}
1 \text{ original} \,, \nonumber\\
3 \text{ Seiberg} \,, \nonumber\\
8 \text{ CSST}
\label{selfdualitiesT0}
\end{gather}
dualities.
\\

\begin{figure}[t]
	\centering
	\makebox[\linewidth][c]{
  	\includegraphics[scale=1.2]{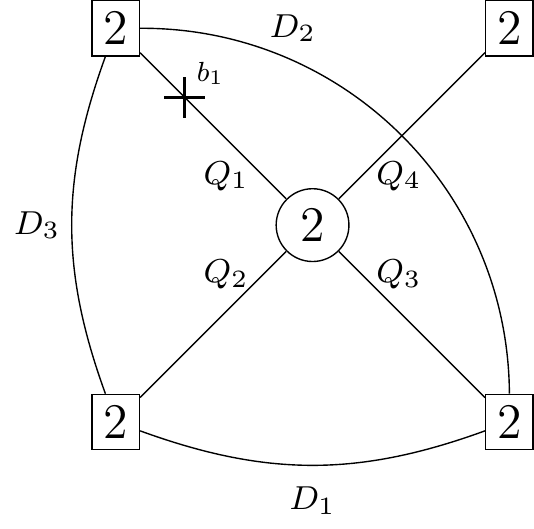} }
  	\caption{Quiver diagram for the $\mathcal{T}_1$ model.}
  	\label{quiverb1_fields}
\end{figure}

Another  model  exhibiting  $SO(8) \times U(1)^3$ enhancement is $\mathcal{T}_1$, obtained including the singlet $b_1$ with the superpotential\footnote{Given the symmetry of the quiver in figure \ref{quiverb1_fields} this is equivalent to introducing the singlet $b_2$.}
\begin{align}
\label{w1}
\mathcal W_1=\mathcal W_0+\Gd\mathcal{W}_1,\qquad \Gd\mathcal{W}_1=b_1\Tr_g \Tr_1 ( Q_1 Q_1)
\end{align}
where $\mathcal W_0$ is given by
\begin{align}
\mathcal W_0 = \Tr_g \Tr_2 \Tr_3 (D_1 Q_2 Q_3) + \Tr_g \Tr_3 \Tr_1 (D_2 Q_3 Q_1) + \Tr_g \Tr_1 \Tr_2 (D_3 Q_1 Q_2)\,.
\end{align}
 The matter content of the theory is now summarized in the quiver diagram of figure \ref{quiverb1_fields}. 
Notice that the new $b_1$ field is flipping the meson $\Tr_g (Q_1 Q_1)$ corresponding to the operator $({\bf 1},{\bf 1},{\bf 1},{\bf 1})_{2,2,2}$ in the notation of section \ref{m8}.

Performing $a$-maximization we find the following approximate values of the mixing coefficients:
\begin{align}
R_1\simeq\frac{15}{13},\qquad R_2=R_3\simeq\frac{14}{13}\,.
\end{align}
The index computed with this R-symmetry is then given by
\begin{align}
\mathcal{I}_1&=1+v^{-6}(pq)^{\frac{9}{26}}(1+p+q)+v^{-2}u^{-2}t^{-2}(pq)^{\frac{11}{26}}(1+p+q)+({\bf 8}_vv^{-2}u^{-2}+\nn\\
&+{\bf 8}_sv^{-2}u\,t^{-1})(pq)^{\frac{23}{52}}(1+p+q)+{\bf 8}_cv^{-2}u\,t\,(pq)^{\frac{6}{13}}(1+p+q)+(v^2u^{-4}+v^2u^2t^{-2})(pq)^{\frac{7}{13}}+\nn\\
&+v^{-12}(pq)^{\frac{9}{13}}+v^{-8}u^{-2}t^{-2}(pq)^{\frac{10}{13}}+({\bf 8}_vv^{-8}u^{-2}+{\bf 8}_sv^{-8}t^{-1}u)(pq)^{\frac{41}{52}}+{\bf 8}_cv^{-8}u\,t\,(pq)^{\frac{21}{26}}+\nn\\
&+v^{-4}u^{-4}t^{-4}(pq)^{\frac{11}{13}}+({\bf 8}_vv^{-4}u^{-4}t^{-2}+{\bf 8}_sv^{-4}u^{-1}t^{-3})(pq)^{\frac{45}{52}}+(({\bf 35}_v+1)v^{-4}u^{-4}+\nn\\
&+({\bf 35}_s+1)v^{-4}u^2t^{-2}+({\bf 56}_c+{\bf 8}_c)v^{-4}u^{-1}t^{-1})(pq)^{\frac{23}{26}}+({\bf 56}_vv^{-4}u^2+{\bf 56}_sv^{-4}u^{-1}t)(pq)^{\frac{47}{52}}+\nn\\
&+{\bf 35}_cv^{-4}u^2t^2(pq)^{\frac{12}{13}}+(t^{-4}+u^{-6}t^{-2})(pq)^{\frac{25}{26}}+({\bf 8}_vu^{-6}+{\bf 8}_su^3t^{-3})p^{\frac{51}{52}}q^{\frac{51}{52}}{-\color{blue}({\bf 28}+3)}pq+\cdots\,,
\end{align}
Again the index organises into characters of $SO(8)$ and at order $pq$ we can see the contribution of the $SO(8)\times U(1)^3$ current highlighted in blue.

We can explain the enhancement of theory $\mathcal{T}_1$ in terms of self-dualities exactly in the same way as for theory $\mathcal{T}_0$. Indeed, the operator $\Tr_g(Q_1 Q_1)$ is trivially mapped to itself under all the self-dualities \eqref{selfdualitiesT0}. Hence, these are also self-dualities of theory $\mathcal{T}_1$ and the same counting we did for $\mathcal{T}_0$ explains the  $SO(8)$ enhancement  for $\mathcal{T}_1$.

The singlet $b_4$ is also a spectator from the point of view of the self-dualities \eqref{selfdualitiesT0}. This means that   theories $\widehat{\mathcal{T}}_0$ and $\widehat{\mathcal{T}}_1$, where we also turn on   $b_4\Tr_g \Tr_4(Q_4 Q_4)$ in the superpotential,  will still exhibit the $SO(8)\times U(1)^3$ enhancement.
\\

\subsubsection{$SO(10) \times U(1)^2$ model}
\label{sec3.1.2}

\begin{figure}[t]
	\centering
	\makebox[\linewidth][c]{
  	\includegraphics[scale=1.2]{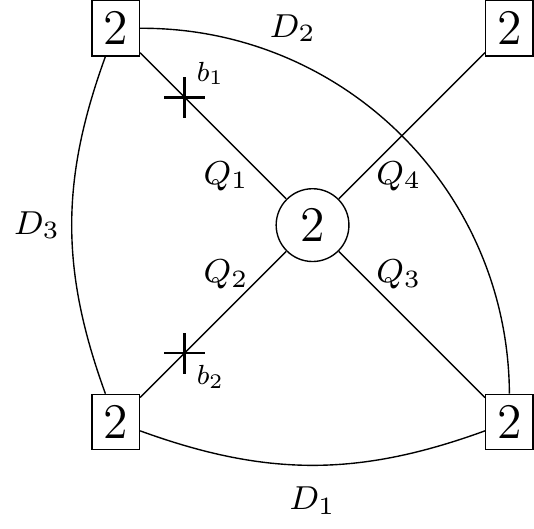} }
  	\caption{Quiver diagram for the $\mathcal{T}_2$ model.}
  	\label{quiverb1b2_fields}
\end{figure}

Now we consider a model with the IR $SO(10) \times U(1)^2$ symmetry. We denote this model by $\mathcal{T}_2$ as, in addition to the usual $D_1$, $D_2$, $D_3$ singlets, we also introduce  the singlets $b_1, \,b_2$. The superpotential is
\begin{align}
\mathcal W_2 = \mathcal W_1+  \Gd\mathcal{W}_2,\qquad \Gd\mathcal{W}_2= b_2\Tr_g \Tr_2 (Q_2 Q_2)
\label{w2}
\end{align}
where $\mathcal W_1$ is given by
\begin{align}
\mathcal W_1 &= \Tr_g \Tr_2 \Tr_3 (D_1 Q_2 Q_3) + \Tr_g \Tr_3 \Tr_1 (D_2 Q_3 Q_1) + \Tr_g \Tr_1 \Tr_2 (D_3 Q_1 Q_2) \nonumber \\
&\quad +b_1\Tr_g \Tr_1 (Q_1 Q_1) \,.
\end{align}
The matter content of the theory is summerized in the quiver diagram of figure \ref{quiverb1b2_fields}, but the full manifest UV global symmetry is actually $SU(4) \times SU(2)_3 \times SU(2)_4 \times U(1)^2$,
since $SU(2)_1$, $SU(2)_2$ and one $U(1)$, specifically $2(U(1)_{x_1}-U(1)_{x_2})=U(1)_t$, recombine to form $SU(4)$.
In particular 
$D_3$ and $b_1, \, b_2$ form the antisymmetric representation of $SU(4)$, which flip the mesonic operators 
\begin{align}
\Tr_g \left(Q_1 Q_2\right) \,, \quad \Tr_g \left(Q_1 Q_1\right) \,, \quad \Tr_g \left(Q_2 Q_2\right) \,,
\end{align}
corresponding to the operator $({\bf 6},{\bf 1},{\bf 1})_{2,2}$ of  section \ref{m10}. Indeed, as it can be seen combining the data contained in table \ref{tab:charges} and the map of the charges \eqref{3u1m}, their charges are compatible with those expected from the branching rule \eqref{BRSU4}.
The singlets $D_1, \, D_2$ also recombine to form the bifundamental representation between $SU(4) \times SU(2)_3$, which flips
\begin{align}
\Tr_g (Q_i Q_3) \,, \qquad i = 1,2 \,,
\end{align}
corresponding to the operator $({\bf 4},{\bf 2},{\bf 1})_{-1,2}$. 

Performing $a$-maximization we find the following approximate values of the mixing coefficients:
\begin{align}
R_1=R_2\simeq\frac{8}{9},\qquad R_3\simeq\frac{21}{20}\,.
\end{align}
Notice that equal $R_1$ and $R_2$ imply $U(1)_t$ doesn't mix with the $R$-symmetry, which is consistent with the fact that $U(1)_t$ is part of the nonabelian symmetry $SU(4)$.
The index computed with this R-symmetry is then\footnote{We choose to parametrize the two $U(1)$ symmetries that don't participate in the enhancement with $U(1)_{u,v}$.}
\begin{align}
\mathcal{I}_2&=1+v^{-6}(pq)^{\frac{7}{20}}(1+p+q)+{\bf 10}\,v^{-2}u^{-2}(pq)^{\frac{7}{16}}(1+p+q)+{\bf \overline{16}}\,v^{-2}u\,(pq)^{\frac{73}{160}}(1+p+q)+\nn\\
&+v^2u^{-4}(pq)^{\frac{21}{40}}+v^{-12}(pq)^{\frac{7}{10}}+{\bf 10}\,v^{-8}u^{-2}(pq)^{\frac{63}{80}}+{\bf \overline{16}}\,v^{-8}u\,(pq)^{\frac{129}{160}}+({\bf 54}+1)v^{-4}u^{-4}(pq)^{\frac{7}{8}}+\nn\\
&+{\bf 144}\,v^{-4}u^{-1}(pq)^{\frac{143}{160}}+{\bf 126}\,v^{-4}u^2(pq)^{\frac{73}{80}}+{\bf 10}\,u^{-6}(pq)^{\frac{77}{80}}{\color{blue}-({\bf 45}+2)}pq+\cdots\,.
\end{align}
Each number is the character of an $SO(10)$ representation and, in particular, the term $-({\bf 45}+2)pq$ highlighted in blue reflects the current multiplet, which is in the adjoint representation  of $SO(10) \times U(1)^2$.

Let's now discuss the self-duality frames responsible for the  enhancement. The size of the Weyl group of $SO(10)$ is given by
\begin{align}
\left|W(SO(10))\right| = 5! \times 2^4\,,
\end{align}
where $\left|W(SU(4) \times SU(2)^2)\right| = 4! \, 2! \, 2!$ is manifest. Thus, $20 = 5! \, 2^4/(4! \, 2! \, 2!)$ self-dual frames including the original theory itself are expected. 
Also in this case to find the self-dual frames we specialize the $SU(8)$ fugacities vector $\vec{u}$ defined in \eqref{chargesvector} according to the breaking of $SU(8)\to SU(4)\times SU(2)_3\times SU(2)_4\times U(1)_u\times U(1)_v$
\begin{align}
u=(v\,u\,w_1,v\,u\,w_2,v\,u\,w_3,v\,u\,w_4,v\,u^{-2}y_3,v\,u^{-2}y_3^{-1},pq\,v^{-3}y_4,pq\,v^{-3}y_4^{-1})\,,
\end{align}
where $w_i$ are the $SU(4)$ fugacities subjected to the constraint $\prod_{i=1}^4 w_i=1$.
In this case we collect a subset of the transformations \eqref{us}, \eqref{ucsst}, \eqref{uip} corresponding to Seiberg, CSST and IP dualities respectively for which we get a transformed vector associated to a collection of chirals with the same R-charge and $U(1)_{u,v}$ charges, as well as the same set of singlets. In the end we have found that the 20 self-dual frames are realized by
\begin{gather}
1 \text{ original} \,, \nonumber\\
7 \text{ Seiberg} \,, \nonumber\\
12 \text{ CSST}
\label{selfdualitiesT2}
\end{gather}
dualities.

Also in this case the singlet $b_4$ is a spectator from the point of view of the self-dualities \eqref{selfdualitiesT2}, so the theory $\widehat{\mathcal{T}}_2$,  where $b_4\Tr_g\Tr_4(Q_4 Q_4)$ is turned on in the superpotential,
 still enjoys the $SO(10)\times U(1)^2$ enhancement.\\

\subsubsection{$E_6 \times U(1)$ model}
\label{sec3.1.3}

\begin{figure}[t]
	\centering
	\makebox[\linewidth][c]{
  	\includegraphics[scale=1.2]{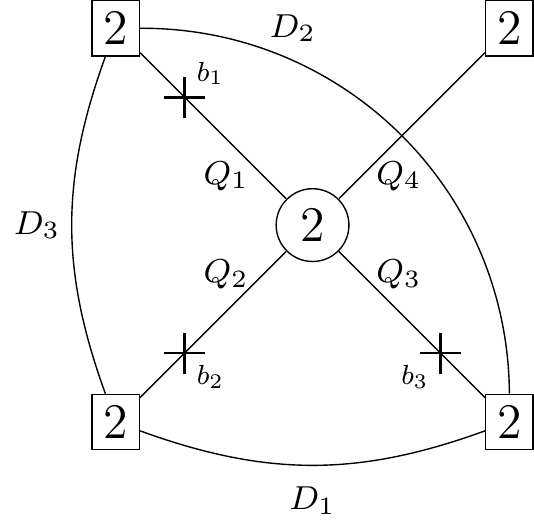} }
  	\caption{Quiver diagram for the $\mathcal{T}_3$ model.}
  	\label{quiverb1b2b3_fields}
\end{figure}

The last model we consider is the one exhibiting the $E_6 \times U(1)$ symmetry enhancement of  \cite{Razamat:2017wsk}.\footnote{Such an $E_6$ model can be also obtained from the compactification of the E-string theory in 6d on a sphere with flux breaking the $E_8$ symmetry of E-string into $E_6 \times U(1) \times SU(2)$ \cite{Hwang:2021xyw}. The last $SU(2)$ symmetry only acts on the decoupled operators in the IR and the interacting SCFT is described by the $E_6$ model of \cite{Razamat:2017wsk}.} Following the same nomenclature of the previous cases, we call this theory $\widehat{\mathcal{T}}_3$ as on top of the singlets $D_1$, $D_2$, $D_3$ we also introduce all the singlets $b_1$, $b_2$, $b_3$, $b_4$ with superpotential
\begin{align}
\widehat{\mathcal W_3} = \mathcal W_2+ \Gd\mathcal{W}_3+\Gd\mathcal{W}_4,\qquad \Gd\mathcal{W}_3=b_3\Tr_g \Tr_3 ( Q_3 Q_3),\qquad \Gd\mathcal{W}_4=b_4\Tr_g \Tr_4 ( Q_4 Q_4)
\end{align}
where $\mathcal W_2$ is given by
\begin{align}
\mathcal W_2 &= \Tr_g \Tr_2 \Tr_3 (D_1 Q_2 Q_3) + \Tr_g \Tr_3 \Tr_1 (D_2 Q_3 Q_1) + \Tr_g \Tr_1 \Tr_2 (D_3 Q_1 Q_2) \nonumber \\
&\quad +b_1\Tr_g \Tr_1 (Q_1 Q_1)+b_2\Tr_g \Tr_2 (Q_2 Q_2) \,.
\end{align}
In this case we have to include also the singlet $b_4$ since otherwise the operator $\Tr_g \Tr_4 (b_4 Q_4 Q_4)$ would be free in the IR.
The matter content of the theory is summerized in the quiver diagram of figure \ref{quiverb1b2b3_fields}, but the full manifest UV global symmetry is actually $SU(6) \times SU(2)_4 \times U(1)$,
since $SU(2)_1$, $SU(2)_2$, $SU(2)_3$ and two $U(1)$ out of $U(1)^3$, specifically $U(1)_t$ and $U(1)_u$, recombine into $SU(6)$. In particular the singlets 
$D_1, \, D_2, \, D_3$ and $b_1, \, b_2, \, b_3$ form the antisymmetric representation of $SU(6)$, which flips the mesonic operators
\begin{align}
\Tr_g \left(Q_1 Q_2\right) \,, \quad \Tr_g \left(Q_1 Q_3\right) \,, \quad \Tr_g \left(Q_2 Q_3\right) \,, &  \nn\\
\Tr_g \left( Q_1 Q_1\right) \,, \quad \Tr_g \left( Q_2 Q_2\right) \,, \quad \Tr_g \left( Q_3 Q_3\right) \,, & 
\end{align}
corresponding to the operator $({\bf 15},{\bf 1})_2$ of section \ref{m6}. Indeed, as it can been seen combining the data contained in table \ref{tab:charges} and the map of the charges \eqref{3u1m}, their charges are compatible with those expected from the branching rules \eqref{eq:(15,1)}-\eqref{BRSU4}.
The singlet $b_4$ instead flips the meson $\Tr_g (Q_4 Q_4)$ corresponding to 
 $({\bf 1},{\bf 1})_{-6}$.
Performing $a$-maximization we find the following values of the mixing coefficients\footnote{Performing $a$-maximization in the theory without the singlet $b_4$ we would find a value of the mixing coefficients for which the operator $\Tr_g (Q_4 Q_4)$ is below the unitarity bound, meaning that it becomes a decoupled free field in the IR.}:
\begin{align}
R_1= R_2=R_3=\frac{10}{9}\,,
\end{align}
which indicates $U(1)_t$ and $U(1)_u$ do not mix with the $R$-symmetry because they are part of the nonabelian symmetry $SU(6)$.
The index computed with this R-symmetry is then\footnote{We choose to parametrize the $U(1)$ symmetry that doesn't participate in the enhancement with $U(1)_{v}$.}
\begin{align}
\widehat{\mathcal I}_3=1+{\bf \overline{27}}\,v^{-2}(pq)^{\frac{4}{9}}(1+p+q)+v^6(pq)^{\frac{2}{3}}+{\bf 351}\,v^4(pq)^{\frac{8}{9}}{\color{blue}-({\bf 78}+1)}pq+\cdots\,.
\end{align}
Each number is the character of an $E_6$ representation and, in particular, the term $-({\bf 78}+1)pq$ highlighted in blue reflects the current multiplet, which is in the adjoint representation  of $E_6 \times U(1)_v$.

In \cite{Razamat:2017wsk} this enhancement was explained by studying the self-dualities of the model, similarly to what we did for the previous cases.
The size of the Weyl group of $E_6$ is given by
\begin{align}
\left|W(E_6)\right| = 2^7\times 3^4\times 5=51840\,,
\end{align}
where $\left|W(SU(6) \times SU(2))\right| = 6! \times 2!$ is manifest. The remaining $36 = 51840/(6! \, 2!)$ should be  realized as self-dualities. In order to determine which of the 72 frames correspond to self-dualities, we specialize the $SU(8)$ fugacities vector $\vec{u}$ defined in \eqref{chargesvector} according to the breaking of $SU(8)\to SU(6)\times SU(2)_4\times U(1)_v$
\begin{align}
u=(v\, w_1,v\,w_2,v\,w_3,v\,w_4, v\,w_5,v\,w_6,pq\,v^{-3}y_4, pq\,v^{-3}y_4^{-1} )\,,
\end{align}
where $w_i$ are the $SU(6)$ fugactities subjected to the constraint $\prod_{i=1}^6 w_i=1$. In this case we collect a subset of the transformations \eqref{us}, \eqref{ucsst}, \eqref{uip} corresponding to Seiberg, CSST and IP dualities respectively for which we get a transformed vector associated to a collection of chirals with the same R-charge and $U(1)_v$ charges, as well as the same set of singlets. In the end we have found that the 36 self-dual frames are realized by
\begin{gather}
1 \text{ original} \,, \nonumber\\
15 \text{ Seiberg} \,, \nonumber\\
20 \text{ CSST}
\label{selfdualitiesT6}
\end{gather}
dualities.
\\

\subsection{Symmetry breaking deformations}\label{deformations}

So far we considered singlets $b_i$ coupling to $SU(2)$ flavor singlet mesons $\Tr_g\Tr_i (Q_i Q_i)$
which ensures that the manifest symmetry includes $SU(N)$ groups. On the other hand we can introduce extra deformations of the form $b_i \, \Tr_g\Tr_j (Q_j Q_j)$ with $i \neq j$.

In the $E_6 \times U(1)$  model $\mathcal{\hat{T}}_3$ all the $SU(2)$ flavor singlet mesons are already flipped and thus trivial in the chiral ring, so this model cannot be deformed  in this way. 
We will thus focus on the deformations of the $SO(10) \times U(1)^2$ model  $\mathcal{T}_2$ and the $SO(8) \times U(1)^3$ model $\mathcal{T}_1$ (the theory $\mathcal{T}_0$ can't be deformed in this way as it doesn't contain any of the $b_i$ singlets).
\\

\subsubsection{$SO(10) \rightarrow SO(9) \rightarrow  F_4$ deformations}
\label{sec3.2.1}

We begin with the $\mathcal{T}_2$ model. Looking at  $\mathcal{W}_2$  in eq.~\eqref{w2}
we see that $\Tr_g\Tr_1 (Q_1 Q_1)$ and $\Tr_g\Tr_2 (Q_2 Q_2)$ are  trivial in the chiral ring while $\Tr_g\Tr_3 (Q_3 Q_3)$ is not. Thus, we can obtain a new theory $\mathcal{T}'_2$ by introducing a stable\footnote{Here ``stable" is meant with respect to the chiral ring stability criterion of \cite{Benvenuti:2017lle}.} deformation term\footnote{Using the superconformal R-charge of theory $\mathcal{T}_2$ we can check that this deformation has R-charge $R[b_i \Tr_g\Tr_3(Q_3 Q_3)]\simeq1.9173<2$ so it is a relevant deformation.}
\begin{align}
\label{eq:def1}
\mathcal{W}'_2=\mathcal{W}_2+\Gd\mathcal{W}_2',\qquad \Delta \mathcal{W}_2' = b_i \Tr_g\Tr_3(Q_3 Q_3)
\end{align}
where $\mathcal W_2$ is given by
\begin{align}
\mathcal W_2 &= \Tr_g \Tr_2 \Tr_3 (D_1 Q_2 Q_3) + \Tr_g \Tr_3 \Tr_1 (D_2 Q_3 Q_1) + \Tr_g \Tr_1 \Tr_2 (D_3 Q_1 Q_2) \nonumber \\
&\quad +b_1\Tr_g \Tr_1 (Q_1 Q_1)+b_2\Tr_g \Tr_2 (Q_2 Q_2) \,.
\end{align}
Here $i$ can be either 1 or 2. For definiteness, we will take $i = 2$ and we sahll denote the deformed theory by $\mathcal{T}_2'$. This deformation breaks the manifest UV symmetry $SU(4) \times SU(2)_3 \times SU(2)_4 \times U(1)^2$ of theory $\mathcal{T}_2$ to
\begin{align}
SU(2)_1 \times USp(4) \times SU(2)_4 \times U(1)^2
\end{align}
This can be seen as follows. The deformation \eqref{eq:def1} breaks one combination of   $U(1)_{x_1}\times U(1)_{x_2}\times U(1)_{x_3}$.
Specifically, at the level of fugacities it imposes the constraint
\be
x_2^{-1}x_3=1\qquad\Longleftrightarrow\qquad t^2u^{-6}=1\,,
\ee
which means that a combination of $U(1)_{x_2,x_3}$ or equivalently of $U(1)_{t,u}$ is broken. We decide to parametrize the surviving combination of these two $U(1)$, which we shall denote by $U(1)_{\tilde{u}}$, as
\be
u=\tilde{u},\qquad t=\tilde{u}^3\,.
\label{utildefug}
\ee
At the level of the charges this means
\begin{align}
\mathcal{Q}_{\tilde u} &= 4 (2 \mathcal{Q}_1-\mathcal{Q}_2-Q_3) \,, \nn\\
\mathcal{Q}_v &= 2 (\mathcal{Q}_1+\mathcal{Q}_2+\mathcal{Q}_3) \,.
\label{utildecharge}
\end{align}
One can then check for example that the fields $Q_2$ and $Q_3$ have the same R-charge and $U(1)_{\tilde{u},v}$ charges and can thus be organized into the fundamental representation of $USp(4)$. Similarly, the singlets
$D_1$ and $b_2$ can be organized into the traceless antisymmetric representation of $USp(4)$. Finally, the singlets $D_2$ and $D_3$ can be collected into the bifundamental representation of $SU(2)_1 \times USp(4)$.

The Weyl group $W(USp(4) \times SU(2)^2)$ of the manifest symmetry is of size $2! \times 2^2 \times 2! \times 2!$. Similarly to what we did in the previous subsection, one can check that out of the 20 self-dualities \eqref{selfdualitiesT2} of the original $\mathcal{T}_2$ theory only 12 map $\Delta \mathcal W_2'$ to itself and remain self-dualities of the deformed theories. These correspond exactly to the self-dualities \eqref{selfdualitiesT0} of the $SO(8)$ models, which were
\begin{gather}
1 \text{ original} \,,\nn \\
3 \text{ Seiberg} \,,\nn \\
8 \text{ CSST}\,.
\end{gather}
Therefore, the expected size of the Weyl group of the enhanced symmetry is now
\begin{align}
&\left|W(USp(4) \times SU(2)^2)\right| \times (1+11 \text{ self-dualities}) \nonumber \\
&= (2! \times 2^2 \times 2! \times 2!) \times 12 \nonumber \\
&= 4! \times 2^4 \nonumber \\
&= \left|W(SO(9))\right| \,.
\end{align}
Thus, we expect that the $SO(10) \times U(1)^2$ model $\mathcal{T}_2$  deformed by $\eqref{eq:def1}$ has the $SO(9) \times U(1)^2$ IR symmetry with  the UV symmetry $SU(2)_1 \times USp(4) \times SU(2)_4$ recombining into  $SO(9)$ in the IR.

This is  confirmed  by  the superconformal index expansion. Performing $a$-maximization we find the following approximate values of the mixing coefficients:
\begin{align}
R_1\simeq\frac{8}{9},\qquad R_2=R_3\simeq\frac{11}{10}\,.
\end{align}
The index computed with this R-symmetry is then\footnote{We choose to parametrize the $U(1)$ symmetries, which don't participate in the enhancement, with $U(1)_{\tilde{u},v}$.}
\begin{align}
\mathcal{I}^{'}_2 &= 1+v^{-6}(pq)^{\frac{27}{80}}(1+p+q)+v^{-2}\tilde{u}^{-8}(pq)^{\frac{7}{16}}(1+p+q)+{\bf 16}\,v^{-2}\tilde{u}^{-2}(pq)^{\frac{71}{160}}(1+p+q)+\nn\\
&+{\bf 9}\,v^{-2}\tilde{u}^4(pq)^{\frac{9}{20}}+v^2\tilde{u}^{-4}(pq)^{\frac{11}{20}}+v^{-12}(pq)^{\frac{27}{40}}+v^{-8}\tilde{u}^{-8}(pq)^{\frac{31}{40}}+{\bf 16}\,v^{-8}\tilde{u}^{-2}(pq)^{\frac{25}{32}}+\nn\\
&+{\bf 9}\,v^{-8}\tilde{u}^4(pq)^{\frac{63}{80}}+v^{-4}\tilde{u}^{-16}(pq)^{\frac{7}{8}}+v^{-4}\tilde{u}^{-10}(pq)^{\frac{141}{160}}+({\bf 126}+{\bf9}+1)v^{-4}\tilde{u}^{-4}(pq)^{\frac{71}{80}}+\nn\\
&+{\bf 128}\,v^{-4}\tilde{u}^2(pq)^{\frac{143}{160}}+{\bf 44}\,v^{-4}\tilde{u}^8(pq)^{\frac{9}{10}}+\tilde{u}^{-12}(pq)^{\frac{79}{80}}{\color{blue}-({\bf 36}+2)}pq+\cdots\,.
\end{align}
Each number is the character of an $SO(9)$ representation and, in particular, the term $-({\bf 36}+2)pq$ highlighted in blue reflects the current multiplet, which is in the adjoint representation  of $SO(9) \times U(1)^2$. Thus, the deformation \eqref{eq:def1} added to the $SO(10) \times U(1)^2$ model $\mathcal{T}_2$ leads to a new theory exhibiting the IR symmetry enhancement
\begin{align}
USp(4) \times SU(2)^2 \times U(1)^2 \qquad \longrightarrow \qquad SO(9) \times U(1)^2 \,.
\end{align}
\\

We can further deform the $\mathcal{T}'_2$ theory. From $\mathcal{W}'_2$ in eq.~\eqref{eq:def1} we see that one linear combination of $\Tr_g\Tr_2 (Q_2 Q_2)$ and $\Tr_g\Tr_3(Q_3 Q_3)$ is flipped by $b_2$, but there is another linearly independent combination, say $\Tr_g\Tr_3 (Q_3 Q_3)$, which is non-trivial in the chiral ring. Thus, we can obtain a new theory $\mathcal{T}''_2$ by introducing another stable deformation term\footnote{Using the superconformal R-charge of theory $\mathcal{T}'_2$ we can check that this deformation has R-charge $R[b_1 \Tr_g\Tr_3(Q_3 Q_3)]\simeq 1.96101<2$ so it is a relevant deformation.}
\begin{align}
\label{eq:def2}
\mathcal{W}''_2=\mathcal{W}'_2+\Gd\mathcal{W}''_2,\qquad \Delta \mathcal{W}''_2 = b_1 \, \Tr_g\Tr_3 (Q_3 Q_3) \,
\end{align}
where $\mathcal{W}'_2$ is given in \eqref{eq:def1}.
This deformation breaks the manifest UV symmetry $SU(2)_1 \times USp(4) \times SU(2)_4 \times U(1)^2$ of theory $\mathcal{T}'_2$ to
\begin{align}
USp(6) \times SU(2)_4 \times U(1) \,.
\end{align}
This can be seen as follows. The deformation \eqref{eq:def2} breaks one combination of the two $U(1)$ symmetries. Specifically, at the level of fugacities it imposes the constraint
\be
x_1^{-1}x_3=1\qquad\Longleftrightarrow\qquad \tilde{u}^{-12}=1\,,
\ee
which means that the $U(1)_{\tilde{u}}$ symmetry is broken. Hence, the surviving abelian symmetry is $U(1)_v$, which we recall was defined in \eqref{3u1m}-\eqref{3u1mfug}.
One can then check for example that the fields $Q_{2,3}$, which we already collected into the fundamental representation of $USp(4)$, have also the same R-charge and $U(1)_{v}$ charge of $Q_1$ and we can thus organize them into the fundamental representation of $USp(6)$. Similarly, the singlets
$D_1$, $D_2$, $D_3$ and $b_1$, $b_2$ can be organized into the traceless antisymmetric representation of $USp(6)$.


The manifest Weyl group is $W(USp(6) \times SU(2))$, whose size is $3! \times 2^3 \times 2!$. In addition, all the 12 self-dualities of the $SO(9) \times U(1)^2$ model still remain self-dualities after the deformation \eqref{eq:def2}. The expected size of the Weyl group of the enhanced symmetry is then
\begin{align}
&\left|W(USp(6) \times SU(2))\right| \times (1+11 \text{ self-dualities}) \nonumber \\
&= (3! \times 2^3 \times 2!) \times 12 \nonumber \\
&= 1152 \nonumber \\
&= \left|W(F_4)\right| \,.
\end{align}
Thus, we expect the $SO(9) \times U(1)^2$ model $\mathcal{T}'_2$ deformed by $\eqref{eq:def2}$ to have the IR symmetry $F_4 \times U(1)$ with the $USp(6) \times SU(2)$ UV symmetry recombining into $F_4$ in the IR.

This is  confirmed by the superconformal index expansion. Performing $a$-maximization we find the following approximate value of the mixing coefficient:
\begin{align}
R_1=R_2=R_3\simeq\frac{21}{19}\,.
\end{align}
The index computed with this R-symmetry is then
\begin{align}
\mathcal{I}''_2 &= 1+v^{-6}(pq)^{\frac{13}{38}}(1+p+q)+{\bf 26}\,v^{-2}(pq)^{\frac{17}{38}}(1+p+q)+v^2(pq)^{\frac{21}{38}}+\nn\\
&+v^{-12}(pq)^{\frac{13}{19}}+{\bf 26}\,v^{-8}(pq)^{\frac{15}{19}}+({\bf 324}+1)v^{-4}(pq)^{\frac{17}{19}}{\color{blue}-({\bf 52}+1)}pq+\cdots \,.
\end{align}
Each number is the character of an $F_4$ representation and, in particular, the term $-({\bf 52}+1)pq$ highlighted in blue reflects the current multiplet, which is in the adjoint representation  of $F_4 \times U(1)$. Thus, the deformation \eqref{eq:def2} added to the $SO(9) \times U(1)^2$ model $\mathcal{T}'_2$ leads to a new theory exhibiting the IR symmetry enhancement
\begin{align}
USp(6) \times SU(2) \times U(1) \qquad \longrightarrow \qquad F_4 \times U(1) \,.
\end{align}

Note that $\mathcal T_2''$ can be also obtained from $\mathcal T_3$ by integrating out the trace part of the antisymmetric representation of flavor $SU(6)$ constituted by $(D_1,D_2,D_3,b_1,b_2,b_3)$, which breaks $SU(6)$ into $USp(6)$. This can be done by introducing an additional singlet $c$ with a superpotential $\Delta \mathcal W = c \, (b_1+b_2+b_3)$, which makes both $c$ and the trace part $b_1+b_2+b_3$ massive. Once those massive fields are integrated out, the resulting theory is exactly $\mathcal T_2''$ with the superpotential \eqref{eq:def2}. The enhanced IR symmetry $E_6 \times U(1)$ is then partially broken to $F_4 \times U(1)$.
\\

\subsubsection{$SO(8) \rightarrow SO(9)\rightarrow SO(8)$ deformations}
\label{sec3.2.2}

Now we consider the $\mathcal{T}_1$ model with the $SO(8) \times U(1)^3$ symmetry. While the $SO(8) \times U(1)^3$ symmetry can be realized either with or without $b_1$, we stick to the model with $b_1$ because we need at least one $b_i$ field to deform the theory.

Looking at  $\mathcal{W}_1$  in eq.~\eqref{w1}
we can see that both $\Tr_g\Tr_2 (Q_2 Q_2)$ and $\Tr_g\Tr_3 (Q_3 Q_3)$ are non-trivial in the chiral ring. Thus, we can obtain a new theory $\mathcal{T}'_1$ by introducing a stable deformation term\footnote{Using the superconformal R-charge of theory $\mathcal{T}_1$ we can check that this deformation has R-charge $R[b_1 \Tr_g\Tr_i(Q_i Q_i)]\simeq1.91822<2$ so it is a relevant deformation.}
\begin{align}
\label{eq:def3}
\mathcal{W}'_1=\mathcal{W}_1+\Gd\mathcal{W}'_1,\qquad \Gd\mathcal{W}'_1 = b_1 \Tr_g\Tr_i(Q_i Q_i)
\end{align}
where $\mathcal W_1$ is given by
\begin{align}
\mathcal W_1 &= \Tr_g \Tr_2 \Tr_3 (D_1 Q_2 Q_3) + \Tr_g \Tr_3 \Tr_1 (D_2 Q_3 Q_1) + \Tr_g \Tr_1 \Tr_2 (D_3 Q_1 Q_2) \nonumber \\
&\quad +b_1\Tr_g \Tr_1 (Q_1 Q_1) \,.
\end{align}
Here $i$ can be either 2 or 3. For definiteness, we will take $i = 2$ and we shall denote the deformed theory by $\mathcal{T}_1'$. This deformation modifies the manifest UV symmetry $SU(2)_1\times SU(2)_2 \times SU(2)_3 \times SU(2)_4 \times U(1)^2$ of theory $\mathcal{T}_1$ to
\begin{align}
USp(4)\times SU(2)_3 \times SU(2)_4 \times U(1)^2 \,.
\end{align}
This can be seen as follows. The deformation \eqref{eq:def3} breaks one combination of the three $U(1)$ symmetries. Specifically, at the level of fugacities it imposes the constraint
\be
x_1^{-1}x_2=1\qquad\Leftrightarrow\qquad t^{-4}=1\,,
\ee
which means that the symmetry $U(1)_t$ is broken. Hence, the surviving abelian symmetries are $U(1)_{u,v}$, which we recall were defined in \eqref{3u1m}-\eqref{3u1mfug}. One can then check for example that the fields $Q_1$ and $Q_2$ have the same R-charge and $U(1)_{u,v}$ charges and can thus be organized into the fundamental representation of $USp(4)$. Similarly, the singlets
$D_3$ and $b_1$ can be organized into the traceless antisymmetric representation of $USp(4)$. Finally, the singlets $D_1$ and $D_2$ can be collected into the bifundamental representation of $USp(4)\times SU(2)_3$.

The Weyl group $W(USp(4) \times SU(2)^2)$ of the manifest symmetry is of size $2! \times 2^2 \times 2! \times 2!$. Similarly to what we did in the previous subsection, one can check that all of the 12 self-dualities \eqref{selfdualitiesT0} of the original $\mathcal{T}_1$ theory are still self-dualities of the deformed theory. Therefore, the expected size of the Weyl group of the enhanced symmetry is now
\begin{align}
&\left|W(USp(4) \times SU(2)^2)\right| \times (1+11 \text{ self-dualities}) \nonumber \\
&= (2! \times 2^2 \times 2! \times 2!) \times 12 \nonumber \\
&= 4! \times 2^4 \nonumber \\
&= \left|W(SO(9))\right| \,.
\end{align}
Thus, we expect that the $SO(8) \times U(1)^3$ model $\mathcal{T}_1$  deformed by \eqref{eq:def3} has  IR symmetry $SO(9) \times U(1)^2$.

This is  confirmed by the superconformal index expansion. Performing $a$-maximization we find the following approximate values of the mixing coefficients:
\begin{align}
R_1=R_2\simeq\frac{10}{9},\qquad R_3\simeq\frac{15}{14}\,.
\end{align}
The index computed with this R-symmetry is then\footnote{We choose to parametrize the $U(1)$ symmetries, which don't participate in the enhancement, with $U(1)_{u,v}$.}
\begin{align}
\mathcal{I}'_1 &= 1+v^{-6}(pq)^{\frac{89}{252}}(1+p+q)+{\bf 9}\,v^{-2}u^{-2}(pq)^{\frac{4}{9}}(1+p+q)+{\bf 16}\,v^{-2}u(pq)^{\frac{229}{504}}(1+p+q)+\nn\\
&+v^2u^{-4}(pq)^{\frac{15}{28}}+v^2u^2(pq)^{\frac{5}{9}}+v^{-12}(pq)^{\frac{89}{126}}+{\bf 9}\,v^{-8}u^{-2}(pq)^{\frac{67}{84}}+{\bf 16}\,v^{-8}u(pq)^{\frac{407}{504}}+\nn\\
&+{\bf 45}\,v^{-4}u^{-4}(pq)^{\frac{8}{9}}+{\bf 128}\,v^{-4}u^{-1}(pq)^{\frac{151}{168}}+({\bf 126}+1)v^{-4}u^{-2}(pq)^{\frac{229}{252}}+{\bf 9}\,u^{-6}(pq)^{\frac{247}{252}}+\nn\\
&{\color{blue}-({\bf 36}+2)}pq+\cdots\,.
\end{align}
Each number is the character of an $SO(9)$ representation and, in particular, the term $-({\bf 36}+2)pq$ highlighted in blue reflects the current multiplet, which is in the adjoint representation  of $SO(9) \times U(1)^2$. Thus, the deformation \eqref{eq:def3} added to the $SO(8) \times U(1)^3$ model $\mathcal{T}_1$ leads to a new theory exhibiting the IR symmetry enhancement
\begin{align}
USp(4) \times SU(2)^2 \times U(1)^2 \qquad \longrightarrow \qquad SO(9) \times U(1)^2 \,.
\end{align}

This $\mathcal T_1'$ model can be also obtained from $\mathcal T_2$ by integrating out the trace part of the antisymmetric presentation of flavor $SU(4)$ constituted by $(D_1, D_2, b_1, b_2)$, which breaks $SU(4)$ into $USp(4)$. This also partially breaks the enhanced IR symmetry $SO(10) \times U(1)^2$ into $SO(9) \times U(1)^2$.
\\

We can further deform the $\mathcal{T}'_1$ theory. From $\mathcal{W}_1'$ in eq.~\eqref{eq:def3} we see that the operator $\Tr_g\Tr_3 (Q_3 Q_3)$ is still non-trivial in the chiral ring. Thus, we can obtain a new theory $\mathcal{T}''_1$ by introducing another stable deformation term\footnote{Using the superconformal R-charge of theory $\mathcal{T}'_2$ we can check that this deformation has R-charge $R[b_1 \Tr_g\Tr_3(Q_3 Q_3)]\simeq 1.95598<2$ so it is a relevant deformation.}
\begin{align}
\label{eq:def4}
\mathcal{W}''_1=\mathcal{W}'_1+\Gd\mathcal{W}''_1,\qquad \Delta \mathcal{W}''_1 = b_1 \, \Tr_g\Tr_3 (Q_3 Q_3) \,
\end{align}
where $\mathcal W_1'$ is given in \eqref{eq:def3}.
This deformation breaks the manifest UV symmetry $USp(4) \times SU(2)_3\times SU(2)_4 \times U(1)^2$ of theory $\mathcal{T}'_2$ to
\begin{align}
SU(2)_1\times SU(2)_2\times SU(2)_3 \times SU(2)_4 \times U(1)
\end{align}
Indeed, the deformation \eqref{eq:def4} breaks one combination of the two $U(1)$ symmetries. Specifically, at the level of fugacities it imposes the constraint
\be
x_1^{-1}x_3=1\qquad\Longleftrightarrow\qquad u^{-6}=1\,,
\ee
which means that the  $U(1)_{u}$ symmetry is broken. Hence, the surviving abelian symmetry is $U(1)_v$, which we recall was defined in \eqref{3u1m}-\eqref{3u1mfug}. The superpotential now doesn't preserve the $USp(4)$ symmetry  since the  singlet $b_1$, which in  $\mathcal{T}'_1$  formed a $USp(4)$ together with  $D_3$, 
now appears (without $D_3$) in the  deformation \eqref{eq:def4}.
The manifest Weyl group is $W(SU(2)^4)$, whose size is $(2!)^2$. In addition, all the 12 self-dualities of the $SO(9) \times U(1)^2$ model $\mathcal{T}'_1$ still remain self-dualities after the deformation \eqref{eq:def4}. The expected size of the Weyl group of the enhanced symmetry is then
\begin{align}
&\left|W(SU(2)^4)\right| \times (1+11 \text{ self-dualities}) \nonumber \\
&= (2! \times 2! \times 2! \times 2!) \times 12 \nonumber \\
&= 4! \times 2^3 \nonumber \\
&= \left|W(SO(8))\right| \,.
\end{align}
Thus, we expect the $SO(9) \times U(1)^2$ model $\mathcal{T}'_1$ deformed by $\eqref{eq:def4}$ to have the IR symmetry $SO(8) \times U(1)$.

Indeed, we have confirmed it by looking at the superconformal index. Performing $a$-maximization we find the following approximate value of the mixing coefficient:
\begin{align}
R_1=R_2=R_3\simeq\frac{11}{10}\,.
\end{align}
The index computed with this R-symmetry is then
\begin{align}
\mathcal{I}''_1 &= 1+v^{-6}(pq)^{\frac{7}{20}}(1+p+q)+({\bf 8}_v+{\bf 8}_s+{\bf 8}_c+1)v^{-2}(pq)^{\frac{9}{20}}(1+p+q)+v^{-2}(pq)^{\frac{11}{20}}+\nn\\
&+v^{-12}(pq)^{\frac{7}{10}}+({\bf 8}_v+{\bf 8}_s+{\bf 8}_c+1)v^{-8}(pq)^{\frac{4}{5}}+({\bf 56}_v+{\bf 56}_s+{\bf 56}_c+{\bf 35}_v+{\bf 35}_s+{\bf 35}_c+\nn\\
&+{\bf 8}_v+{\bf 8}_s+{\bf 8}_c+3)v^{-4}(pq)^{\frac{9}{10}}{\color{blue}-({\bf 28}+1)}pq+\cdots
\end{align}
Each number is the character of an $SO(8)$ representation and, in particular, the term $-({\bf 28}+1)pq$ highlighted in blue reflects the current multiplet, which is in the adjoint representation  of $SO(8) \times U(1)$. Thus, the deformation \eqref{eq:def4} added to the $SO(9) \times U(1)^2$ model $\mathcal{T}'_1$ leads to a new theory exhibiting the IR symmetry enhancement
\begin{align}
SU(2)^4 \times U(1) \qquad \longrightarrow \qquad SO(8) \times U(1) \,.
\end{align}

\section{Higher rank theories}\label{hrt}

In section \ref{sec2} we have argued that enhanced symmetries are expected for $USp(2 N)$ theories once we flip a set of operators breaking the $SU(8)$ UV global symmetry to particular subgroups. Since the argument holds regardless of the gauge rank, we have an infinite family of theories for a given enhanced symmetry in the IR.\footnote{Recently the $USp(2 N)$ theories with one traceless antisymmetric and eight fundamental chirals but without the extra singlets and the superpotential, thus exhibiting $SU(8)$ global symmetry, are shown to have dense spectrum in the large $N$ limit and satisfy a version of Weak Gravity Conjecture for large enough $N$ \cite{Agarwal:2020pol}. It would be interesting to ask similar questions to our models, which exhibit different global symmetries in the presence of the extra singlets and the nontrivial superpotential.} For higher rank theories there can be  multiple operators in a given representation of the global symmetry and we need to flip all of them to realise the enhanced symmetry.

In this section, we show how this works explicitly in some examples. We consider 4d $\mathcal N = 1$ $USp(2 N)$ theories with one antisymmetric chiral $A$ and 8 fundamental chirals, and extra gauge singlets $D_{1,2,3}^i$ and a subset of $b_{1,2,3,4}^i$ for $i = 0, \dots, N-1$, whose global charges are shown in table \ref{tab:higher rank charges}.
\begin{table}[tbp]
\centering
\begin{tabular}{c|cccccccc|c}
{} & $SU(2)_1$ & $SU(2)_2$ & $SU(2)_3$ & $SU(2)_4$ & $U(1)_{x_1}$ & $U(1)_{x_2}$ & $U(1)_{x_3}$ & $U(1)_a$ & $U(1)_{R_0}$ \\ \hline
$Q_1$ & $\bf 2$ & $\bullet$ & $\bullet$ & $\bullet$ & $\frac{1}{2}$ & $0$ & $0$ & $-\frac{N-1}{4}$ & $0$ \\
$Q_2$ & $\bullet$ & $\bf 2$ & $\bullet$ & $\bullet$ & $0$ & $\frac{1}{2}$ & $0$ & $-\frac{N-1}{4}$ & $0$ \\
$Q_3$ & $\bullet$ & $\bullet$ & $\bf 2$ & $\bullet$ & $0$ & $0$ & $\frac{1}{2}$ & $-\frac{N-1}{4}$ & $0$ \\
$Q_4$ & $\bullet$ & $\bullet$ & $\bullet$ & $\bf 2$ & $-\frac{1}{2}$ & $-\frac{1}{2}$ & $-\frac{1}{2}$ & $-\frac{N-1}{4}$ & $2$ \\
$A$ & $\bullet$ & $\bullet$ & $\bullet$ & $\bullet$ & $1$ & $0$ & $0$ & $0$ & $0$ \\
$D_1^i$ & $\bullet$ & $\bf 2$ & $\bf 2$ & $\bullet$ & $0$ & $-\frac{1}{2}$ & $-\frac{1}{2}$ & $\frac{N-1}{2}-i$ & $2$  \\
$D_2^i$ & $\bf 2$ & $\bullet$ & $\bf 2$ & $\bullet$ & $-\frac{1}{2}$ & $0$ & $-\frac{1}{2}$ & $\frac{N-1}{2}-i$ & $2$ \\
$D_3^i$ & $\bf 2$ & $\bf 2$ & $\bullet$ & $\bullet$ & $-\frac{1}{2}$  & $-\frac{1}{2}$ & $0$ & $\frac{N-1}{2}-i$ & $2$ \\
$b_1^i$ & $\bullet$ & $\bullet$ & $\bullet$ & $\bullet$ & $-1$ & $0$ & $0$ & $\frac{N-1}{2}-i$ & $2$ \\
$b_2^i$ & $\bullet$ & $\bullet$ & $\bullet$ & $\bullet$ & $0$ & $-1$ & $0$ & $\frac{N-1}{2}-i$ & $2$ \\
$b_3^i$ & $\bullet$ & $\bullet$ & $\bullet$ & $\bullet$ & $0$ & $0$ & $-1$ & $\frac{N-1}{2}-i$ & $2$ \\
$b_4^i$ & $\bullet$ & $\bullet$ & $\bullet$ & $\bullet$ & $1$ & $1$ & $1$ & $\frac{N-1}{2}-i$ & $-2$ \\
$a_i$ & $\bullet$ & $\bullet$ & $\bullet$ & $\bullet$ & $-i$ & $0$ & $0$ & $0$ & $2$ \\
\end{tabular}
\caption{The representations of the chiral fields under the symmetry groups for higher rank theories.}
\label{tab:higher rank charges}
\end{table}
The $D^i_{1,2,3}$ singlets couple to the fundamental chirals via the following superpotential terms:
\begin{align}
\label{eq:higher rank superpotential}
\mathcal W^N_0 = \sum_{i = 0}^{N-1} \left[\Tr_g \Tr_2 \Tr_3 (D_1^i A^i Q_2 Q_3) + \Tr_g \Tr_3 \Tr_1 (D_2^i A^i Q_3 Q_1) + \Tr_g \Tr_1 \Tr_2 (D_3^i A^i Q_1 Q_2)\right]\,,
\end{align}
while the $b^i_{1,2,3,4}$ singlets, if present couple through
\begin{align}
\sum_{i = 0}^{N-1} b_l^i \Tr_g \Tr_l \left(A^i Q_l Q_l\right)\,.
\label{higherrankb}
\end{align}
In such higher rank cases we can also construct gauge invariant operators of the form $\Tr_g \left(A^i\right)$. These turn out to fall below the unitarity bound in all the examples we are going to consider, so we will also need additional singlets $a_i$ that flip them through the superpotential terms
\begin{align}
a_i \Tr_g \left(A^i\right)\,.
\label{higherranka}
\end{align}\\
Similarly to the rank one case, we will denote by $\mathcal{T}_i^N$ with $i=1,2,3$ the theory where the interactions involving respectively  $b_1$ or $b_1,b_2$ or $b_1,b_2, b_3$ are turned on and by $\widehat{\mathcal T}^N_i$ the theory where also the interaction involving $b_4$ is turned on, where now the upper index denotes the rank of the gauge group.
\\

\subsection{$E_6 \times U(1)^2$ model}

In section \ref{m6} we have shown that to realize the $E_6$ symmetry we have to flip the operators either in the representation $({\bf 6},{\bf 2})_{-2}$ or in the representation $(15,1)_2$ of $SU(6) \times SU(2) \times U(1)_v \subset SU(8)$. For definiteness we may take the latter, which are then given by
\begin{align}
\Tr_g \left(A^i Q_1 Q_2\right) \,, \quad \Tr_g \left(A^i Q_1 Q_3\right) \,, \quad \Tr_g \left(A^i Q_2 Q_3\right) \,, & \qquad i = 0, \dots, N-1\,, \nn\\
\Tr_g \left(A^i Q_1 Q_1\right) \,, \quad \Tr_g \left(A^i Q_2 Q_2\right) \,, \quad \Tr_g \left(A^i Q_3 Q_3\right) \,, & \qquad i = 0, \dots, N-1\,.
\end{align}
Furthermore, we also flip 
\begin{align}
\Tr_g \Tr_4 \left(A^i Q_4 Q_4\right) \,, & \qquad i = 0, \dots, N-1 \,, \nn\\
\Tr_g \left(A^i\right) \,, & \qquad i = 2, \dots, N
\end{align}
because those with low powers of $A$, violate the unitarity bound. Thus, we introduce the following flipping fields in total:
\begin{align}
D_1^i \,, \quad D_2^i \,, \quad D_3^i \,, \quad b_1^i \,, \quad b_2^i \,, \quad b_3^i \,, \quad b_4^i \,,
& \qquad i = 0, \dots N-1 \,, \nn\\
a_i \,, & \qquad i = 2, \dots, N
\label{higherrankE6sing}
\end{align}
with the superpotential given in \eqref{eq:higher rank superpotential}-\eqref{higherrankb}-\eqref{higherranka}. In the following, we will denote this theory by $\widehat{\mathcal{T}}_3^N$, where the lower index represents the number of towers $b$-singlets included, while the upper index is the rank of the gauge group.

Once those operators are flipped, the manifest UV symmetry is given by
\begin{align}
SU(6) \times SU(2)_4 \times U(1)_v \times U(1)_a \,,
\end{align}
where, similarly to what we discussed in section \ref{sec3.1.3} for the rank one case, $SU(6)$ is formed by $SU(2)_1 \times SU(2)_2 \times SU(2)_3$ and the two combinations of $U(1)_{x_1,x_2,x_3}$ corresponding to the $U(1)_{t,u}$ symmetries we defined in \eqref{3u1m}-\eqref{3u1mfug}.
According to the argument in section \ref{m6}, this is supposed to be enhanced in the IR to
\begin{align}
E_6 \times U(1)_v \times U(1)_a\,,
\end{align}
where $U(1)_v$ is the remaining combination of $U(1)_{x_1,x_2,x_3}$ defined in \eqref{3u1m}-\eqref{3u1mfug}.

This enhancement can be checked by computing the superconformal index of the theory for low values of $N$, for example $N=2$. In this case, performing $a$-maximization we find the following approximate values of the mixing coefficients of the R-symmetry with $U(1)_{x_1}\times U(1)_{x_2}\times U(1)_{x_3}\times U(1)_a$:
 \begin{align}
R_1 =R_2 =R_3 \simeq \frac{12}{11} \,, \quad R_a \simeq \frac{1}{4} \,.
\end{align}
The superconformal index of the theory computed with these R-charges then reads
\begin{align}
\widehat{\mathcal{I}}_3^{N=2} &=1+ {\bf\overline{27}}\,v^{-2}a^{-\frac{1}{2}}(pq)^{\frac{69}{176}}(1+p+q)+{\bf\overline{27}}\,v^{-2}a^{\frac{1}{2}}(pq)^{\frac{91}{176}}+v^6a^{-\frac{1}{2}}(pq)^{\frac{101}{176}}v^6+\nn\\
&+a^{\frac{1}{2}}(pq)^{\frac{123}{176}}+a^{-2}(pq)^{\frac{3}{4}}+({\bf 351}'+{\bf 27})v^{-4}a^{-\frac{1}{2}}(pq)^{\frac{69}{88}}+({\bf 351}'+{\bf 351})v^{-4}(pq)^{\frac{10}{11}}+\nn\\
&+{\bf\overline{27}}\,v^4a^{-1}(pq)^{\frac{85}{88}}{\color{blue}-({\bf 78}+2)pq}+\cdots\,.
\end{align}
Each number is the character of an $E_6$ representation and, in particular, the term $-({\bf 78}+2)pq$ highlighted in blue reflects the current multiplet, which is in the adjoint representation  of $E_6 \times U(1)_v\times U(1)_a$. 


It is worth mentioning that the existence of the $N = 2$ copies of the gauge singlets listed in \eqref{higherrankE6sing} is crucial to realize the $E_6$ symmetry. Let us look at the second term ${\bf\overline{27}}\,v^{-2}a^{-\frac{1}{2}}(pq)^{\frac{69}{176}}$. ${\bf\overline{27}}$ is decomposed into the representations of $SU(6) \times SU(2)_4$ as follows:
\begin{align}
{\bf\overline{27}}\ \quad \rightarrow \quad \left(\mathbf{6},\mathbf{2}\right)+\left(\mathbf{\overline{15}},\mathbf{1}\right) .
\end{align}
$(\mathbf{6},\mathbf{2})$, the bifundamental representation of $SU(6) \times SU(2)_4$, is further decomposed into\footnote{We omit $U(1)$ charges in this discussion for simplicity.}
\begin{align}
(\mathbf{6},\mathbf{2}) \quad \rightarrow \quad (\mathbf{2},\mathbf{1},\mathbf{1},\mathbf{2})+(\mathbf{1},\mathbf{2},\mathbf{1},\mathbf{2})+(\mathbf{1},\mathbf{1},\mathbf{2},\mathbf{2})
\end{align}
of $SU(2)_1 \times SU(2)_2 \times SU(2)_3 \times SU(2)_4$, which are provided by the operators
\begin{align}
\Tr_g (Q_1 Q_4) \,, \quad \Tr_g (Q_2 Q_4) \,, \quad \Tr_g (Q_3 Q_4)
\end{align}
respectively. $(\mathbf{\overline{15}},\mathbf{1})$, the antisymmetric representation of $SU(6) \times SU(2)_4$, is also decomposed into
\begin{align}
(\mathbf{\overline{15}},\mathbf{1}) \quad \rightarrow \quad (\mathbf{1},\mathbf{2},\mathbf{2},\mathbf{1})+(\mathbf{2},\mathbf{1},\mathbf{2},\mathbf{1})+(\mathbf{2},\mathbf{2},\mathbf{1},\mathbf{1})+3 \times (\mathbf{1},\mathbf{1},\mathbf{1},\mathbf{1})
\end{align}
of $SU(2)_1 \times SU(2)_2 \times SU(2)_3 \times SU(2)_4$, which are provided by
\begin{align}
D_1^1 \,, \quad D_2^1 \,, \quad D_3^1 \,, \quad b_1^1 \,, \quad b_2^1 \,, \quad b_3^1
\end{align}
respectively. Similarly, the third term ${\bf\overline{27}}v^{-2}a^{\frac{1}{2}}(pq)^{\frac{91}{176}}$ is contributed by
\begin{align}
\Tr_g (A Q_1 Q_4) \,, \quad \Tr_g (A Q_2 Q_4) \,, \quad \Tr_g (A Q_3 Q_4) \,,
\end{align}
which form $\left(\mathbf{6},\mathbf{2}\right)$ of $SU(6) \times SU(2)_4$, and
\begin{align}
D_1^0 \,, \quad D_2^0 \,, \quad D_3^0 \,, \quad b_1^0 \,, \quad b_2^0 \,, \quad b_3^0 \,,
\end{align}
which form $\left(\mathbf{\overline{15}},\mathbf{1}\right)$ of $SU(6) \times SU(2)_4$. Thus, one can see that $D_{1,2,3}^i$ and $b_{1,2,3}^i$ for both $i = 0$ and $i = 1$ are crucially involved in the enhancement to $E_6$. On the other hand, the singlets $b_4^i$ for $i = 0, \, 1$ don't play any role in the $E_6$ enhancement. Nevertheless, we include them in the theory, especially $b_4^0$, because $b_4^0$ flips the operator $\Tr_g \Tr_4 \left(Q_4 Q_4\right)$ which also violates the unitarity bound.
\\

\subsection{$SO(10) \times U(1)^3$ model and $SO(10) \rightarrow SO(9) \rightarrow F_4$ deformations}
\label{sec:T2}

The next example is the model $\widehat{\mathcal{T}}_2^N$ exhibiting the $SO(10) \times U(1)^3$ global symmetry. As we discussed in section \ref{m10}, in order to obtain this model we need to flip, for example,  the operators in $(6,1,1)_{2,2}$ and those in $(4,2,1,)_{-1,2}$, which are representations of $SU(4) \times SU(2)^2 \times U(1)_u\times U(1)_v \subset SU(8)$. The corresponding operators are
\begin{align}
\Tr_g \left(A^i Q_1 Q_2\right) \,, & \nn\\
\Tr_g \Tr_m \left(A^i Q_m Q_m\right) \,, & \qquad m = 1,2 \,,
\end{align}
which construct $({\bf 6},{\bf 1},{\bf 1})_{2,2}$ for each $i = 0, \dots, N-1$ and
\begin{align}
&\Tr_g \Tr_3 \left(A^i Q_m Q_3\right) \,, \qquad m = 1,2 \,,
\end{align}
which construct $({\bf 4},{\bf 2},{\bf 1})_{-1,2}$ for each $i = 0, \dots, N-1$. In addition, we need to flip
\begin{align}
\Tr_g \Tr_4 \left(A^i Q_4 Q_4\right) \,, & \qquad i = 0, \dots, N-1 \,, \nn\\
\Tr_g \left(A^i\right) \,, & \qquad i = 2, \dots, N
\end{align}
because those with low $i$ fall below the unitarity bound. Thus, we introduce the following flipping fields:
\begin{align}
D_1^i \,, \quad D_2^i \,, \quad D_3^i \,, \quad b_1^i \,, \quad b_2^i \,, \quad b_4^i \,,
& \qquad i = 0, \dots N-1 \,, \nn\\
a_i \,, & \qquad i = 2, \dots, N
\label{higherrankSO10sing}
\end{align}
with the superpotential given in \eqref{eq:higher rank superpotential}-\eqref{higherrankb}-\eqref{higherranka}. 

Once those flipping fields are taken into account, the UV symmetry is broken to
\begin{align}
SU(4) \times SU(2)_3 \times SU(2)_4 \times U(1)_u \times U(1)_v \times U(1)_a \,,
\end{align}
where, similarly to what we discussed in section \ref{sec3.1.2} for the rank one case, $SU(4)$ is formed by $SU(2)_1 \times SU(2)_2$ and the combination of the abelian symmetries $2(U(1)_{x_1}-U(1)_{x_2})=U(1)_t$.
According to the argument in section \ref{m10}, this is supposed to be enhanced in the IR to
\begin{align}
SO(10) \times U(1)_u \times U(1)_v \times U(1)_a\,,
\end{align}
where $U(1)_{u,v}$ are the remaining combinations of $U(1)_{x_1,x_2,x_3}$ defined in \eqref{3u1m}-\eqref{3u1mfug}.

This enhancement of the global symmetry can be checked using the superconformal index for low values of the rank of the gauge group. For instance, for $N=2$ we find the following approximate values of the mixing coefficients from $a$-maximization:
\begin{align}
R_1 =R_2 \simeq \frac{11}{10} \,, \quad R_3 \simeq \frac{34}{33} \,, \quad R_a \simeq \frac{4}{15} \,.
\end{align}
The superconformal index of the theory computed with these R-charges then reads\footnote{We choose to parametrize the two $U(1)$ symmetries that don't participate in the enhancement with $U(1)_{u,v,a}$.}
\begin{align}
\widehat{\mathcal{I}}_2^{N=2} &=1+{\bf 10}\,v^{-2}u^{-2}a^{-\frac{1}{2}}(pq)^{\frac{23}{60}}+{\bf \overline{16}}\,v^{-2}u\,a^{-\frac{1}{2}}(pq)^{\frac{529}{1320}}+v^2u^{-4}a^{-\frac{1}{2}}(pq)^{\frac{74}{165}}+\nn\\
&+{\bf 10}\,v^{-2}u^{-2}a^{\frac{1}{2}}(pq)^{\frac{31}{60}}+{\bf\overline{16}}\,v^{-2}u\,a^{\frac{1}{2}}(pq)^{\frac{47}{88}}+v^6a^{-\frac{1}{2}}(pq)^{\frac{181}{330}}+v^2u^{-4}a^{\frac{1}{2}}(pq)^{\frac{32}{55}}+\nn\\
&+v^6a^{\frac{1}{2}}(pq)^{\frac{15}{22}}+a^{-2}(pq)^{\frac{11}{15}}+\cdots{\color{blue}-({\bf 45}+3)pq}+\cdots\,.
\end{align}
Each number is the character of an $SO(10)$ representation and, in particular, the term $-({\bf 45}+3)pq$ highlighted in blue reflects the current multiplet, which is in the adjoint representation  of $SO(10)\times U(1)_u \times U(1)_v\times U(1)_a$. 
\\

Analogously to the $SU(2)$ case, one can deform the theory by introducing extra superpotential terms of the form $b_m^i \, \Tr_g \Tr_n (A^i Q_n Q_n)$ with $m \neq n$. The first term we introduce is\footnote{Using the superconformal R-charge of theory $\widehat{\mathcal{T}}_2^{N=2}$ we can check that this deformation has R-charge $R[b_2^i \, \Tr_g \Tr_3 \left(A^i Q_3 Q_3\right)]\simeq1.93153<2$ for $i=0,1$, so it is a relevant deformation.}
\begin{align}
\label{eq:SO(9) deformation}
\Delta \mathcal W_2'{}^N = \sum_{i = 0}^{N-1} b_2^i \, \Tr_g \Tr_3 \left(A^i Q_3 Q_3\right)
\end{align}
and we label the theory obtained from this deformation as $\widehat{\mathcal{T}}_2'{}^N$.
Similarly to what happened in the rank one case of section \ref{sec3.2.1}, this deformation breaks the global symmetry of theory $\widehat{\mathcal{T}}_2^N$ from $SU(4) \times SU(2)_3 \times SU(2)_4 \times U(1)_u \times U(1)_v \times U(1)_a$ to
\begin{align}
SU(2)_1 \times USp(4) \times SU(2)_4 \times U(1)_{\tilde u} \times U(1)_v \times U(1)_a\,,
\end{align}
where again we are parametrizing the two surviving combinations of $U(1)_{x_1,x_2,x_3}$ with $U(1)_{\tilde{u},v}$ which are defined in \eqref{utildefug}-\eqref{utildecharge}.
Note, for example, that $D_1^i$ and $b_2^i$ construct the traceless antisymmetric representation of $USp(4)$.

Given the approximate mixing coefficients of $U(1)_{x_1} \times U(1)_{x_2} \times U(1)_{x_3} \times U(1)_a$ with the R-symmetry of the theory for rank $N=2$
\begin{align}
R_1 \simeq \frac{11}{10} \,, \quad R_2 = R_3 \simeq \frac{16}{15} \,, \quad R_a \simeq \frac{3}{11} \,,
\end{align}
the superconformal index for $N = 2$ is given by\footnote{We choose to parametrize the $U(1)$ symmetries, which don't participate in the enhancement, with $U(1)_{\tilde{u},v,a}$.}
\begin{align}
\widehat{\mathcal{I}}_2'{}^{N=2} &=1+ v^{-2}\tilde{u}^{-8}a^{-\frac{1}{2}}(pq)^{\frac{21}{55}}+{\bf 16}\,v^{-2}\tilde{u}^{-2}a^{-\frac{1}{2}}(pq)^{\frac{103}{264}}+{\bf 9}\,v^{-2}\tilde{u}^4a^{-\frac{1}{2}}(pq)^{\frac{263}{660}}+v^2\tilde{u}^{-4}a^{-\frac{1}{2}}(pq)^{\frac{307}{660}}+\nn\\
&+v^{-2}\tilde{u}^{-8}a^{\frac{1}{2}}(pq)^{\frac{57}{110}}+{\bf 16}\,v^{-2}\tilde{u}^{-2}a^{\frac{1}{2}}(pq)^{\frac{139}{264}}+{\bf 9}\,v^{-2}\tilde{u}^4a^{\frac{1}{2}}(pq)^{\frac{353}{660}}+v^6a^{-\frac{1}{2}}(pq)^{\frac{181}{330}}+\nn\\
&+v^2\tilde{u}^{-4}a^{\frac{1}{2}}(pq)^{\frac{397}{330}}+v^6a^{\frac{1}{2}}(pq)^{\frac{113}{165}}+a^{-2}(pq)^{\frac{8}{11}}+\cdots+({\bf 9}{\color{blue}-{\bf 36}-3})pq+\cdots\,.
\end{align}
Each number is the character of an $SO(9)$ representation and, in particular, the term $-({\bf 36}+3)pq$ highlighted in blue reflects the current multiplet, which is in the adjoint representation  of $SO(9)\times U(1)_{\tilde{u}} \times U(1)_v\times U(1)_a$.


We should comment that the index can be also written in terms of $SO(8)$ characters because any $SO(9)$ representation can be decomposed into $SO(8)$ representations. In that case, the term $\left(\mathbf{9}-\mathbf{36}-3\right)pq$ would be written as $\left(-\mathbf{28}_{SO(8)}-2\right) pq$, where $\mathbf{28}_{SO(8)}$ is the character of the adjoint representation of $SO(8)$. Indeed, if there is no marginal operator that can be constructed, the conserved current must be in the representation $\mathbf{28}_{SO(8)}$ and the enhanced non-abelian symmetry is $SO(8)$ rather than $SO(9)$. Thus, we have to show that the marginal operators in the representation $\mathbf{9}$ exist in order to claim that the enhanced symmetry includes $SO(9)$.

Let us list the operators contributing to order $pq$ except some boson-fermion pairs trivially canceled. We first define the following single trace operators organized into representations of the manifest symmetry $SU(2)_1 \times USp(4) \times SU(2)_4$, where we omit the charges under the abelian symmetries:
\begin{align}
\begin{aligned}
\Sigma_i = \left(D_1^i,b_2^i\right) \quad &\rightarrow \quad ({\bf 1},{\bf 5},{\bf 1}) \,, \\
\Pi_i = \Tr_g \left(A^i Q_1 Q_4\right) \quad &\rightarrow \quad ({\bf 2},{\bf 1},{\bf 2}) \,, \\
P_i = \Tr_g \left(A^i Q_{2,3} Q_4\right) \quad &\rightarrow \quad ({\bf 1},{\bf 4},{\bf 2}) \,, \\
L_i = \Tr_g \left(A^i Q_1 Q_{2,3}\right) \quad &\rightarrow \quad ({\bf 2},{\bf 4},{\bf 1}) \,, \\
M_i = \Tr_g \left(A^i Q_{2,3} Q_{2,3}\right) \quad &\rightarrow \quad ({\bf 1},{\bf 1},{\bf 1})+({\bf 1},{\bf 5},{\bf 1}) \,, \\
\sigma_i = \left(\overline \Psi_{D_1^i},\overline \Psi_{b_2^i}\right) \quad &\rightarrow \quad ({\bf 1},{\bf 5},{\bf 1}) \,, \\
\tau_i = \left(\overline \Psi_{D_3^i},\overline \Psi_{D_2^i}\right) \quad &\rightarrow \quad ({\bf 2},{\bf 4},{\bf 1}) \,, \\
\xi = \Tr_g \left(A \overline \Psi_A\right) \quad &\rightarrow \quad ({\bf 1},{\bf 1},{\bf 1}) \,, \\
\pi = \Tr_g \left(Q_1 \overline \Psi_1\right) \quad &\rightarrow \quad ({\bf 1},{\bf 1},{\bf 1})+({\bf 3},{\bf 1},{\bf 1}) \,, \\
\rho = \Tr_g \left(Q_4 \overline \Psi_4\right) \quad &\rightarrow \quad ({\bf 1},{\bf 1},{\bf 1})+({\bf 1},{\bf 1},{\bf 3}) \,, \\
\mu = \Tr_g \left(Q_{2,3} \overline \Psi_{2,3}\right) \quad &\rightarrow \quad ({\bf 1},{\bf 1},{\bf 1})+({\bf 1},{\bf 5},{\bf 1})+({\bf 1},{\bf 10},{\bf 1}) \,.
\end{aligned}
\end{align}
Then the bosonic operators contributing to order $pq$ are given by
\begin{align}
\begin{aligned}
\label{eq:SO(9) marginal}
\Tr_g \left(\lambda^2\right) \quad &\rightarrow \quad ({\bf 1},{\bf 1},{\bf 1}) \,, \\
M_0 \Pi_1 \quad &\rightarrow \quad ({\bf 2},{\bf 1},{\bf 2})+({\bf 2},{\bf 5},{\bf 2}) \,, \\
M_1 \Pi_0 \quad &\rightarrow \quad ({\bf 2},{\bf 1},{\bf 2})+({\bf 2},{\bf 5},{\bf 2}) \,, \\
L_0 P_1 \quad &\rightarrow \quad ({\bf 2},{\bf 1},{\bf 2})+({\bf 2},{\bf 5},{\bf 2})+({\bf 2},{\bf 10},{\bf 2}) \,, \\
L_1 P_0 \quad &\rightarrow \quad ({\bf 2},{\bf 1},{\bf 2})+({\bf 2},{\bf 5},{\bf 2})+({\bf 2},{\bf 10},{\bf 2}) \,, \\
M_0 \Sigma_0 \quad &\rightarrow \quad ({\bf 1},{\bf 1},{\bf 1})+({\bf 1},{\bf 5},{\bf 1})+({\bf 1},{\bf 10},{\bf 1})+({\bf 1},{\bf 14},{\bf 1}) \,, \\
M_1 \Sigma_1 \quad &\rightarrow \quad ({\bf 1},{\bf 1},{\bf 1})+({\bf 1},{\bf 5},{\bf 1})+({\bf 1},{\bf 10},{\bf 1})+({\bf 1},{\bf 14},{\bf 1})
\end{aligned}
\end{align}
where $\lambda$ is  the gaugino and the operators $M_0 \Pi_1, \, M_1 \Pi_0, \, L_0 P_1$ and $L_1 P_0$ are not all independent but satisfy 24 relations which come in the representations
\begin{align}
\label{eq:SO(9) relations}
({\bf 2},{\bf 1},{\bf 2})+({\bf 2},{\bf 5},{\bf 2}) \,.
\end{align}
 In appendix \ref{appPL} we give an argument for this based on the analysis of the plethystic logarithm of the superconformal index. 
On the other hand, the fermionic operators contributing to order $pq$ are
\begin{align}
\begin{aligned}
\label{eq:SO(9) fermions}
\sigma_0 \Pi_1 \quad &\rightarrow \quad ({\bf 2},{\bf 5},{\bf 2}) \,, \\
\sigma_1 \Pi_0 \quad &\rightarrow \quad ({\bf 2},{\bf 5},{\bf 2}) \,, \\
\tau_0 P_0 \quad &\rightarrow \quad ({\bf 2},{\bf 1},{\bf 2})+({\bf 2},{\bf 5},{\bf 2})+({\bf 2},{\bf 10},{\bf 2}) \,, \\
\tau_1 P_1 \quad &\rightarrow \quad ({\bf 2},{\bf 1},{\bf 2})+({\bf 2},{\bf 5},{\bf 2})+({\bf 2},{\bf 10},{\bf 2}) \,, \\ 
\sigma_0 \Sigma_0 \quad &\rightarrow \quad ({\bf 1},{\bf 1},{\bf 1})+({\bf 1},{\bf 10},{\bf 1})+({\bf 1},{\bf 14},{\bf 1}) \,, \\
\sigma_1 \Sigma_1 \quad &\rightarrow \quad ({\bf 1},{\bf 1},{\bf 1})+({\bf 1},{\bf 10},{\bf 1})+({\bf 1},{\bf 14},{\bf 1}) \,, \\ 
\xi \quad &\rightarrow \quad ({\bf 1},{\bf 1},{\bf 1}) \,, \\
\pi \quad &\rightarrow \quad ({\bf 1},{\bf 1},{\bf 1})+({\bf 3},{\bf 1},{\bf 1}) \,, \\
\rho \quad &\rightarrow \quad ({\bf 1},{\bf 1},{\bf 1})+({\bf 1},{\bf 1},{\bf 3}) \,, \\
\mu \quad &\rightarrow \quad ({\bf 1},{\bf 1},{\bf 1})+({\bf 1},{\bf 5},{\bf 1})+({\bf 1},{\bf 10},{\bf 1}) \,.
\end{aligned}
\end{align}

Now let us look at the bosonic operators $M_0 \Pi_1, \, M_1 \Pi_0, \, L_0 P_1$ and $L_1 P_0$, which satisfy the relations \eqref{eq:SO(9) relations} as well as the F-term conditions realized by parts of the fermionic operators $\sigma_0 \Pi_1, \, \sigma_1 \Pi_0, \, \tau_0 P_0$ and $\tau_1 P_1$. The remaining independent one is only $({\bf 2},{\bf 1},{\bf 2})$, which cannot be lifted because there is no fermionic operator to be paired up. Similarly, $M_0 \Sigma_0$ and $M_1 \Sigma_1$ can be paired up with $\sigma_0 \Sigma_0$ and $\sigma_1 \Sigma_1$ respectively, leaving
\begin{align}
\label{eq:trace}
\Tr_{USp(4)} (M_0) \Sigma_0 \,, \qquad \Tr_{USp(4)} (M_1) \Sigma_1\,.
\end{align}
Note that those cancelations reflect the $F$-term conditions from the superpotential.

The remaining operators in \eqref{eq:trace} are in the representation $2 \times ({\bf 1},{\bf 5},{\bf 1})$. One combination of them can become a long multiplet being paired up with the traceless antisymmetric part of $\mu$. On the other hand, the other combination still remains short and combines with the remaining $({\bf 2},{\bf 1},{\bf 2})$ of $M_0 \Pi_1, \, M_1 \Pi_0, \, L_0 P_1, \, L_1 P_0$ into $\mathbf{9}$ of $SO(9)$. Therefore, we have found 9 marginal operators consisting of $({\bf 2},{\bf 1},{\bf 2})$ from $M_0 \Pi_1, \, M_1 \Pi_0, \, L_0 P_1, \, L_1 P_0$ and $({\bf 1},{\bf 5},{\bf 1})$ from \eqref{eq:trace}. Moreover, the remaining fermionic operators constitute the supersymmetric partners of the conserved current in the adjoint representation of $SO(9)\times U(1)_{\tilde{u}} \times U(1)_v\times U(1)_a$.
\\

The second deformation we introduce is\footnote{Using the superconformal R-charge of theory $\widehat{\mathcal{T}}_2'{}^{N=2}$ we can check that this deformation has R-charge $R[b_1^i \, \Tr_g \Tr_3 \left(A^i Q_3 Q_3\right)]\simeq1.9671<2$ for $i=0,1$, so it is a relevant deformation.}
\begin{align}
\Delta \mathcal {W}_2''{}^N = \sum_{i = 0}^{N-1} b_1^i \, \Tr_g \Tr_3 \left(A^i Q_3 Q_3\right)
\end{align}
and we label the theory obtained from this deformation as $\widehat{\mathcal{T}}_2''{}^N$.
Similarly to what happened in the rank one case of section \ref{sec3.2.1}, due to this deformation $U(1)_{\tilde u}$ is broken whereas $SU(2)_1 \times USp(4)$ gets enhanced to $USp(6)$. Thus, the entire UV global symmetry is now given by
\begin{align}
USp(6) \times SU(2)_4 \times U(1)_v \times U(1)_a
\end{align}
where once again $U(1)_v$ is defined as in \eqref{3u1m}-\eqref{3u1mfug}. Note, for example, that $D_1^i, \, D_2^i, \, D_3^i$ and $b_1^i, \, b_2^i$ are organized into the traceless antisymmetric representation of $USp(6)$.

Given the approximate mixing coefficients of $U(1)_{x_1} \times U(1)_{x_2} \times U(1)_{x_3} \times U(1)_a$ with the R-symmetry of the theory for rank $N=2$
\begin{align}
R_1 = R_2 = R_3 \simeq \frac{14}{13} \,, \quad R_a \simeq \frac{3}{11} \,,
\end{align}
we have the superconformal index for $N = 2$ as follows:
\begin{align}
\widehat{\mathcal{I}}_2''{}^{N=2} &=1+ {\bf 26}\,v^{-2}a^{-\frac{1}{2}}(pq)^{\frac{225}{572}}+v^2a^{-\frac{1}{2}}(pq)^{\frac{269}{572}}+{\bf 26}\,v^{-2}a^{\frac{1}{2}}(pq)^{\frac{303}{572}}+v^6a^{-\frac{1}{2}}(pq)^{\frac{313}{572}}+\nn\\
&+v^2a^{\frac{1}{2}}(pq)^{\frac{347}{572}}+v^6a^{\frac{1}{2}}(pq)^{\frac{391}{572}}+a^{-2}(pq)^{\frac{8}{11}}+\cdots+({\bf 26}{\color{blue}-{\bf 52}-2})pq+\cdots\,.
\end{align}
Each number is the character of an $F_4$ representation and, in particular, the term $-({\bf 52}+2)pq$ highlighted in blue reflects the current multiplet, which is in the adjoint representation  of $F_4\times U(1)_v\times U(1)_a$.


Again we need to check if the marginal operators in the representation $\mathbf{26}$ really exist. We first define the following single trace operators organized into representations of the manifest symmetry $USp(6) \times SU(2)_4$:
\begin{align}
\begin{aligned}
\Pi_i = \Tr_g \left(A^i Q_{1,2,3} Q_4\right) \quad &\rightarrow \quad ({\bf 6},{\bf 2}) \,, \\
\Sigma_i = \left(D_1^i,D_2^i,D_3^i,b_1^i,b_2^i\right) \quad &\rightarrow \quad ({\bf 14},{\bf 1}) \,, \\
M_i = \Tr_g \left(A^i Q_{1,2,3} Q_{1,2,3}\right) \quad &\rightarrow \quad ({\bf 1},{\bf 1})+({\bf 14},{\bf 1}) \,, \\
\xi = \Tr_g \left(A \overline \Psi_A\right) \quad &\rightarrow \quad ({\bf 1},{\bf 1}) \,, \\
\pi = \Tr_g \left(Q_4 \overline \Psi_4\right) \quad &\rightarrow \quad ({\bf 1},{\bf 1})+({\bf 1},{\bf 3}) \,, \\
\sigma_i = \left(\overline \Psi_{D_1^i},\overline \Psi_{D_2^i},\overline \Psi_{D_3^i},\overline \Psi_{b_1^i},\overline \Psi_{b_2^i}\right) \quad &\rightarrow \quad ({\bf 14},{\bf 1}) \,, \\
\mu = \Tr_g \left(Q_{1,2,3} \overline \Psi_{1,2,3}\right) \quad &\rightarrow \quad ({\bf 1},{\bf 1})+({\bf 14},{\bf 1})+({\bf 21},{\bf 1})
\end{aligned}
\end{align}
where we have used the same names of the operators as those in the previous $SO(9)$ case.
The operators contributing to order $pq$ are now given by
\begin{align}
\begin{aligned}
\label{eq:F4 bosons}
\Tr_g \left(\lambda^2\right) \quad &\rightarrow \quad ({\bf 1},{\bf 1}) \,, \\
M_0 \Pi_1 \quad &\rightarrow \quad 2 \times ({\bf 6},{\bf 2})+({\bf 14'},{\bf 2})+({\bf 64},{\bf 2}) \,, \\
M_1 \Pi_0 \quad &\rightarrow \quad 2 \times ({\bf 6},{\bf 2})+({\bf 14'},{\bf 2})+({\bf 64},{\bf 2}) \,, \\
M_0 \Sigma_0 \quad &\rightarrow \quad ({\bf 1},{\bf 1})+2 \times ({\bf 14},{\bf 1})+({\bf 21},{\bf 1})+({\bf 70},{\bf 1})+({\bf 90},{\bf 1}) \,, \\
M_1 \Sigma_1 \quad &\rightarrow \quad ({\bf 1},{\bf 1})+2 \times ({\bf 14},{\bf 1})+({\bf 21},{\bf 1})+({\bf 70},{\bf 1})+({\bf 90},{\bf 1})
\end{aligned}
\end{align}
for bosonic ones and
\begin{align}
\begin{aligned}
\label{eq:F4 fermions}
\sigma_0 \Pi_1 \quad &\rightarrow \quad ({\bf 6},{\bf 2})+({\bf 14'},{\bf 2})+({\bf 64},{\bf 2}) \,, \\
\sigma_1 \Pi_0 \quad &\rightarrow \quad ({\bf 6},{\bf 2})+({\bf 14'},{\bf 2})+({\bf 64},{\bf 2}) \,, \\
\sigma_0 \Sigma_0 \quad &\rightarrow \quad ({\bf 1},{\bf 1})+({\bf 14},{\bf 1})+({\bf 21},{\bf 1})+({\bf 70},{\bf 1})+({\bf 90},{\bf 1}) \,, \\
\sigma_1 \Sigma_1 \quad &\rightarrow \quad ({\bf 1},{\bf 1})+({\bf 14},{\bf 1})+({\bf 21},{\bf 1})+({\bf 70},{\bf 1})+({\bf 90},{\bf 1}) \,, \\ 
\xi \quad &\rightarrow \quad ({\bf 1},{\bf 1}) \,, \\
\pi \quad &\rightarrow \quad ({\bf 1},{\bf 1})+({\bf 1},{\bf 3}) \,, \\
\mu \quad &\rightarrow \quad ({\bf 1},{\bf 1})+({\bf 14},{\bf 1})+({\bf 21},{\bf 1})
\end{aligned}
\end{align}
for fermionic ones. Similarly to the previous case, $M_0 \Pi_1$ and $M_1 \Pi_0$ are not all independent, but satisfy 40 relations
\begin{align}
\label{eq:F4 relations}
({\bf 6},{\bf 2})+({\bf 14'},{\bf 2}) \,.
\end{align}
In appendix \ref{appPL} we check this looking at the plethystic logarithm of the superconformal index.

Comparing \eqref{eq:F4 bosons} and \eqref{eq:F4 fermions}, we find that there are exactly 26 bosonic operators in the representation $({\bf 6},{\bf 2})+({\bf 14},{\bf 1})$ that cannot be paired up with any of the fermionic operators in \eqref{eq:F4 fermions}. Therefore, the theory has the marginal operators in the representation $({\bf 6},{\bf 2})+({\bf 14},{\bf 1})$, which is $\mathbf{26}$ of $F_4$. Moreover, the remaining fermionic operators constitute the supersymmetric partners of the conserved current in the adjoint representation of $F_4\times U(1)_v\times U(1)_a$.
\\

\subsection{$SO(8) \times U(1)^4$ models and $SO(8) \rightarrow SO(9) \rightarrow SO(8)$ deformations}

The last two examples are the models exhibiting the $SO(8) \times U(1)^4$ global symmetry. As we discussed in section \ref{m8}, in order to obtain this model we need to flip, for example,  the operators in $({\bf 2},{\bf 2},{\bf 1},{\bf 1})_{0,2,2}+({\bf 1},{\bf 2},{\bf 2},{\bf 1})_{-1,-1,2}+({\bf 2},{\bf 1},{\bf 2},{\bf 1})_{1,-1,2}+({\bf 1},{\bf 1},{\bf 1},{\bf 1})_{2,2,2}$, which are representations of $SU(2)^4 \times U(1)_t\times U(1)_u\times U(1)_t\subset SU(8)$. The corresponding operators are
\begin{align}
\Tr_g \left(A^i Q_1 Q_2\right) \,, & \nn\\
\Tr_g \left(A^i Q_2 Q_3\right) \,, & \nn\\
\Tr_g \left(A^i Q_1 Q_3\right) \,, & \nn\\
\Tr_g \Tr_1 \left(A^i Q_1 Q_1\right)
\end{align}
for $i = 0, \dots, N-1$ respectively. In addition, we need to flip
\begin{align}
\Tr_g \Tr_4 \left(A^i Q_4 Q_4\right) \,, & \qquad i = 0, \dots, N-1 \,, \nn\\
\Tr_g \left(A^i\right) \,, & \qquad i = 2, \dots, N
\end{align}
because those with low $i$ violate the unitarity bound. Thus, we introduce the following flipping fields:
\begin{align}
D_1^i \,, \quad D_2^i \,, \quad D_3^i \,, \quad b_1^i \,, \quad b_4^i \,,
& \qquad i = 0, \dots N-1 \,, \nn\\
a_i \,, & \qquad i = 2, \dots, N
\label{higherrankSO8sing}
\end{align}
with the superpotential given in \eqref{eq:higher rank superpotential}-\eqref{higherrankb}-\eqref{higherranka}. We denote this theory by $\widehat{\mathcal{T}}_1^N$.

Once those flipping fields are taken into account, the UV symmetry is broken to
\begin{align}
SU(2)_1 \times SU(2)_2 \times SU(2)_3 \times SU(2)_4 \times U(1)_t \times U(1)_u \times U(1)_v \times U(1)_a \,,
\end{align}
which is supposed to be enhanced in the IR to
\begin{align}
SO(8) \times U(1)_t \times U(1)_u \times U(1)_v \times U(1)_a\,,
\end{align}
where $U(1)_{t,u,v}$ are again defined as in \eqref{3u1m}-\eqref{3u1mfug}.

This enhancement of the global symmetry can be checked using the superconformal index for low values of the rank of the gauge group. For instance, for $N=2$ we find the following approximate values of the mixing coefficients from $a$-maximization:
\begin{align}
R_1 \simeq \frac{10}{9} \,, \quad R_2=R_3 \simeq \frac{22}{11} \,, \quad R_a \simeq \frac{3}{11} \,.
\end{align}
The superconformal index of the theory computed with these R-charges then reads\footnote{Since none of the $U(1)$ symmetries participates in the enhancement it is equivalent to parametrize them with $U(1)_{x_1,x_2,x_3,a}$ or $U(1)_{t,u,v,a}$ when computing the index. We decide to use the latter parametrization. }
\begin{align}
\widehat{\mathcal{I}}_1^{N=2} &=1+v^{-2}u^{-2}t^{-2}a^{-\frac{1}{2}}(pq)^{\frac{149}{396}}+({\bf 8}_vv^{-2}u^{-2}a^{-\frac{1}{2}}+{\bf 8}_sv^{-2}u\,t^{-1}a^{-\frac{1}{2}}(pq)^{\frac{1087}{2772}})+\nn\\
&+{\bf 8}_cv^{-2}u\,t\,a^{-\frac{1}{2}}(pq)^{\frac{377}{924}}+(v^2u^{-4}a^{-\frac{1}{2}}+v^2u^2t^{-2}a^{-\frac{1}{2}})(pq)^{\frac{421}{924}}+v^{-2}u^{-2}t^{-2}a^{\frac{1}{2}}(pq)^{\frac{203}{396}}+\nn\\
&+({\bf 8}_vv^{-2}u^{-2}a^{\frac{1}{2}}+{\bf 8}_sv^{-2}u\,t^{-1}a^{\frac{1}{2}})(pq)^{\frac{1465}{2772}}+v^6a^{-\frac{1}{2}}(pq)^{\frac{1483}{2772}}+{\bf 8}_cv^{-2}u\,t\,a^{\frac{1}{2}}(pq)^{\frac{503}{924}}+\nn\\
&+(v^2u^{-4}a^{\frac{1}{2}}+v^2u^2t^{-2}a^{\frac{1}{2}})(pq)^{\frac{547}{924}}+v^6a^{\frac{1}{2}}(pq)^{\frac{1861}{2772}}+a^{-2}(pq)^{\frac{8}{11}}+\nn\\
&+\cdots+(\mathbf 8_c u^{-3}t+\mathbf 8_c  u^3t^{-1}{\color{blue}-{\bf 28}-4})pq+\cdots\,.
\end{align}
Each number is the character of an $SO(8)$ representation and, in particular, the term $-({\bf 28}+4)pq$ highlighted in blue reflects the current multiplet, which is in the adjoint representation  of $SO(8)\times U(1)_t\times U(1)_u \times U(1)_v\times U(1)_a$. 
\\

Analogously to the $SU(2)$ case, one can deform the theory by introducing extra superpotential terms of the form $b_1^i \, \Tr_g \Tr_l (A^i Q_l Q_l)$. The first term we introduce is\footnote{Using the superconformal R-charge of theory $\widehat{\mathcal{T}}_1^{N=2}$ we can check that this deformation has R-charge $R[b_1^i \, \Tr_g \Tr_2 \left(A^i Q_2 Q_2\right)]\simeq1.93212<2$ for $i=0,1$, so it is a relevant deformation.}
\begin{align}
\Delta {\mathcal W_1'}^N = \sum_{i = 0}^{N-1} b_1^i \, \Tr_g \Tr_2 \left(A^i Q_2 Q_2\right)
\end{align}
and we label the theory obtained from this deformation as $\widehat{\mathcal{T}}_1'{}^N$.
Similarly to what happened in the rank one case of section \ref{sec3.2.2}, this deformation breaks $U(1)_t$ and makes $SU(2)_1 \times SU(2)_2$ recombine into $USp(4)$. Therefore, the manifest symmetry is now given by
\begin{align}
USp(4) \times SU(2)_3 \times SU(2)_4 \times U(1)_u \times U(1)_v \times U(1)_a
\end{align}
where as usual $U(1)_{u,v}$ are defined in \eqref{3u1m}-\eqref{3u1mfug}.
Note, for example, that $D_1^i$ and $b_1^i$ construct the traceless antisymmetric representation of $USp(4)$.

Given the approximate mixing coefficients of $U(1)_{x_1} \times U(1)_{x_2} \times U(1)_{x_3} \times U(1)_a$ with the R-symmetry of the theory for rank $N=2$
\begin{align}
R_1 = R_2 \simeq \frac{13}{12} \,, \quad R_3 \simeq \frac{23}{22} \,, \quad R_a \simeq \frac{3}{11} \,,
\end{align}
the superconformal index for $N = 2$ is given by\footnote{We choose to parametrize the $U(1)$ symmetries, which don't participate in the enhancement, with $U(1)_{u,v,a}$.}
\begin{align}
\widehat{\mathcal{I}}_1'{}^{N=2} &=1+{\bf 9}\,v^{-2}u^{-2}a^{-\frac{1}{2}}(pq)^{\frac{103}{264}}+{\bf 16}\,v^{-2}u\,a^{-\frac{1}{2}}(pq)^{\frac{221}{528}}+v^2u^{-4}a^{-\frac{1}{2}}(pq)^{\frac{5}{11}}+v^2u^2a^{-\frac{1}{2}}(pq)^{\frac{125}{264}}+\nn\\
&+{\bf 9}\,v^{-2}u^{-2}a^{\frac{1}{2}}(pq)^{\frac{139}{264}}+{\bf 16}\,v^{-2}u\,a^{\frac{1}{2}}(pq)^{\frac{283}{528}}+v^6a^{-\frac{1}{2}}(pq)^{\frac{71}{132}}+v^2u^{-4}a^{\frac{1}{2}}(pq)^{\frac{13}{22}}+\nn\\
&+v^2u^2a^{\frac{1}{2}}(pq)^{\frac{161}{264}}+v^6a^{\frac{1}{2}}(pq)^{\frac{89}{132}}+a^{-2}(pq)^{\frac{8}{11}}+\cdots+({\bf 9}{\color{blue}-{\bf 36}-3})pq+\cdots\,.
\end{align}
Each number is the character of an $SO(9)$ representation and, in particular, the term $-({\bf 36}+3)pq$ highlighted in blue reflects the current multiplet, which is in the adjoint representation  of $SO(9)\times U(1)_{u} \times U(1)_v\times U(1)_a$. Note that there exist the marginal operators in the representation $\mathbf{9}$, which can be explicitly constructed in a similar way to the $SO(9) \times U(1)^3$ model in the previous subsection.
\\

The second deformation we introduce is\footnote{Using the superconformal R-charge of theory $\widehat{\mathcal{T}}_1'{}^{N=2}$ we can check that this deformation has R-charge $R[b_1^i \, \Tr_g \Tr_3 \left(A^i Q_3 Q_3\right)]\simeq1.96449<2$ for $i=0,1$, so it is a relevant deformation.}
\begin{align}
\Delta {\mathcal {W}_1''}^N = \sum_{i = 0}^{N-1} b_1^i \, \Tr_g \Tr_3 \left(A^i Q_3 Q_3\right) \,,
\end{align}
and we label the theory obtained from this deformation as $\widehat{\mathcal{T}}_1''{}^N$. Similarly to what happened in the rank one case of section \ref{sec3.2.2}, this deformation breaks $U(1)_u$ as well as $USp(4)$ into $SU(2)_1 \times SU(2)_2$. Hence, the entire UV global symmetry is now given by
\begin{align}
SU(2)_1 \times SU(2)_2 \times SU(2)_3 \times SU(2)_4 \times U(1)_v \times U(1)_a
\end{align}
where as usual $U(1)_{v}$ is defined in \eqref{3u1m}-\eqref{3u1mfug}.

Given the approximate mixing coefficients of $U(1)_1 \times U(1)_2 \times U(1)_3 \times U(1)_a$ with the R-symmetry of the theory of rank $N=2$
\begin{align}
R_1 = R_2 = R_3 \simeq \frac{15}{14} \,, \quad R_a \simeq \frac{3}{11} \,,
\end{align}
we have the superconformal index for $N = 2$ as follows:
\begin{align}
\widehat{\mathcal{I}}_1''{}^{N=2} &=1+({\bf 8}_v+{\bf 8}_s+{\bf 8}_c+1)v^{-2}a^{-\frac{1}{2}}(pq)^{\frac{61}{154}}+2v^2a^{-\frac{1}{2}}(pq)^{\frac{36}{77}}+({\bf 8}_v+{\bf 8}_s+{\bf 8}_c+\nn\\
&+1)v^{-2}a^{\frac{1}{2}}(pq)^{\frac{41}{77}}+v^6a^{-\frac{1}{2}}(pq)^{\frac{83}{154}}+2v^2a^{\frac{1}{2}}(pq)^{\frac{93}{154}}+v^6a^{\frac{1}{2}}(pq)^{\frac{52}{77}}+a^{-2}(pq)^{\frac{8}{11}}+\nn\\
&+\cdots+[2({\bf 8}_v+{\bf 8}_s+{\bf 8}_c+1){\color{blue}-{\bf 28}-2}]pq+\cdots\,.
\end{align}
Each number is the character of an $SO(8)$ representation and, in particular, the term $-({\bf 28}+2)pq$ highlighted in blue reflects the current multiplet, which is in the adjoint representation  of $SO(8) \times U(1)_v\times U(1)_a$.


One has to check if the marginal operators in two copies of the representations $\mathbf{8}_s+\mathbf{8}_v+\mathbf{8}_c+2$ exist. Although only $SU(2)^4 \times U(1)^2$ is manifest in this model, we will see that the operators are organized in a very similar way to the $F_4 \times U(1)^2$ model in the previous section, whose manifest symmetry was $USp(6) \times SU(2) \times U(1)^2$. Thus, here we also use a similar notation for the single trace operators, which are now in the representations of $SU(2)_1 \times SU(2)_2 \times SU(2)_3 \times SU(2)_4$:
\begin{align}
\begin{aligned}
\Pi_i = \Tr_g \left(A^i Q_{1,2,3} Q_4\right) \quad &\rightarrow \quad ({\bf 2},{\bf 1},{\bf 1},{\bf 2})+({\bf 1},{\bf 2},{\bf 1},{\bf 2})+({\bf 1},{\bf 1},{\bf 2},{\bf 2}) \,, \\
\Sigma_i = \left(b_1^i,D_1^i,D_2^i,D_3^i\right) \quad &\rightarrow \quad ({\bf 1},{\bf 1},{\bf 1},{\bf 1})+({\bf 2},{\bf 2},{\bf 1},{\bf 1})+({\bf 1},{\bf 2},{\bf 2},{\bf 1})+({\bf 2},{\bf 1},{\bf 2},{\bf 1}) \,, \\
M_i = \Tr_g \left(A^i Q_{1,2,3} Q_{1,2,3}\right) \quad &\rightarrow \quad 3 \times ({\bf 1},{\bf 1},{\bf 1},{\bf 1})+({\bf 2},{\bf 2},{\bf 1},{\bf 1})+({\bf 1},{\bf 2},{\bf 2},{\bf 1})+({\bf 2},{\bf 1},{\bf 2},{\bf 1}) \,, \\
\xi = \Tr_g \left(A \overline \Psi_A\right) \quad &\rightarrow \quad ({\bf 1},{\bf 1},{\bf 1},{\bf 1}) \,, \\
\pi = \Tr_g \left(Q_4 \overline \Psi_4\right) \quad &\rightarrow \quad ({\bf 1},{\bf 1},{\bf 1},{\bf 1})+({\bf 1},{\bf 1},{\bf 1},{\bf 3}) \,, \\
\sigma_i = \left(\overline \Psi_{b_1^i},\overline \Psi_{D_1^i},\overline \Psi_{D_2^i},\overline \Psi_{D_3^i}\right) \quad &\rightarrow \quad ({\bf 1},{\bf 1},{\bf 1},{\bf 1})+({\bf 2},{\bf 2},{\bf 1},{\bf 1})+({\bf 1},{\bf 2},{\bf 2},{\bf 1})+({\bf 2},{\bf 1},{\bf 2},{\bf 1}) \,, \\
\mu = \Tr_g \left(Q_{1,2,3} \overline \Psi_{1,2,3}\right) \quad &\rightarrow \quad 3 \times ({\bf 1},{\bf 1},{\bf 1},{\bf 1})+2 \times ({\bf 2},{\bf 2},{\bf 1},{\bf 1})+2 \times ({\bf 1},{\bf 2},{\bf 2},{\bf 1})+2 \times ({\bf 2},{\bf 1},{\bf 2},{\bf 1}) \\
&\qquad +({\bf 3},{\bf 1},{\bf 1},{\bf 1})+({\bf 1},{\bf 3},{\bf 1},{\bf 1})+({\bf 1},{\bf 1},{\bf 3},{\bf 1})\,.
\end{aligned}
\end{align}
The operators contributing to order $pq$ are now given by
\begin{align}
\begin{aligned}
\label{eq:SO(8) bosons}
\Tr_g \left(\lambda^2\right) \quad &\rightarrow \quad ({\bf 1},{\bf 1},{\bf 1},{\bf 1}) \,, \\
M_0 \Pi_1 \quad &\rightarrow \quad 5 \times ({\bf 2},{\bf 1},{\bf 1},{\bf 2})+5 \times ({\bf 1},{\bf 2},{\bf 1},{\bf 2})+5 \times ({\bf 1},{\bf 1},{\bf 2},{\bf 2}) \\
&\qquad +({\bf 3},{\bf 2},{\bf 1},{\bf 2})+({\bf 3},{\bf 1},{\bf 2},{\bf 2})+({\bf 2},{\bf 3},{\bf 1},{\bf 2})+({\bf 1},{\bf 3},{\bf 2},{\bf 2})+({\bf 2},{\bf 1},{\bf 3},{\bf 2})+({\bf 1},{\bf 2},{\bf 3},{\bf 2}) \\
&\qquad +3 \times ({\bf 2},{\bf 2},{\bf 2},{\bf 2}) \,, \\
M_1 \Pi_0 \quad &\rightarrow \quad 5 \times ({\bf 2},{\bf 1},{\bf 1},{\bf 2})+5 \times ({\bf 1},{\bf 2},{\bf 1},{\bf 2})+5 \times ({\bf 1},{\bf 1},{\bf 2},{\bf 2}) \\
&\qquad +({\bf 3},{\bf 2},{\bf 1},{\bf 2})+({\bf 3},{\bf 1},{\bf 2},{\bf 2})+({\bf 2},{\bf 3},{\bf 1},{\bf 2})+({\bf 1},{\bf 3},{\bf 2},{\bf 2})+({\bf 2},{\bf 1},{\bf 3},{\bf 2})+({\bf 1},{\bf 2},{\bf 3},{\bf 2}) \\
&\qquad +3 \times ({\bf 2},{\bf 2},{\bf 2},{\bf 2}) \,, \\
M_0 \Sigma_0 \quad &\rightarrow \quad 6 \times ({\bf 1},{\bf 1},{\bf 1},{\bf 1})+6 \times ({\bf 2},{\bf 2},{\bf 1},{\bf 1})+6 \times ({\bf 1},{\bf 2},{\bf 2},{\bf 1})+6 \times ({\bf 2},{\bf 1},{\bf 2},{\bf 1}) \\
&\qquad +2 \times ({\bf 3},{\bf 1},{\bf 1},{\bf 1})+2 \times ({\bf 1},{\bf 3},{\bf 1},{\bf 1})+2 \times ({\bf 1},{\bf 1},{\bf 3},{\bf 1}) \\
&\qquad +({\bf 3},{\bf 3},{\bf 1},{\bf 1})+({\bf 1},{\bf 3},{\bf 3},{\bf 1})+({\bf 3},{\bf 1},{\bf 3},{\bf 1}) \\
&\qquad +2 \times ({\bf 3},{\bf 2},{\bf 2},{\bf 1})+2 \times ({\bf 2},{\bf 3},{\bf 2},{\bf 1})+2 \times ({\bf 2},{\bf 2},{\bf 3},{\bf 1}) \,, \\
M_1 \Sigma_1 \quad &\rightarrow \quad 6 \times ({\bf 1},{\bf 1},{\bf 1},{\bf 1})+6 \times ({\bf 2},{\bf 2},{\bf 1},{\bf 1})+6 \times ({\bf 1},{\bf 2},{\bf 2},{\bf 1})+6 \times ({\bf 2},{\bf 1},{\bf 2},{\bf 1}) \\
&\qquad +2 \times ({\bf 3},{\bf 1},{\bf 1},{\bf 1})+2 \times ({\bf 1},{\bf 3},{\bf 1},{\bf 1})+2 \times ({\bf 1},{\bf 1},{\bf 3},{\bf 1}) \\
&\qquad +({\bf 3},{\bf 3},{\bf 1},{\bf 1})+({\bf 1},{\bf 3},{\bf 3},{\bf 1})+({\bf 3},{\bf 1},{\bf 3},{\bf 1}) \\
&\qquad +2 \times ({\bf 3},{\bf 2},{\bf 2},{\bf 1})+2 \times ({\bf 2},{\bf 3},{\bf 2},{\bf 1})+2 \times ({\bf 2},{\bf 2},{\bf 3},{\bf 1})
\end{aligned}
\end{align}
for bosonic ones and
\begin{align}
\begin{aligned}
\label{eq:SO(8) fermions}
\sigma_0 \Pi_1 \quad &\rightarrow \quad 3 \times ({\bf 2},{\bf 1},{\bf 1},{\bf 2})+3 \times ({\bf 1},{\bf 2},{\bf 1},{\bf 2})+3 \times ({\bf 1},{\bf 1},{\bf 2},{\bf 2}) \\
&\qquad +({\bf 3},{\bf 2},{\bf 1},{\bf 2})+({\bf 3},{\bf 1},{\bf 2},{\bf 2})+({\bf 2},{\bf 3},{\bf 1},{\bf 2})+({\bf 1},{\bf 3},{\bf 2},{\bf 2})+({\bf 2},{\bf 1},{\bf 3},{\bf 2})+({\bf 1},{\bf 2},{\bf 3},{\bf 2}) \\
&\qquad +3 \times ({\bf 2},{\bf 2},{\bf 2},{\bf 2}) \,, \\
\sigma_1 \Pi_0 \quad &\rightarrow \quad 3 \times ({\bf 2},{\bf 1},{\bf 1},{\bf 2})+3 \times ({\bf 1},{\bf 2},{\bf 1},{\bf 2})+3 \times ({\bf 1},{\bf 1},{\bf 2},{\bf 2}) \\
&\qquad +({\bf 3},{\bf 2},{\bf 1},{\bf 2})+({\bf 3},{\bf 1},{\bf 2},{\bf 2})+({\bf 2},{\bf 3},{\bf 1},{\bf 2})+({\bf 1},{\bf 3},{\bf 2},{\bf 2})+({\bf 2},{\bf 1},{\bf 3},{\bf 2})+({\bf 1},{\bf 2},{\bf 3},{\bf 2}) \\
&\qquad +3 \times ({\bf 2},{\bf 2},{\bf 2},{\bf 2}) \,, \\
\sigma_0 \Sigma_0 \quad &\rightarrow \quad 4 \times ({\bf 1},{\bf 1},{\bf 1},{\bf 1})+4 \times ({\bf 2},{\bf 2},{\bf 1},{\bf 1})+4 \times ({\bf 1},{\bf 2},{\bf 2},{\bf 1})+4 \times ({\bf 2},{\bf 1},{\bf 2},{\bf 1}) \\
&\qquad +2 \times ({\bf 3},{\bf 1},{\bf 1},{\bf 1})+2 \times ({\bf 1},{\bf 3},{\bf 1},{\bf 1})+2 \times ({\bf 1},{\bf 1},{\bf 3},{\bf 1}) \\
&\qquad +({\bf 3},{\bf 3},{\bf 1},{\bf 1})+({\bf 1},{\bf 3},{\bf 3},{\bf 1})+({\bf 3},{\bf 1},{\bf 3},{\bf 1}) \\
&\qquad +2 \times ({\bf 3},{\bf 2},{\bf 2},{\bf 1})+2 \times ({\bf 2},{\bf 3},{\bf 2},{\bf 1})+2 \times ({\bf 2},{\bf 2},{\bf 3},{\bf 1}) \,, \\
\sigma_1 \Sigma_1 \quad &\rightarrow \quad 4 \times ({\bf 1},{\bf 1},{\bf 1},{\bf 1})+4 \times ({\bf 2},{\bf 2},{\bf 1},{\bf 1})+4 \times ({\bf 1},{\bf 2},{\bf 2},{\bf 1})+4 \times ({\bf 2},{\bf 1},{\bf 2},{\bf 1}) \\
&\qquad +2 \times ({\bf 3},{\bf 1},{\bf 1},{\bf 1})+2 \times ({\bf 1},{\bf 3},{\bf 1},{\bf 1})+2 \times ({\bf 1},{\bf 1},{\bf 3},{\bf 1}) \\
&\qquad +({\bf 3},{\bf 3},{\bf 1},{\bf 1})+({\bf 1},{\bf 3},{\bf 3},{\bf 1})+({\bf 3},{\bf 1},{\bf 3},{\bf 1}) \\
&\qquad +2 \times ({\bf 3},{\bf 2},{\bf 2},{\bf 1})+2 \times ({\bf 2},{\bf 3},{\bf 2},{\bf 1})+2 \times ({\bf 2},{\bf 2},{\bf 3},{\bf 1}) \,, \\ 
\xi \quad &\rightarrow \quad ({\bf 1},{\bf 1},{\bf 1},{\bf 1}) \,, \\
\pi \quad &\rightarrow \quad ({\bf 1},{\bf 1},{\bf 1},{\bf 1})+({\bf 1},{\bf 1},{\bf 1},{\bf 3}) \,, \\
\mu \quad &\rightarrow \quad 3 \times ({\bf 1},{\bf 1},{\bf 1},{\bf 1})+2 \times ({\bf 2},{\bf 2},{\bf 1},{\bf 1})+2 \times ({\bf 1},{\bf 2},{\bf 2},{\bf 1})+2 \times ({\bf 2},{\bf 1},{\bf 2},{\bf 1}) \\
&\qquad +({\bf 3},{\bf 1},{\bf 1},{\bf 1})+({\bf 1},{\bf 3},{\bf 1},{\bf 1})+({\bf 1},{\bf 1},{\bf 3},{\bf 1})
\end{aligned}
\end{align}
for fermionic ones. $M_0 \Pi_1$ and $M_1 \Pi_0$ are not all independent but satisfy 40 relations
\begin{align}
\label{eq:SO(8) relations}
2 \times ({\bf 2},{\bf 1},{\bf 1},{\bf 2})+2 \times ({\bf 1},{\bf 2},{\bf 1},{\bf 2})+2 \times ({\bf 1},{\bf 1},{\bf 2},{\bf 2})+({\bf 2},{\bf 2},{\bf 2},{\bf 2}) \,.
\end{align}
In appendix \ref{appPL} we check this looking at the plethystic logarithm of the superconformal index.

Comparing \eqref{eq:SO(8) bosons} and \eqref{eq:SO(8) fermions}, we find that there are exactly 24 bosonic operators in the representation $2 \times ({\bf 2},{\bf 2},{\bf 1},{\bf 1})+2 \times ({\bf 1},{\bf 2},{\bf 2},{\bf 1})+2 \times ({\bf 2},{\bf 1},{\bf 2},{\bf 1})+2 \times ({\bf 2},{\bf 1},{\bf 1},{\bf 2})+2 \times ({\bf 1},{\bf 2},{\bf 1},{\bf 2})+2 \times ({\bf 1},{\bf 1},{\bf 2},{\bf 2})$ that cannot be paired up with any of the fermionic operators in \eqref{eq:SO(8) fermions}. Moreover, there are 13 bosonic $({\bf 1},{\bf 1},{\bf 1},{\bf 1})$ and 13 fermionic $({\bf 1},{\bf 1},{\bf 1},{\bf 1})$, among which we expect at least two pairs of bosons and fermions do not combine and remain short. This is because we already have two $U(1)$ symmetries in the UV which do not belong to the enhanced $SO(8)$ symmetry in the IR. Thus, we expect at least two U(1) symmetries in the IR, which requires two fermionic $({\bf 1},{\bf 1},{\bf 1},{\bf 1})$ to remain as short multiplets. Hence, there are 26 marginal operators in the representation $2 \times ({\bf 1},{\bf 1},{\bf 1},{\bf 1})+2 \times ({\bf 2},{\bf 2},{\bf 1},{\bf 1})+2 \times ({\bf 1},{\bf 2},{\bf 2},{\bf 1})+2 \times ({\bf 2},{\bf 1},{\bf 2},{\bf 1})+2 \times ({\bf 2},{\bf 1},{\bf 1},{\bf 2})+2 \times ({\bf 1},{\bf 2},{\bf 1},{\bf 2})+2 \times ({\bf 1},{\bf 1},{\bf 2},{\bf 2})$, which is twice $\mathbf{8}_s+\mathbf{8}_v+\mathbf{8}_c+1$ of $SO(8)$. The remaining fermionic operators constitute the supersymmetric partners of the conserved current in the adjoint representation of $SO(8) \times U(1)_v\times U(1)_a$.
\\

We conclude considering the other model  exhibiting the $SO(8)\times U(1)^4$ IR global symmetry. Indeed, similarly to what happens in the rank one case of section \ref{sec:T0}, the operators $\Tr_g \Tr_1 \left(A^i Q_1 Q_1\right)$ in the representation $(1,1,1,1)_{2,2,2}$ of $SU(2)^4 \times U(1)_t\times U(1)_u\times U(1)_t\subset SU(8)$ are spectators from the point of view of the $SO(8)$ enhancement. Hence, we expect that theory $\widehat{\mathcal{T}}_0^N$  including the same singlets \eqref{higherrankSO8sing} that define theory $\widehat{\mathcal{T}}_1^N$ except the $b_1^i$ fields  will have the same enhancement.
More precisely, the manifest UV symmetry
\begin{align}
SU(2)_1 \times SU(2)_2 \times SU(2)_3 \times SU(2)_4 \times U(1)_t \times U(1)_u \times U(1)_v \times U(1)_a \,,
\end{align}
is supposed to be enhanced in the IR to
\begin{align}
SO(8) \times U(1)_t \times U(1)_u \times U(1)_v \times U(1)_a\,,
\end{align}
where $U(1)_{t,u,v}$ are again defined as in \eqref{3u1m}-\eqref{3u1mfug}.

This enhancement of the global symmetry can be checked using the superconformal index for low values of the rank of the gauge group. For instance, for $N=2$ we find the following approximate values of the mixing coefficients from $a$-maximization:
\begin{align}
R_1 = R_2=R_3 \simeq \frac{17}{16} \,, \quad R_a \simeq \frac{2}{7} \,.
\end{align}
The superconformal index of the theory computed with these R-charges then reads\footnote{Since none of the $U(1)$ symmetries participates in the enhancement it is equivalent to parametrize them with $U(1)_{x_1,x_2,x_3,a}$ or $U(1)_{t,u,v,a}$ when computing the index. We decide to use the latter parametrization. }
\begin{align}
\widehat{\mathcal{I}}_0^{N=2} &=1+({\bf 8}_vv^{-2}u^{-2}a^{-\frac{1}{2}}+{\bf 8}_sv^{-2}u\,t^{-1}a^{-\frac{1}{2}}+{\bf 8}_cv^{-2}u\,t\,a^{-\frac{1}{2}})(pq)^{\frac{89}{224}}+(v^2u^{-4}a^{-\frac{1}{2}}+\nn\\
&+v^2u^2t^{-2}a^{-\frac{1}{2}}+v^2u^2t^2a^{-\frac{1}{2}})(pq)^{\frac{103}{224}}+({\bf 8}_vv^{-2}u^{-2}a^{\frac{1}{2}}+{\bf 8}_sv^{-2}u\,t^{-1}a^{\frac{1}{2}}+\nn\\
&+{\bf 8}_cv^{-2}u\,t\,a^{\frac{1}{2}})(pq)^{\frac{121}{224}}+(v^2u^{-4}a^{\frac{1}{2}}+v^2u^2t^{-2}a^{\frac{1}{2}}+v^2u^2t^2a^{\frac{1}{2}})(pq)^{\frac{135}{224}}+\nn\\
&+\cdots+[{\bf 8}_v(2u^{-6}+t^2+t^{-2})+{\bf 8}_s(2u^3t^{-3}+u^3t+u^{-3}t^{-1})+\nn\\
&+{\bf 8}_c(2u^3t^3+u^{-3}t+u^3t^{-1}){\color{blue}-{\bf 28}-4}]pq+\cdots\,.
\end{align}
Each number is the character of an $SO(8)$ representation and, in particular, the term $-({\bf 28}+4)pq$ highlighted in blue reflects the current multiplet, which is in the adjoint representation  of $SO(8)\times U(1)_t\times U(1)_u \times U(1)_v\times U(1)_a$.

\section*{Acknowledgements}

We would like to thank S.~Benvenuti, I.~Garozzo, H.-C.~Kim, N.~Mekareeya, K.-H.~Lee, J.~Song, P.~Yi for useful discussions and   S.~Razamat for discussions and collaboration on related topics. C.H. is grateful to KIAS and APCTP for the kind hospitality where part of this work was done. C.H. is partially supported by the ERC-STG grant 637844-HBQFTNCER, by the STFC consolidated grants ST/P000681/1, ST/T000694/1 and by the INFN. S.P. is partially supported by the ERC-STG grant 637844-HBQFTNCER and  by  the  INFN.   M.S. is partially supported by the ERC-STG grant 637844-HBQFTNCER, by the University of Milano-Bicocca grant 2016-ATESP0586, by the MIUR-PRIN contract 2017CC72MK003 and by the INFN.\\ 

\appendix

\section{Dualities for $USp(2N)$ with one antisymmetric and 8 fundamental chirals}
\label{app1}

In this appendix we briefly review the dualities enjoyed by the $USp(2N)$ gauge theory with antisymmetric and 8 fundamental chirals, focusing in particular on their action on the global symmetries.

We consider the theory without the addition of any flipping fields and we denote by $A$ the antisymmetric chiral and by $Q_a$ the fundamental chirals, where $a=1,\ldots,8$. In this case there is no superpotential
\be
\mathcal{W}=0\,.
\ee
The non-anomalous global symmetry of the theory is $SU(8)_v\times U(1)_x$ and the way the matter fields transform under them can be summarized with the following vector of fugacities:
\be
&&\vec{u}=\left.\{u_1,u_2,u_3,u_4,u_5,u_6,u_7,u_8;x\right.\}=\nn\\
&&\qquad\qquad\qquad=\left\{(pq)^{\frac{1}{4}}x^{-\frac{N-1}{4}}v_1,(pq)^{\frac{1}{4}}x^{-\frac{N-1}{4}}v_2,(pq)^{\frac{1}{4}}x^{-\frac{N-1}{4}}v_3,(pq)^{\frac{1}{4}}x^{-\frac{N-1}{4}}v_4,\right.\nn\\
&&\qquad\qquad\qquad\left.(pq)^{\frac{1}{4}}x^{-\frac{N-1}{4}}v_5,(pq)^{\frac{1}{4}}x^{-\frac{N-1}{4}}v_6,(pq)^{\frac{1}{4}}x^{-\frac{N-1}{4}}v_7,(pq)^{\frac{1}{4}}x^{-\frac{N-1}{4}}v_8;x\right\}\,,
\label{chargesvector}
\ee
where the first 8 entries correspond to the 8 fundamental chirals, while the last entry corresponds to the antisymmetric chiral. In this expression, the power of $pq$ denotes the half of the R-charge under a possible choice of trial non-anomalous R-symmetry, the power of $x$ denotes the charge under $U(1)_x$ and the powers of each $v_a$ denote the charges under the Cartan $\prod_{a=1}^7U(1)_{v_a}\subset SU(8)_v$, where $v_8$ can be determined solving the following constraint coming from requiring that the R-symmetry is not anomalous:
\be
\prod_{a=1}^8v_a=1\quad\Leftrightarrow\quad\prod_{a=1}^8u_a=(pq)^{2}x^{2-2N}\,.
\ee
This theory enjoys three different types of dualities \cite{Spiridonov:2008zr}, which are generalizations of the Intriligator--Pouliot \cite{Intriligator:1995ne}, Seiberg \cite{Seiberg:1994pq} and Csaki--Schmaltz--Skiba--Terning \cite{Csaki:1997cu} dualities for $N=1$.

The generalization of Intriligator--Pouliot duality first appeared in \cite{Csaki:1996eu}. The dual theory is still a $USp(2N)$ gauge theory with one antisymmetric chiral $\hat{A}$ and 8 fundamental chirals $q^a$, but in addition we have $28N$ gauge singlet chiral fields $M_{ab;i}$ with $a<b=1,\ldots,8$ and $i=1,\ldots,N$ interacting with the superpotential
\be
\hat{\mathcal{W}}=\sum_{i=1}^NM_{ab;i}\Tr_g\left(A^{i-1}q^aq^b\right)\,.
\ee
The action of the duality on the global symmetries can be easily expressed in terms of the fugacities we introduced in \eqref{chargesvector}
\be
u_a\to u_a^{-1}\prod_{b=1}^8u_b^\frac14=(pq)^\frac12 x^{\frac{1}{2}(1-N)}u_a^{-1}\,.
\label{uip}
\ee
Accordingly, we have the following operator map:
\be
\Tr_g\left(A^{i-1}Q_aQ_b\right)\quad&\leftrightarrow&\quad M_{ab;N-i}\nn\\
\Tr_gA^{i}\quad&\leftrightarrow&\quad\Tr_g\hat{A}^i\,.
\ee

The generalization of Seiberg duality breaks the manifest $SU(8)_v$ symmetry to the subgroup $SU(4)^2\times U(1)$ in the dual frame. Indeed, the dual theory is again a $USp(2N)$ gauge theory with one antisymmetric $\hat{A}$ and 8 fundamental chirals, but now the fundamentals are naturally divided into two groups of four that we denote by $q^a$ and $p^b$ with $a,b=1,\ldots,4$. This is because we also have additional $16N$ gauge singlets $M_{ab;i}$ with $a,b=1,\ldots,4$ and $i=1,\ldots,N$ interacting with the superpotential
\be
\hat{\mathcal{W}}=\sum_{i=1}^NM_{ab;i}\Tr_g\left(A^{i-1}q^ap^b\right)\,.
\ee
The action of the duality on the global symmetries can again be expressed in terms of the fugacities we introduced in \eqref{chargesvector}. In order to do so, we have to make a choice on how to break $SU(8)_v$ to the subgroup $SU(4)^2\times U(1)$, which is equivalent to choosing how to split the 8 chirals $Q_a$ of the original theory into two groups of four. The most intuitive option is to split $Q_{1,2,3,4}$ from $Q_{5,6,7,8}$. With this choice, we have
\be
\begin{cases}
u_a\to u_+^2u_a^{-1} & a=1,2,3,4 \\
u_a\to u_-^2u_a^{-1}& a=5,6,7,8
\end{cases}
\label{us}
\ee
where we defined $u_+^4=\prod_{a=1}^4u_a$ and $u_-^4=\prod_{a=5}^8u_a$. Accordingly, we have the following operator map\footnote{Here and in the following we use that the two-index antisymmetric representation of $SU(4)$ is real to freely lower its indices. For example explicitly
\be
\Tr_g\left(\hat{A}^{i-1}q_aq_b\right)=\epsilon_{abcd}\Tr_g\left(\hat{A}^{i-1}q^cq^d\right)\,.
\ee}
\be
\Tr_g\left(A^{i-1}Q_aQ_{b+4}\right)\quad&\leftrightarrow&\quad M_{ab;N-i}\nn\\
\Tr_g\left(A^{i-1}Q_aQ_b\right)\quad&\leftrightarrow&\quad \Tr_g\left(\hat{A}^{i-1}q_aq_b\right)\nn\\
\Tr_g\left(A^{i-1}Q_{a+4}Q_{b+4}\right)\quad&\leftrightarrow&\quad \Tr_g\left(\hat{A}^{i-1}p_ap_b\right)\nn\\
\Tr_gA^{i}\quad&\leftrightarrow&\quad\Tr_g\hat{A}^i\,.
\ee
Clearly, this is not the unique choice for splitting 8 chirals into two groups of four. In total we have $\frac{1}{2}{{8}\choose{4}}=35$ different possibilities that will give rise to inequivalent dual frames.

Finally, the generalization of the Csaki--Schmaltz--Skiba--Terning duality also breaks the manifest $SU(8)_v$ symmetry to the subgroup $SU(4)^2\times U(1)$ in the dual frame.
Indeed, the dual theory is once more a $USp(2N)$ gauge theory with one antisymmetric $\hat{A}$ and 8 fundamental chirals, where the fundamentals are naturally divided into two groups of four that we denote by $q_a$ and $p_b$ with $a,b=1,\ldots,4$. This time this is due to the presence of additional $12N$ gauge singlets $\mu^{ab}_i$, $\nu^{ab}_i$ with $a,b=1,\ldots,4$ and $i=1,\ldots,N$ interacting with the superpotential
\be
\hat{\mathcal{W}}=\sum_{i=1}^N\mu^{ab}_i\Tr_g\left(\hat{A}^{i-1}q_aq_b\right)+\nu^{ab}_i\Tr_g\left(\hat{A}^{i-1}p_ap_b\right)\,.
\ee
Also in this case in order to express the action of the duality on the global symmetries in terms of the fugacities of \eqref{chargesvector} we have decide how to split the 8 chirals $Q_a$ of the original frame into two groups of four. Using the most natural decomposition into $Q_{1,2,3,4}$ and $Q_{5,6,7,8}$ we have the transformation
\be
\begin{cases}
u_a\to u_-u_+^{-1}u_a & a=1,2,3,4 \\
u_a\to u_+u_-^{-1}u_a& a=5,6,7,8
\end{cases}
\label{ucsst}
\ee
where recall that we defined $u_+^4=\prod_{a=1}^4u_a$ and $u_-^4=\prod_{a=5}^8u_a$. Accordingly, we have the following operator map:
\be
\Tr_g\left(A^{i-1}Q_aQ_{b+4}\right)\quad&\leftrightarrow&\quad \Tr_g\left(\hat{A}^{i-1}q_ap_b\right)\nn\\
\Tr_g\left(A^{i-1}Q_aQ_b\right)\quad&\leftrightarrow&\quad \mu_{ab;N-i}\nn\\
\Tr_g\left(A^{i-1}Q_{a+4}Q_{b+4}\right)\quad&\leftrightarrow&\quad \nu_{ab;N-i}\nn\\
\Tr_gA^{i}\quad&\leftrightarrow&\quad\Tr_g\hat{A}^i\,.
\ee
Also this duality can be applied in 35 independent ways, corresponding to all the possible splittings of 8 elements into two groups of four. Hence, the $USp(2N)$ gauge theory with one antisymmetric chiral and 8 fundamental chirals possesses 72 inequivalent dual frames, including the original one.\\

\section{$SO(10)\times U(1)^2$ enhancement of $FE[USp(4)]$ and chiral ring stability}\label{feusp}

\begin{figure}[t]
	\centering
	\makebox[\linewidth][c]{
  	\includegraphics[scale=1.5]{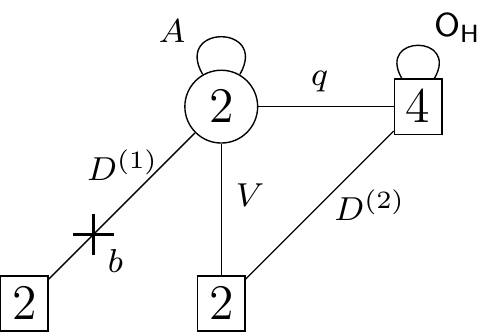} }
  	\caption{Quiver diagram for the $FE[USp(4)]$ theory. The arc on the $SU(2)$ gauge node represents the field $A$ in the antisymmetric representation of the gauge symmetry, i.e. a gauge singlet, while the arc on the $USp(4)$ flavor node represents the field $\mathsf{O_H}$ in the traceless antisymmetric representation of the flavor symmetry.}
  	\label{feusp4}
\end{figure}

In this appendix we revisit the $FE[USp(4)]$ theory part of the family of $FE[USp(2 N)]$ theories discussed in
 \cite{Pasquetti:2019hxf}.\footnote{See also \cite{Garozzo:2020pmz,Hwang:2021xyw} for discussions of the more general $FE[USp(2N)]$ theory and applications in models with global symmetry enhancements.} 
The  $FE[USp(2 N)]$ theories where shown to enjoy $USp(2 N)\times USp(2 N)\times U(1)^2$ global symmetry with
one of the $USp(2 N)$ factor emerging in the IR from $SU(2)^n$.
 As we will see, for $n=2$ we have a further enhancement  with $USp(4) \times USp(4)$ recombining into $SO(10)$.

%

$FE[USp(4)]$ is an $\mathcal{N}=1$ $SU(2)$ gauge theory with 8 fundamental chiral fields and 15 gauge singlets, so it belongs to the class of theories studied in the main text. Following similar conventions to \cite{Hwang:2020wpd}, we split the fundamental chirals in three groups that we denote by $D^{(1)}_\ga$, $V_\ga$ and $q_a$ and the gauge singlets in other four groups that we denote by $b$, $D^{(2)}_{\ga a}$, $A$ and $\mathsf{O}_{\mathsf{H},[ab]}$, with $\ga=1,2$ and $a=1,\cdots,4$ and with $\mathsf{O_H}$ such that $\Tr_4\mathsf{O_H}=0$. The superpotential is
%
\be
\mathcal{W}&=&A\Tr_g\Tr_x\left(q\,q\right)+\Tr_g\Tr_x\left(\mathsf{O_H}\,q\,q\right)+\Tr_g\Tr_x \Tr_{y_2}\left(Vq\,D^{(2)}\right)+b\Tr_g \Tr_{y_1} \left(D^{(1)}D^{(1)}\right)\,,\nn\\
\label{superpotfeusp4}
\ee
where $\Tr_x$ and $\Tr_{y_i}$ are the traces over the $USp(4)_x$ and $SU(2)_{y_i}$ flavor indices respectively.
The matter content is summarized in the quiver diagram of Figure \ref{feusp4}.

The non-anomalous manifest global symmetry is
\be
USp(4)_x\times\prod_{n=1}^2SU(2)_{y_i}\times U(1)_c\times U(1)_t\, .
\ee
In \cite{Pasquetti:2019hxf} it was argued that the symmetries $\prod_{n=1}^2SU(2)_{y_i}$ of the saw are enhanced to a second $USp(4)_y$ symmetry in the IR.

\begin{table}[t]
\centering
\scalebox{1}{
\begin{tabular}{c|ccccc|c}
{} & $USp(4)_x$ & $SU(2)_{y_1}$ & $SU(2)_{y_2}$ & $U(1)_c$ & $U(1)_t$ & $U(1)_{R_0}$ \\ \hline
$q$ & $\bf 4$ & $\bullet$ & $\bullet$ & $0$ & $\frac{1}{2}$ & $0$ \\
$V$ & $\bullet$ & $\bullet$ & $\bf 2$ & $-1$ & $-\frac{1}{2}$ & $2$ \\
$D^{(1)}$ & $\bullet$ & $\bf 2$ & $\bullet$ & $1$ & $-\frac{1}{2}$ & $0$ \\
$D^{(2)}$ & $\bf 4$ & $\bullet$ & $\bf 2$ & $1$ & $0$ & $0$ \\
$b$ & $\bullet$ & $\bullet$ & $\bullet$ & $-2$ & $1$ & $2$ \\
$A$ & $\bullet$ & $\bullet$ & $\bullet$ & $0$ & $-1$ & $2$ \\
$\mathsf{O_H}$ & $\bf 5$ & $\bullet$ & $\bullet$ & $0$ & $-1$ & $2$
\end{tabular}}
\caption{Transformation rules under the global symmetries and the trial R-symmetry $U(1)_{R_0}$ of all the chiral fields of the theory.}
\label{chargeschirals}
\end{table}

In order to detect the enhancement we can compute the superconformal index of $FE[USp(4)]$. This can be done using the assignment of charges for the chiral fields summarized in Table \ref{chargeschirals}. With this parametrization, we get from $a$-maximization the following values for the mixing coefficients of $U(1)_c$ and $U(1)_t$ with $U(1)_{R_0}$:
\be
R_c\simeq 0.912487,\quad R_t\simeq 1.12931
\ee
In computing the index we approximate these mixing coefficients with $R_c\simeq\frac{91}{100}$ and $R_t\simeq\frac{113}{100}$. In the result we immediately recognize characters of $SO(10)$ at each order
\be
\mathcal{I}&=&1+{\bf 16}\,c(pq)^{\frac{91}{200}}+{\bf 10}\,t^{-1}(pq)^{\frac{87}{200}}+t\,c^{-2}(pq)^{\frac{131}{200}}
-{\bf{47}}pq +t^{-1}c^{-2}(pq)^{\frac{21}{20}}+\cdots
\ee
We can also see at order $pq$ the contribution of $SO(10)$ conserved current.
Indeed  we find $-{\bf 45}-2$, which is the character of the  adjoint representation of $SO(10)\times U(1)_c\times U(1)_t$.

This enhancement of the global symmetry is quite peculiar, as $USp(4)^2$ is a maximal subgroup of $SO(10)$ but its rank is lower than the one of $SO(10)$. 
This curious  behaviour can be explained as follows.
It is useful to work in a basis of fields that makes manifest the $SU(2)^2$ subgroup of $USp(4)$, similarly to what we did in section \ref{sec3}. Specifically, we split the four fundamental chirals $q$ into two groups of two that we denote by $Q_1$ and $Q_2$. In these new conventions, the first two terms in the superpotential \eqref{superpotfeusp4} read
\be
\mathcal{W}&\supset& A(Q_1^2-Q_2^2)+\mathsf{O}_{\mathsf{H},12}Q_1^2+\mathsf{O}_{\mathsf{H},13}(Q_1Q_2)_{22}+\mathsf{O}_{\mathsf{H},14}(Q_1Q_2)_{21}+\nn\\
&+&\mathsf{O}_{\mathsf{H},23}(Q_1Q_2)_{12}+\mathsf{O}_{\mathsf{H},24}(Q_1Q_2)_{11}
\label{superpotunstable}
\ee
where we wrote explicitly all the traces $\Tr_x$, $\Tr_{y_1}$, $\Tr_{y_2}$  over the $USp(4)$, $SU(2)_{y_1}$, $SU(2)_{y_2}$ flavor indices and we are omitting $\Tr_g$ for the contraction of color indices. From this expression, it is clear that this superpotential violates the chiral ring stability criterion \cite{Benvenuti:2017lle}. Suppose that we deform the theory by removing the first term $A\,Q_1^2$. The equation of motion of the field $\mathsf{O}_{\mathsf{H},12}$ set the operator $Q_1^2$ to zero, meaning that the operator $A\,Q_1^2$ that we removed vanishes in the chiral ring of the deformed theory. Hence, this term is unstable in the superpotential \eqref{superpotunstable} and should be dropped. Notice that this operation doesn't really modify the theory, since it can be realized with a trivial linear field redefinition
\be
\mathsf{O}_{\mathsf{H},12}\to \mathsf{O}_{\mathsf{H},12}-A\,.
\ee

It is easy to check that with such a modification we recover nothing but the $\mathcal{T}_2$ theory of section \ref{sec3.1.2} with the inclusion of the $b_4$ singlet\footnote{Remember from the discussion in section \ref{sec3.1.2} that this field is a spectator from the point of view of the $SO(10)$ enhancement.}, where the fundamentals $V$, $D^{(1)}$ are identified with $Q_3$, $Q_4$, the singlet $b$ is identified with $b_4$, the singlets $D^{(2)}$ are identified with $D_1$, $D_2$ and the singlets $\mathsf{O}_{\mathsf{H},13}$, $\mathsf{O}_{\mathsf{H},14}$, $\mathsf{O}_{\mathsf{H},23}$, $\mathsf{O}_{\mathsf{H},24}$ are identified with $D_3$ up to a linear transformation, while
 $\mathsf{O}_{\mathsf{H},12}$ and $A$ are identified with $b_1$ and $b_2$.

 Hence, after the stabilization the manifest UV symmetry is actually the $SU(4)\times SU(2)^2\times U(1)^3$ symmetry of $\mathcal{T}_2$, which we know is enhanced in the IR to $SO(10)\times U(1)^2$.\\

Furthermore, it has been discussed that the compactification of $FE[USp(2 N)]$ on a circle and its real mass deformations lead to many interesting 3d theories exhibiting similar properties of $FE[USp(2 N)]$ \cite{Pasquetti:2019hxf,Pasquetti:2019tix}. This is certainly true for $FE[USp(4)]$. For example, the direct reduction of $FE[USp(4)]$ gives rise to the 3d theory with the same matter contents and a monopole superpotential, which exhibits the same enhancement of the global symmetry into $SO(10) \times U(1)^2$. In addition, one can also take a subsequent real mass deformation, as explained in \cite{Pasquetti:2019hxf}, such that each $USp(2 n)$ factor is broken to $U(n)$. With a certain traceless condition imposed, the resulting theory is $FM[SU(2)]$ proposed in \cite{Pasquetti:2019tix}. While the manifest UV symmetry of $FM[SU(2)]$ is $SU(2) \times U(1)^3$, we have checked that its superconformal index exhibits the characters of $SO(6) \times U(1)^2$. Thus, we expect $FM[SU(2)]$ enjoys the symmetry enhancement:
\begin{align}
SU(2) \times U(1)^3 \quad \longrightarrow \quad SO(6) \times U(1)^2 \,.
\end{align}
Moreover, this $SO(6)$ enhancement is closely related to an example discussed in \cite{Benini:2018bhk}. Indeed, as explained in \cite{Pasquetti:2019tix}, one can further deform $FM[SU(2)]$ to obtain the theory called $FT[SU(2)]$ \cite{Aprile:2018oau}, which has one less $U(1)$. $FT[SU(2)]$ is basically the same theory as the model in \cite{Benini:2018bhk} exhibiting $SO(6)$ but with an unstable superpotential. After the stabilization of the superpotential as above, we obtain a $U(1)$ gauge theory with two flavors $(Q_\alpha,\tilde Q_\beta)$, four gauge singlets $\eta_{\alpha\beta}$ for $\alpha,\beta = 1,2$ and the superpotential
\begin{align}
W = \sum_{\alpha,\beta = 1}^2 \eta_{\alpha \beta} Q_\alpha \tilde Q_\beta \,,
\end{align}
which is exactly the model in \cite{Benini:2018bhk} showing the enhancement of the global symmetry:
\begin{align}
SU(2)^2 \times U(1)^2 \quad \longrightarrow \quad SO(6) \times U(1) \,.
\end{align}

\section{The plethystic logarithm and the representations of the relations}
\label{appPL}

In section \ref{hrt}, we have seen that higher rank theories mostly have marginal operators satisfying some relations. The correct identification of such relations is important to argue the existence of the independent marginal operators. In this appendix, we explain how to read the relations of the marginal operators by examining the superconformal index.
\\

We introduce the plethystic logarithm \cite{Benvenuti:2006qr}
\begin{align}
\mathrm{PL}[g(t)] = \sum_{k = 1}^\infty \frac{\mu(k)}{k} \log(g(t^k)) \,,
\end{align}
which is the inverse function of the plethystic exponential
\begin{align}
\mathrm{PE}[f(t)] = \exp\left[\sum_{k = 1}^\infty \frac1k f(t^k)\right] \,.
\end{align}
The coefficient $\mu(k)$ is the Möbius function defined by
\begin{align}
\mu(k) = \left\{\begin{array}{ll}
0 \,, & \qquad \text{$k$ has repeated prime factors,} \\
1 \,, & \qquad k = 1, \\
(-1)^n \,, & \qquad \text{$k$ is a product of $n$ distinct primes.}
\end{array}\right.
\end{align}
If we take the plethystic log of the superconformal index, it will give the generating function of the single trace operators, either bosonic and fermionic, as well as the relations among them, which reflect the interaction of the theory. For example, $N$ free chiral multiplets have the PL index
\begin{align}
\mathrm{PL}[I_\text{free}] = \frac{\sum_{i = 1}^N \left(a_i (pq)^\frac13-a_i^{-1} (pq)^\frac23\right)}{(1-p) (1-q)}
\end{align}
where $a_i$ is the fugacity for each $U(1)$ rotating each chiral multiplet. There will be extra terms if the theory is interacting. In addition, the power of $pq$ will be varied by the shift $a_i \rightarrow a_i (pq)^{\epsilon_i}$ where $\epsilon_i$ is determined by the interaction.

In fact, one should remember that there are not only the relations of bosonic and fermionic operators but also those of relations themselves. Here, however, we focus on the relations of bosonic operators, which then give negative contributions to the PL index. One should note that such relations of bosonic operators appear in the PL index in two different ways: one is realized by the negative contribution of a fermionic single trace operator while the other is the negative contribution corresponding to the absence of a bosonic operator in the original index. For example, let us consider an F-term condition coming from a superpotential
\begin{align}
W = \mathcal S \mathcal O
\end{align}
where $\mathcal S$ is a gauge singlet and $\mathcal O$ is some gauge invariant bosonic operator. The F-term condition demands that
\begin{align}
\mathcal O = 0 \,.
\end{align}
This relation for the bosonic operator is realized in the index as the contribution of the fermionic operator $\overline \Psi_{\mathcal S}$, which cancels the contribution of $\mathcal O$ such that it vanishes in the index.

On the other hand, the other type of relation does not come up with a fermionic operator. In order to understand this, let us consider the example of the $SU(2)$ gauge theory with 8 fundamental chirals and no superpotential that we reviewed in section \ref{sec2}. Recall that in this theory the chiral ring generators are the mesons $m_{0,ij} = \Tr_{g} \left(Q_{i} Q_{j}\right)$ which transform in the antisymmetric representation of the $SU(8)$ flavor symmetry. Out of them we can construct the marginal operators
\begin{align}
m_{0,ij} m_{0,kl}
\end{align}
which we explained are subject to the relation
\begin{align}
m_{0,[ij} m_{0,kl]} = 0 \,.
\end{align}
This is the kind of relation that is not realized by fermionic operators. Instead, the corresponding operators are absent in the superconformal index in the first place. Computing the PL of the unrefined superconformal index \eqref{indexpuresu2w8} of this theory we find
\be
\mathrm{PL}\left[\mathcal{I}\right]=28(pq)^{\frac{1}{2}}(1+p+q)-133pq+\cdots\,.
\label{PLsu2w8}
\ee
The term $28(pq)^{\frac{1}{2}}$ represents the chiral ring generators $m_{0,ij}$, while the term $-133pq$ corresponds to the sum of 70 relations $m_{0,[ij} m_{0,kl]} = 0$ and of the fermionic superpartners of the conserved current in the adjoint representation of $SU(8)$.
In order to distinguish the two types of relations, we can introduce a fictitious fugacity $F_i$ in the numerator of the 1-loop determinant of each matter multiplet. Then the contributions involving any matter fermion will come up with extra factor $\prod_i F_i^{n_i}$ with some power $n_i$. Such contributions can be either the independent fermionic operators or the fermionic operators corresponding to relations. On the other hand, we said that there are also the relations that are not realized by fermionic operators. The contributions of those relations do not include any factor of $F_i$.

If we go back to our example of the $SU(2)$ gauge theory with 8 fundamental chirals and turn on the same fictitious fugacity $F$ for all the fermions contained in the chirals, we obtain the following PL of the index
\be
\mathrm{PL}\left[\mathcal{I}\right]=28(pq)^{\frac{1}{2}}(1+p+q)-(69+64F)pq+\cdots\,,
\ee
which reduces to \eqref{PLsu2w8} in the limit $F\to 1$. The order $pq$ of this result should be actually understood as
\be
-(69+64F)pq=(1-70-64F)pq\,.
\ee
The positive term correspons indeed to the contribution of $\Tr_g \left(\lambda^2\right)$, which recombines with one of the 64 fermionic operators into a long multiplet in the true index that we get in the limit $F\to 1$. The remaining 63 fermionic operators correspond to the superparteners of the $SU(8)$ flavor current, while the 70 relations do not carry any power of $F$, meaning that they do not come from fermionic operators as we anticipated before.\\

Now let us apply this strategy to some of the models we considered in section \ref{hrt}. Consider for example the rank-2 $\widehat{\mathcal{T}}_2'{}^{N = 2}$ model in section \ref{sec:T2}. This has the manifest symmetry
\begin{align}
SU(2)_1 \times USp(4) \times SU(2)_4 \times U(1)^3 \,,
\end{align}
which is enhanced to 
\begin{align}
SO(9) \times U(1)^3
\end{align}
in the IR. In order to prove the symmetry enhancement to $SO(9) \times U(1)^3$, it was important to argue the existence of the marginal operators in the representation $\mathbf{9}$ of $SO(9)$. Those independent marginal operators can be found by constructing the candidate marginal operators and their relations, which are either realized by fermionic operators or not. In particular, we have claimed in \eqref{eq:SO(9) relations} that the relations not realized by fermionic operators are in the representation
\begin{align}
({\bf 2},{\bf 1},{\bf 2})+({\bf 2},{\bf 5},{\bf 2})
\end{align}
of $SU(2)_1 \times USp(4) \times SU(2)_4$. Here we explain how to read this representation of the relations from the PL index.

Turning on a fugacity $F_i = F$ for each matter fermion, the plethystic log of the superconformal index is given by
\begin{align}
& (1-p)(1-q) \, \mathrm{PL}\left[\widehat{\mathcal{I}}_2'{}^{N=2}\right] \nonumber \\
&= (1-F) \, a^2 (p q)^{3/11}+(1-F) \, v^{-6} a^{-1/2} (p q)^{52/165}+u^{-8} v^{-2} a^{-1/2} (p q)^{21/55} \nonumber \\
&\quad +16 \, u^{-2} v^{-2} a^{-1/2} (p q)^{103/264}+9 \, u^4 v^{-2} a^{-1/2} (p q)^{263/660}+(1-F) \, v^{-6} a^{1/2} (p q)^{149/330} \nonumber \\
&\quad +(6-5 F) \, u^{-4} v^2 a^{-1/2} (p q)^{307/660}+(8-8 F) \, u^2 v^2 a^{-1/2} (p q)^{125/264}+(1-F) \, u^8 v^2 a^{-1/2} (p q)^{53/110} \nonumber \\
&\quad +(26-26 F) \, (p q)^{1/2}+u^{-8} v^{-2} a^{1/2} (p q)^{57/110}+16 \, u^{-2} v^{-2} a^{1/2} (p q)^{139/264}+9 \, u^4 v^{-2} a^{1/2} (p q)^{353/660} \nonumber \\
&\quad +v^6 a^{-1/2} (p q)^{181/330}+(6-5 F) \, u^{-4} v^{2} a^{1/2} (p q)^{397/660}+(8-8 F) \, u^2 v^2 a^{1/2} (p q)^{161/264} \nonumber \\
&\quad +(1-F) \, u^8 v^2 a^{1/2} (p q)^{34/55}+v^6 a^{1/2} (p q)^{113/165}+a^{-2} (p q)^{8/11}+(-6-4 F) \, u^{-4} v^{-4} (p q)^{11/12} \nonumber \\
&\quad +(-8-8 F) \, u^2 v^{-4} (p q)^{37/40}-u^8 v^{-4} (p q)^{14/15}+(-8-8 F) \, u^{-6} (p q)^{119/120}+(-23-25 F) \, pq \nonumber \\
&\quad +\mathcal O((p q)^{69/11})\,,
\end{align}
where we have turned off the fugacities for $SU(2)_1 \times USp(4) \times SU(2)_4$ for simplicity. If they are turned on, all the numeric coefficients are written as the characters of $SU(2)_1 \times USp(4) \times SU(2)_4$, which are enhanced to those of $SO(9)$ in the $F \rightarrow 1$ limit. For the first several terms, one can read bosonic single trace operators as well as their relations, especially the F-term conditions, realized by fermionic operators. For example,  the first term $(1-F) \, a^2 r^{18/11}$ indicates the operator
\begin{align}
\Tr_g \left(A^2\right)
\end{align}
and its F-term relation
\begin{align}
\Tr_g \left(A^2\right) = 0
\end{align}
due to the superpotential term $a_2 \, \Tr_g \left(A^2\right)$.

The term of our interest here is $-(23+25 F) \, pq$. As we mentioned, the refined version of this term is written in terms of the characters of $SU(2)_1 \times USp(4) \times SU(2)_4$:
\begin{align}
\left(\chi_{({\bf 1},{\bf 1},{\bf 1})}-\chi_{({\bf 2},{\bf 1},{\bf 2})}-\chi_{({\bf 2},{\bf 5},{\bf 2})}-F \left(4 \, \chi_{({\bf 1},{\bf 1},{\bf 1})}+\chi_{({\bf 3},{\bf 1},{\bf 1})}+\chi_{({\bf 1},{\bf 1},{\bf 3})}+\chi_{({\bf 1},{\bf 5},{\bf 1})}+\chi_{({\bf 1},{\bf 10},{\bf 1})}\right)\right) pq \,,
\end{align}
which indicates that there is a single trace marginal operator, 24 relations of marginal operators that are \emph{not realized by fermionic operators} and 25 fermionic single trace operators which are either relations or independent fermionic operators. Note that those independent fermionic operators belong to the current multiplet.

Indeed, the single trace marginal operator is $\Tr_g \left(\lambda^2\right)$ in \eqref{eq:SO(9) marginal} and 24 relations are those satisfied by $M_0 \Pi_1, \, M_1 \Pi_0, \, L_0 P_1$ and $L_1 P_0$ shown in \eqref{eq:SO(9) relations}. On the other hand, 25 fermionic single trace operators correspond to $\xi, \, \pi, \, \rho$ and $\mu$ in \eqref{eq:SO(9) fermions}. A flavor singlet part $(1,1,1)$ among $\xi, \, \pi, \, \rho, \, \mu$ gives rise to the relation
\begin{align}
\Tr_g \left(\lambda^2\right) = 0\,,
\end{align}
whereas the traceless antisymmetric part of $\mu$ gives the relation in the representation
\begin{align}
({\bf 1},{\bf 5},{\bf 1})
\end{align}
for a combination of $\Tr_{USp(4)} \left(M_0\right) \Sigma_0$ and $\Tr_{USp(4)} \left(M_1\right) \Sigma_1$. The others then give the independent fermionic operators belonging to the conserved current multiplet; together with a combination of $\sigma_0 \Pi_1, \, \sigma_1 \Pi_0, \tau_0 P_0$ and $\tau_1 P_1$ in $({\bf 2},{\bf 5},{\bf 2})$, those remaining operators of $\xi, \, \pi, \, \rho$ and $\mu$ give 39 fermionic operators in the representation $\mathbf{36}+3 \times \mathbf 1$ of $SO(9)$, which are the supersymmetric partners of the conserved currents of $SO(9) \times U(1)^3$.
\\

In similar ways, one can also obtain the PL indices of  $\widehat{\mathcal{T}}_2''{}^{N = 2}$,  $\widehat{\mathcal{T}}_1'{}^{N = 2}$ and  $\widehat{\mathcal{T}}_1''{}^{N = 2}$ in the presence of a fictitious fugacity $F$.
\begin{itemize}
\item theory $\widehat{\mathcal{T}}_2''{}^{N = 2}$
\begin{align}
&(1-p) (1-q) \, \mathrm{PL}\left[\widehat{\mathcal{I}}_2''{}^{N=2}\right] =\nonumber \\
&= \dots+\left(\chi_{({\bf 1},{\bf 1})}-\chi_{({\bf 6},{\bf 2})}-\chi_{({\bf 14'},{\bf 2})}+F \left(3 \, \chi_{({\bf 1},{\bf 1})}+\chi_{({\bf 1},{\bf 3})}+\chi_{({\bf 14},{\bf 1})}+\chi_{({\bf 21},{\bf 1})}\right)\right) pq+\dots\,,
\end{align}
where $\chi_{({\bf m},{\bf n})}$ is the character of the $USp(6) \times SU(2)_4$ representation $({\bf m},{\bf n})$. One can find the relation \eqref{eq:F4 relations} in the representation
\begin{align}
({\bf 6},{\bf 2})+({\bf 14'},{\bf 1})
\end{align}
for the redundant marginal operators $M_0 \Pi_1$ and $M_1 \Pi_0$. In addition, there are 41 fermionic single trace operators corresponding to $\xi, \, \pi$ and $\mu$ in \eqref{eq:F4 fermions}.
\item theory $\widehat{\mathcal{T}}_1'{}^{N = 2}$
\begin{align}
&(1-p) (1-q) \, \mathrm{PL}\left[\widehat{\mathcal{I}}_1'{}^{N=2}\right] =\nonumber \\
&= \dots+\left(\chi_{({\bf 1},{\bf 1},{\bf 1})}-\chi_{({\bf 1},{\bf 2},{\bf 2})}-\chi_{({\bf 5},{\bf 2},{\bf 2})}\right. \nonumber \\
&\quad \qquad \left.-F \left(4 \, \chi_{({\bf 1},{\bf 1},{\bf 1})}+\chi_{({\bf 1},{\bf 3},{\bf 1})}+\chi_{({\bf 1},{\bf 1},{\bf 3})}+\chi_{({\bf 5},{\bf 1},{\bf 1})}+\chi_{({\bf 10},{\bf 1},{\bf 1})}\right)\right) pq+\dots\,,
\end{align}
where $\chi_{({\bf l},{\bf m},{\bf n})}$ is the character of the $USp(4) \times SU(2)_3 \times SU(2)_4$ representation $({\bf l},{\bf m},{\bf n})$. One can find the relation in the representation
\begin{align}
({\bf 1},{\bf 2},{\bf 2})+({\bf 5},{\bf 2},{\bf 2})
\end{align}
for the redundant marginal operators $M_0 \Pi_1, \, M_1 \Pi_0, \, L_0 P_1$ and $L_1 P_0$ and the fermionic single trace operators $\xi, \, \pi, \, \rho$ and $\mu$ in the representation
\begin{align}
4 \times ({\bf 1},{\bf 1},{\bf 1})+({\bf 1},{\bf 3},{\bf 1})+({\bf 1},{\bf 1},{\bf 3})+({\bf 5},{\bf 1},{\bf 1})+({\bf 10},{\bf 1},{\bf 1}) \,.
\end{align}
\item theory $\widehat{\mathcal{T}}_1''{}^{N = 2}$
\begin{align}
&(1-p) (1-q) \, \mathrm{PL}\left[\widehat{\mathcal{I}}_1''{}^{N=2}\right] =\nonumber \\
&= \dots+\left(\chi_{({\bf 1},{\bf 1},{\bf 1},{\bf 1})}-2 \, \chi_{({\bf 2},{\bf 1},{\bf 1},{\bf 2})}-2 \, \chi_{({\bf 1},{\bf 2},{\bf 1},{\bf 2})}-2 \, \chi_{({\bf 1},{\bf 1},{\bf 2},{\bf 2})}-\chi_{({\bf 2},{\bf 2},{\bf 2},{\bf 2})}\right. \nonumber \\
&\quad \qquad \left.-F \left(5 \, \chi_{({\bf 1},{\bf 1},{\bf 1},{\bf 1})}+2 \, \chi_{({\bf 2},{\bf 2},{\bf 1},{\bf 1})}+2 \, \chi_{({\bf 1},{\bf 2},{\bf 2},{\bf 1})}+2 \, \chi_{({\bf 2},{\bf 1},{\bf 2},{\bf 1})}\right.\right. \nonumber \\
&\qquad \qquad \left.\left.+\chi_{({\bf 3},{\bf 1},{\bf 1},{\bf 1})}+\chi_{({\bf 1},{\bf 3},{\bf 1},{\bf 1})}+\chi_{({\bf 1},{\bf 1},{\bf 3},{\bf 1})}+\chi_{({\bf 1},{\bf 1},{\bf 1},{\bf 3})}\right)\right) pq+\dots\,,
\end{align}
where $\chi_{({\bf k},{\bf l},{\bf m},{\bf n})}$ is the character of the $SU(2)_1 \times SU(2)_2 \times SU(2)_3 \times SU(2)_4$ representation $({\bf k},{\bf l},{\bf m},{\bf n})$. One can find the relation \eqref{eq:SO(8) relations} in the representation
\begin{align}
2 \times ({\bf 2},{\bf 1},{\bf 1},{\bf 2})+2 \times ({\bf 1},{\bf 2},{\bf 1},{\bf 2})+2 \times ({\bf 1},{\bf 1},{\bf 2},{\bf 2})+({\bf 2},{\bf 2},{\bf 2},{\bf 2})
\end{align}
for the redundant marginal operators $M_0 \Pi_1$ and $M_1 \Pi_0$. There are also 41 fermionic single trace operators corresponding to $\xi, \, \pi$ and $\mu$ in \eqref{eq:SO(8) fermions}.
\end{itemize}

\bibliographystyle{ytphys}
\bibliography{refs}

\end{document}